\documentclass[
reprint,
superscriptaddress,
amsmath,
amssymb,
aps,
prxquantum,                        
twocolumn,
nofootinbib
]{revtex4-2}

\usepackage{graphicx}
\usepackage{dcolumn}
\usepackage{bm}
\usepackage{hyperref}
\usepackage{xcolor}
\usepackage{booktabs}
\usepackage{xspace}
\usepackage{subcaption}
\usepackage{tikz}
\usepackage{orcidlink}

\IfFileExists{quantikz2.sty}{\usepackage{quantikz2}}{\usepackage{quantikz}}
\usetikzlibrary{arrows.meta, positioning, shapes.geometric, fit, backgrounds}

\newcommand{\Zb}{\langle Z_\beta \rangle}
\newcommand{\Zth}{\langle Z_\beta \rangle_{\boldsymbol{\theta}}}
\newcommand{\Zda}{\langle Z_\beta \rangle_{\mathrm{data}}}
\newcommand{\dZ}{\delta\!\langle Z_\beta \rangle}
\newcommand{\dz}{\delta\bm{z}}
\newcommand{\Kpsck}{K_{\mathrm{PSCK}}}

\newcommand{\Lpsck}{\mathcal{L}_{\mathrm{PSCK}}}
\newcommand{\Lhe}{\mathcal{L}_{\mathrm{heat}}}
\newcommand{\Lrho}{\mathcal{L}_{\rho\text{-tan}}}
\newcommand{\rhog}{\rho_{fg}}
\newcommand{\Jstar}{\mathsf{J}}
\newcommand{\moiqp}{\textsc{MoIQP}\xspace}
\newcommand{\ciqp}{\textsc{cIQP}\xspace}
\newcommand{\psck}{\textsc{PSCK}\xspace}
\newcommand{\lwmmd}{\textsc{LW-MMD}\xspace}

\newcommand{\nfeat}{n}
\newcommand{\Dfeat}{D}
\newcommand{\Bbits}{B}
\newcommand{\Lcomp}{L}
\newcommand{\maerho}{\mathrm{MAE}_\rho}
\newcommand{\maez}{\mathrm{MAE}_z}
\newcommand{\rfit}{r_\rho}
\newcommand{\enfloor}[1]{\mathrm{MAE}_\rho^{\mathrm{enc},\,#1}}

\begin{document}
	
	\title{An IQP Born Machine for Calorimeter Image Generation at 64 Qubits with Compiled-IQP Deployment on Superconducting Hardware}
	
	\author{Jamal Slim\orcidlink{0000-0002-9418-8459}}
	\email{jamal.slim@desy.de}
	
	\affiliation{Deutsches Elektronen-Synchrotron DESY, 22603 Hamburg, Germany}
	
	\author{Saverio Monaco\orcidlink{0000-0001-8784-5011}}
	\affiliation{Deutsches Elektronen-Synchrotron DESY, 22603 Hamburg, Germany}
	\affiliation{RWTH Aachen University, 52062 Aachen, Germany}
	
	\author{Florian Rehm\orcidlink{0000-0002-8337-0239}}
	\affiliation{European Organization for Nuclear Research (CERN), 1211 Geneva, Switzerland}
	
	\author{Dirk Kr\"ucker\orcidlink{0000-0003-1610-8844}}
	\affiliation{Deutsches Elektronen-Synchrotron DESY, 22603 Hamburg, Germany}
	
	\author{Kerstin Borras\orcidlink{0000-0003-1111-249X}}
	\affiliation{Deutsches Elektronen-Synchrotron DESY, 22603 Hamburg, Germany}
	\affiliation{RWTH Aachen University, 52062 Aachen, Germany}

	\date{\today}

	%
	
	\begin{abstract}
		The challenge to scaling quantum generative models on near-term hardware is training. Variational circuit Born machines require repeated quantum sampling and are prone to barren plateaus. Instantaneous Quantum Polynomial-time (IQP) Born machines sidestep both, since their loss is built from low-order Pauli-$Z$ correlators that admit an unbiased classical estimator, while sampling from worst-case circuits in the class is conjectured to be classically hard. We take this train-on-classical, deploy-on-quantum workflow to a real high-energy-physics generative task, learning calorimeter shower profiles at $64$ qubits and running the trained model on an IBM Heron~r2 superconducting processor at $67$ physical qubits. Three ingredients make it work. A uniform mixture of IQP circuits (\moiqp{}) widens the model class at single-circuit training cost. The Pearson-Stabilized Correlation Kernel (\psck{}) biases descent toward the pairwise correlations that carry the shower-development physics, which the standard heat kernel systematically compresses. An exact deferred-measurement compilation collapses the mixture into a single IQP circuit, realized on hardware as a constant-depth dynamic circuit with zero SWAP insertions on the device's native heavy-hex graph. The trained model reconstructs the correlation structure to within $0.016$ of the floor imposed by the encoding itself. Raw device samples reproduce the per-cell energy spectra and the full pairwise correlation structure at $\rfit = 0.989$, up to a single global amplitude compression of depolarizing origin. A Gaussian copula fitted to the same training split matches the pairwise target more accurately than the quantum model at negligible cost. The contribution is therefore the classical trainability, exact compilation, and hardware deployability of a quantum generative model at this scale, not superiority over classical surrogates.
	\end{abstract}
	
	\maketitle

	\section{Introduction}
	
	Calorimeter shower simulation is one of the more demanding parts of the LHC analysis pipeline and is projected to consume millions of CPU-years per year during the high-luminosity run~\cite{calochallenge2024,atlas_atlfast3}. Classical generative surrogates have made a dent in this over the last five years. GANs~\cite{paganini2018calogan,chang2024qgan_clic}, normalizing flows~\cite{krause2023caloflow,buckley2024icaloflow,buss2025allshowers}, and diffusion models~\cite{mikuni2024caloscore_v2,favaro2025calodream} reach near-Geant4 fidelity with two to three orders of magnitude speed-up, and several are already in experiment software stacks. Quantum generative models have been proposed as a further alternative~\cite{delgado2022qcbm_hep,kiss2022conditional_born,chang2024qgan_clic,hoque2024caloqvae}, the motivation being that the Born-rule probability representation can express certain correlation structures with fewer parameters than a classical parameterization~\cite{coyle2020born_supremacy}.
	
	The obstacle to scaling quantum generative models on near-term hardware is that variational training of generic circuit Born machines requires repeated quantum sampling and is prone to barren plateaus~\cite{mcclean2018bp,cerezo2021costfn}. A useful way around this was opened by restricting the model class to \emph{instantaneous quantum polynomial-time} (IQP) circuits~\cite{shepherd2009iqp,bremner2011iqp_hardness,bremner2016bjs}. Two results make this class well suited to scalable generative modeling. Van den Nest~\cite{vandennest2010classical_sim} showed that expectation values of Pauli-$Z$ words at the output of an IQP circuit can be estimated classically in time linear in the number of qubits and gates by a Fourier Monte Carlo algorithm. Rudolph et al.~\cite{rudolph2024trainability} then observed that the MMD$^2$ loss with a Walsh-diagonal kernel decomposes as a mixture of such Pauli-$Z$ expectations. Together these give a fully classical training procedure for IQP Born machines. Sampling from the trained circuit, on the other hand, is believed to be classically intractable~\cite{bremner2011iqp_hardness,marshall2024iqp}, which is the point of the \emph{train on classical, deploy on quantum} workflow~\cite{recio2025train_classical_deploy_quantum}.
	
	That workflow has been explored recently on synthetic and biological data. Recio-Armengol et al.~\cite{recio2025train_classical_deploy_quantum} trained models up to $1000$ qubits in classical simulation. Lerch et al.~\cite{kumar2026data_dependent_init} trained $150$-qubit models on genomic data with a data-dependent initialization scheme. Ball\'o-Gimbernat et al.~\cite{ballo2025shallow_iqp_graph} ran trained shallow-IQP graph generators on superconducting hardware at up to $153$ qubits. In high-energy physics the situation is different. Prior IQP and generic quantum-circuit Born-machine work has stayed at 8 to 12 qubits, typically modeling joint distributions of two or three observables~\cite{delgado2022qcbm_hep,kiss2022conditional_born}. We are not aware of a prior demonstration that combines a real HEP shower-generation task with the roughly $10^2$-qubit scale required to encode a full longitudinal profile at realistic amplitude resolution, either in classical training or in hardware execution. This paper provides both.
	
	A second point worth flagging is that all existing IQP-MMD work uses the Liu and Wang heat kernel~\cite{liu2018differentiable}, a bandwidth-mixed Walsh-diagonal Gaussian that spreads its Fourier weight fairly evenly across low-order Pauli-$Z$ correlators. That uniform weighting is not ideal when the signal of physical interest is concentrated in pairwise correlation structure, as it is for calorimeter showers, where the correlation between energy depositions at different shower depths encodes the shower-development physics that downstream reconstruction and particle identification rely on. In our hands the Liu and Wang baseline saturates at a correlation reconstruction error about twice the encoding-fidelity floor and shows a systematic $\sim\!30\%$ amplitude compression on recovered Pearson correlations.
	
	\begin{figure*}[t]
		\centering
\begin{tikzpicture}[
    >=latex,
    every node/.style={font=\small},
    stagedata/.style={draw, rounded corners=2pt, fill=blue!10,
                      minimum width=2.2cm, minimum height=0.9cm, align=center},
    stageproc/.style={draw, rounded corners=2pt, fill=gray!10,
                      minimum width=2.2cm, minimum height=0.9cm, align=center},
    stagetrain/.style={draw, rounded corners=2pt, fill=red!12,
                       minimum width=2.4cm, minimum height=1.2cm, align=center},
    stagedeploy/.style={draw, rounded corners=2pt, fill=orange!14,
                        minimum width=2.4cm, minimum height=1.2cm, align=center},
    stagesample/.style={draw, rounded corners=2pt, fill=green!12,
                        minimum width=2.2cm, minimum height=0.9cm, align=center}]

\node[stagedata]   (raw)    at (0,   0)   {raw calorimeter\\$\mathbf{r}\in\mathbb{R}^{D}$};
\node[stageproc]   (enc)    at (3.1, 0)   {quantile\\binary encoding\\$\to \mathbf{x}\in\{0,1\}^{n}$};
\node[stagetrain]  (moiqp)  at (6.7, 0)   {\textbf{MoIQP}\\train on classical\\via PSCK-MMD\\$\to\boldsymbol\Theta^{\star}$};
\node[stagedeploy] (ciqp)   at (10.3, 0)  {\textbf{cIQP}\\compile to single IQP\\on $n{+}a$ qubits};
\node[stagesample] (sample) at (13.6, 0)  {sample on quantum\\device\\$\mathbf{x}\sim p^{\mathrm{MoIQP}}$};

\draw[->] (raw.east)   -- (enc.west);
\draw[->] (enc.east)   -- (moiqp.west);
\draw[->] (moiqp.east) -- (ciqp.west);
\draw[->] (ciqp.east)  -- (sample.west);

\draw[dashed, gray!60] (8.5, -1.0) -- (8.5, 1.0);
\node[font=\footnotesize\itshape, color=red!70!black]
     at (6.7, -1.4)  {fully classical training};
\node[font=\footnotesize\itshape, color=orange!70!black]
     at (12.0, -1.4) {quantum deployment};

\end{tikzpicture}
%
		\caption{Overall pipeline. Raw calorimeter shower data $\mathbf{r} \in \mathbb{R}^D$ is quantile-binned and binary-encoded to bitstrings $\mathbf{x} \in \{0,1\}^{n}$ with $n = D \cdot B$. The \moiqp{} model is trained on classical hardware through Van den Nest Fourier Monte Carlo~\cite{vandennest2010classical_sim} evaluation of its low-order Pauli-$Z$ correlators with the \psck{} MMD objective. After training, the $\Lcomp$ components are compiled into a single IQP circuit on $n + a$ qubits with $a = \lceil \log_2 \Lcomp \rceil$ Walsh--Hadamard ancillas (\ciqp{}). This compiled circuit is the object deployed on quantum hardware to generate calorimeter samples. The classical to quantum boundary separates training (Van den Nest MC) from deployment (sampling from the trained IQP circuit).}
		\label{fig:pipeline-overview}
	\end{figure*}
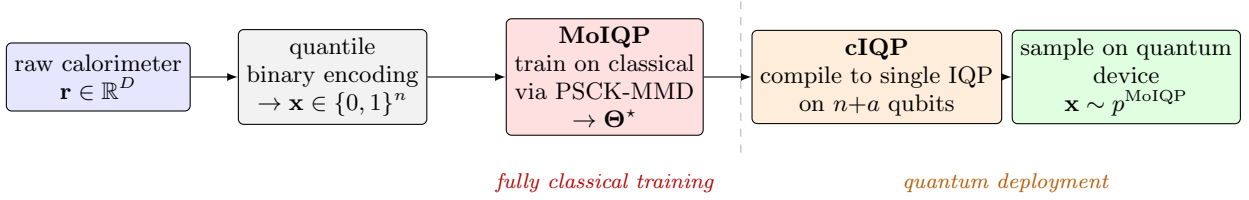
	
	This paper presents a method to train IQP Born machines on real HEP non-binary data at scale. Fig.~\ref{fig:pipeline-overview} summarizes the pipeline. Our method first proposes a Mixture-of-IQP (\moiqp{}) architecture in which $\Lcomp$ IQP components share a fixed Erd\H{o}s--R\'enyi gate graph but have independent trainable angles and are mixed uniformly at the output. Because mixture $Z$-correlators are linear in the per-component $Z$-correlators, the Van den Nest training machinery carries over with no changes. We also show that the \moiqp{} can be compiled exactly to a single IQP circuit on $\nfeat + \lceil \log_2 \Lcomp \rceil$ qubits through deferred measurement of a Walsh--Hadamard ancilla register. We call this compiled circuit \ciqp{}. The compiled-vs-mixture disagreement sits below the Monte Carlo noise floor across five independent training seeds at $\nfeat = 64$, consistent with exact agreement at the MC precision (the specific sub-unity ratio is a variance constant of the paired estimators, not a measure of residual error. Appendix~\ref{app:ciqp-derivation}).
	
	The second contribution is the Pearson-Stabilized Correlation Kernel,
	\begin{equation}
		\Kpsck \;=\; \mathrm{diag}(\boldsymbol\omega_{\mathrm{heat}})
		\;+\; \eta\, \Jstar^{\!\top}\Jstar,
		\label{eq:Kpsck-intro}
	\end{equation}
	where $\Jstar$ is the Jacobian of the empirical Pearson correlation matrix with respect to the model $Z$-marginals, evaluated at the data. It is a positive-semidefinite rank-$P$ correction on top of a positive-definite diagonal and is therefore positive-definite for any $\eta > 0$. Training reduces to a Gauss--Newton-style bias toward parameter directions that actually change the correlation matrix. The kernel change touches only the loss landscape, not the deployed circuit, so sampling hardness is untouched.
	
	The third addition is the empirical demonstration. We train on $47\,682$ CLIC-detector electron shower images encoded at $\Bbits = 8$ bits per cell, giving $\nfeat = 64$ qubits. Across five independent seeds at $\Lcomp = 8$ and $1500$ epochs, on a split-correct pipeline in which the quantile encoding and all training targets are fit on the $38\,146$-sample training split only, we measure $\maerho = 0.068 \pm 0.006$ on the training split against a $0.0515$ encoding-fidelity floor, and $0.070 \pm 0.006$ on the untouched $9\,536$-sample test split against a $0.0548$ floor (Table~\ref{tab:headline}). The Liu and Wang baseline at the same configuration plateaus at $\maerho = 0.095$. The aggregate train-test gap of $\approx 0.002$ sits below the encoding-floor train-test gap of $0.0033$, so there is no sign of overfitting beyond sample-statistical fluctuation between the two splits. Per-feature marginal distributions, recovered exactly through Walsh--Hadamard inversion of intra-feature $Z$-correlators and without sampling, split the eight calorimeter cells cleanly into five approximately Gaussian cells matched within a few integer levels and three heavy-tailed inner-shower cells where the residual mismatch is consistent with the absence of weight $\geq 3$ intra-feature observables from the training objective. A gradient scan across $\nfeat \in \{16, 24, 32, 48, 64\}$ shows per-gate gradient variance scaling as $\nfeat^{0.05}$ for \psck{} and $\nfeat^{1.87}$ for Liu and Wang. Both scalings are polynomial. Neither shows a barren plateau in the regime we study.
	
	The fourth addition is quantum deployment. Because the data-register marginal of \ciqp{} equals the classical mixture \emph{exactly} (Appendix~\ref{app:ciqp-derivation}), the trained model admits an exact constant-depth dynamic-circuit realization, dcIQP, in which the ancillas are measured first and classical feedforward selects the component (Sec.~\ref{sec:dciqp}). We retrain the model at the identical headline geometry on the native heavy-hex coupling graph of an IBM Heron~r2 processor, in a Walsh-sparse parameterization whose coherent compiled circuit is a bona fide \emph{degree-3} IQP circuit, and execute dcIQP on \texttt{ibm\_kingston} at $67$ physical qubits with $5\times10^{4}$ shots and zero-SWAP transpilation, evaluated split-correct end to end. The raw device samples reproduce the per-cell spectra (mean KS $0.087$), the total-energy distribution (KS $0.151$ against a $0.006$ encoding ceiling), and the full pairwise Pearson structure at $\rfit = 0.989$ with a single global amplitude compression of $0.647$, the depolarizing signature. The scaling of this protocol to the $10^{2}$-qubit regime on a finer-grained dataset is reported separately. The epistemic status of the hardware result is explicit. It is a faithfulness demonstration for large-scale quantum generative deployment.
	
	The paper is organized as follows. Sec.~\ref{sec:framework} recaps the IQP framework and the Van den Nest algorithm. Sec.~\ref{sec:moiqp} introduces \moiqp{} and the \ciqp{} compilation. Sec.~\ref{sec:psck} derives \psck{}. Sec.~\ref{sec:experiments} reports the calorimeter experiment and the trainability scan. Sec.~\ref{sec:hardware} reports the hardware deployment at the headline geometry on $67$ superconducting qubits. Sec.~\ref{sec:discussion} discusses scope and limitations. Appendices~\ref{app:ciqp-derivation}, \ref{app:variance-advantage} and~\ref{app:jacobian-derivation} give the \ciqp{} compilation, the base-angle variance advantage, and the Pearson-Jacobian derivation. Appendices~\ref{app:bp-supp} and~\ref{app:per-seed} give the supplementary numerics, and Appendix~\ref{app:hardware} the hardware-deployment details. Figure~\ref{fig:pipeline-overview} summarizes the full pipeline, from raw calorimeter data to quantum deployment.

	\section{Framework}
	\label{sec:framework}

	\subsection{IQP Born model}
	
\begin{figure}[t]
		\centering
\begin{tikzpicture}
\node[scale=0.95] {%
\begin{quantikz}[column sep=0.42cm, row sep=0.36cm]
\lstick{$\ket{0}_{1}$}       & \gate{H} & \gate[wires=5]{D(\boldsymbol{\theta})} & \gate{H} & \meter{} \\
\lstick{$\ket{0}_{2}$}       & \gate{H} &                                          & \gate{H} & \meter{} \\
\lstick{$\vdots$}            &          &                                          &          & \vdots   \\
\lstick{$\ket{0}_{n-1}$}     & \gate{H} &                                          & \gate{H} & \meter{} \\
\lstick{$\ket{0}_{n}$}       & \gate{H} &                                          & \gate{H} & \meter{}
\end{quantikz}
};
\end{tikzpicture}
%
		\caption{Single IQP Born-machine circuit. The Hadamard--diagonal--Hadamard sandwich is the defining structure of IQP circuits. The diagonal block $D(\boldsymbol\theta) = \prod_{G \in \mathcal{G}} \exp(i\,\theta_G \prod_{i\in G} Z_i)$ gives one $Z$-phase rotation per gate $G$ of the graph, a single-qubit $R_Z$ for weight-1 gates, a $ZZ$-phase rotation (CNOT, $R_Z$, CNOT, or a native cross-resonance $ZX$) for weight-2 gates. Measurement in the computational basis yields samples from $p_\theta(x) = |\langle x|U(\boldsymbol\theta)|0\rangle|^2$.}
		\label{fig:iqp-circuit}
	\end{figure}
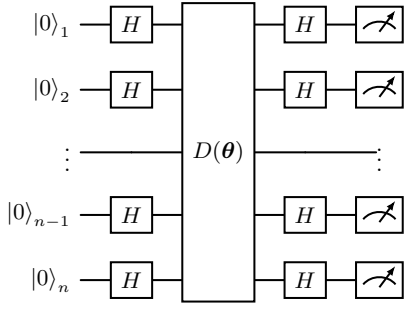

	A parameterized IQP circuit on $\nfeat$ qubits has the form $U(\boldsymbol\theta) = H^{\otimes\nfeat}\, D(\boldsymbol\theta)\,H^{\otimes\nfeat}$, with $D$ diagonal in the computational basis and entries $D_{xx} = \prod_{G \in \mathcal{G}} \exp(i\,\theta_G\,(-1)^{|x \cap G|})$. Equivalently, $U(\boldsymbol\theta) = \prod_G \exp(i\,\theta_G\, X_G)$ with $X_G = \prod_{i \in G} X_i$, where $\mathcal{G}$ is a fixed multiset of qubit subsets (the gate graph). The model defines a Born distribution $p_{\boldsymbol\theta}(x) = |\langle x | U(\boldsymbol\theta)|0\rangle|^2$. See Fig.~\ref{fig:iqp-circuit}.

	We take $\mathcal{G}$ to be a sparse Erd\H{o}s--R\'enyi graph of average degree $6$, a regime that supports both classical training	and conjectured sampling hardness~\cite{hakkaku2026characterizing}.	At $\nfeat = 64$ this gives a base gate count $|\mathcal{G}|$ in the range $256$ to $287$ across the five seed-specific realizations (graph seed $=$ user-seed $+ 1$, as shown Sec.~\ref{sec:experiments}).
	
	For $\beta \subseteq \{0, \dots, \nfeat-1\}$ the Pauli-$Z$ correlator $Z_\beta = \prod_{i \in \beta} Z_i$ has expectation
	\begin{equation}
		\Zth \;=\; \mathbb{E}_{p_{\boldsymbol\theta}}\!\big[\chi_\beta(x)\big],
		\qquad \chi_\beta(x) = (-1)^{\beta \cdot x},
		\label{eq:Zexp}
	\end{equation}
	with $\chi_\beta$ the Walsh character indexed by $\beta$. Van den Nest's algorithm~\cite{vandennest2010classical_sim} provides an unbiased classical Monte Carlo estimator,
	\begin{equation}
		\Zth \;=\;
		\mathbb{E}_{y \sim \mathrm{Unif}(\{0,1\}^\nfeat)}
		\!\Big[ \cos\!\Big( 2 \sum_{G \in \mathrm{act}(\beta)}
		\theta_G\, \xi_G(y) \Big) \Big],
		\label{eq:vdn}
	\end{equation}
	where $\mathrm{act}(\beta)$ collects the gates whose support overlaps $\beta$ in odd parity and $\xi_G(y) = (-1)^{|G \cap y|}$. The cosine acts on the \emph{sum} of the signed angles, with all signs induced by a single shared uniform latent $y$ per sample, because gates share qubits, the signs $\{\xi_G(y)\}$ are correlated random variables, not independent Rademacher draws. (A product of per-gate cosines would, cosine being even, be independent of the signs altogether and hence a deterministic constant rather than an estimator.) The estimator is evaluated with $M$ Monte Carlo samples at cost $\mathcal{O}(M \cdot |\mathrm{act}(\beta)|)$.
	
	\subsection{Walsh-diagonal MMD$^2$}
	
	If the kernel $k(x,y)$ is invariant under bitwise XOR (equivalently, diagonal in the Walsh basis with non-negative spectrum $\hat k_\beta \geq 0$), the squared MMD between $p_\theta$ and $p_{\mathrm{data}}$ takes the closed form
	\begin{align}
		\mathrm{MMD}^2(\boldsymbol\theta)
		&\;=\; \sum_{\beta \neq \emptyset} \hat k_\beta\,
		\big(\Zth - \Zda\big)^2 \nonumber\\
		&\;=\; \dz(\boldsymbol\theta)^{\!\top}\,
		\mathrm{diag}(\hat{\bm{k}})\, \dz(\boldsymbol\theta),
		\label{eq:mmd-walsh}
	\end{align}
	with $\dZ \equiv \Zth - \Zda$. The Liu and Wang heat kernel~\cite{liu2018differentiable} corresponds to $\hat k_\beta^{\mathrm{heat}} = \frac{1}{N_\sigma}\sum_{b=1}^{N_\sigma}\exp(-2 \sigma_b^2 |\beta|)$ for a chosen set of $N_\sigma$ bandwidths $\{\sigma_b\}$ (the symbol $N_\sigma$ is used here to avoid collision with the bit depth $\Bbits$).
	
	\paragraph*{Training procedure.}
	For a finite observable basis $\mathcal{O}_K = \{\beta : 1 \leq |\beta| \leq K\}$ we use classical Adam on Eq.~\eqref{eq:mmd-walsh}, Monte Carlo evaluation of $\Zth$ via Eq.~\eqref{eq:vdn}, exact evaluation of $\Zda$ on the empirical data, reverse-mode automatic differentiation (AD) through Eq.~\eqref{eq:vdn} for parameter gradients, and a gradient update. We use $K = 2$ throughout, which gives $|\mathcal{O}_2| = \nfeat + \binom{\nfeat}{2}$ observables ($2080$ at $\nfeat = 64$).
	
	\paragraph*{Sampling hardness.}
	Training is fully classical, and sampling from a \emph{worst-case} IQP circuit is believed to be classically intractable within multiplicative or additive error~\cite{bremner2011iqp_hardness,bremner2016bjs,marshall2024iqp}. That is a statement about the class, not about any particular trained instance, instance-level tractability is controlled by the gate-graph treewidth, which for the sparse graphs used here is small (Appendix~\ref{app:ciqp-derivation}). Deployment on a quantum device is therefore what makes the model a generator, not what makes it hard.

	\section{Mixture-of-IQP and Compiled-IQP deployment}
	\label{sec:moiqp}

	\subsection{\moiqp{} as a uniform mixture}
	
\begin{figure*}[t]
		\centering
		\scalebox{0.8}{\begin{tikzpicture}[
	>=latex,
	every node/.style={font=\small},
	sharedbox/.style={draw, rounded corners=3pt, fill=red!8,
		minimum width=3.4cm, minimum height=1.3cm,
		align=center, line width=0.4pt},
	branchbox/.style={draw, rounded corners=3pt, fill=red!8,
		minimum width=1.9cm, minimum height=0.80cm,
		align=center, line width=0.4pt},
	pbox/.style={draw, rounded corners=3pt, fill=red!3,
		minimum width=2.8cm, minimum height=0.90cm,
		align=center, line width=0.4pt},
	mixnode/.style={circle, draw, fill=blue!6, line width=0.4pt,
		inner sep=2.5pt, minimum size=0.95cm,
		font=\normalsize},
	outbox/.style={draw, rounded corners=3pt, fill=gray!8,
		minimum width=2.6cm, minimum height=1.0cm,
		align=center, line width=0.4pt}]
	
	\node[font=\small, align=center] (sharedlabel)
	at (0, 2.5) {shared gate graph\\$\mathcal{G}$};
	\node[sharedbox] (shared) at (0, 1.1)
	{$U(\boldsymbol\theta) = H^{\otimes n}\,
		D(\boldsymbol\theta)\,H^{\otimes n}$\\
		on graph $\mathcal{G}$};
	\node[font=\footnotesize, align=center] at (0, -0.2)
	{instantiate with $L$ independent\\angle vectors};
	
	\node[branchbox] (c0)  at (4.2,  2.4) {$U(\boldsymbol\theta^{(0)})$};
	\node[branchbox] (c1)  at (4.2,  0.8) {$U(\boldsymbol\theta^{(1)})$};
	\node           (vd)   at (4.2, -0.5) {$\vdots$};
	\node[branchbox] (cLm) at (4.2, -1.8) {$U(\boldsymbol\theta^{(L-1)})$};
	
	\node[pbox] (p0)  at (9.0,  2.4)
	{$p_{0}(\mathbf{x}) = |\langle\mathbf{x}|U(\boldsymbol\theta^{(0)})|0\rangle|^{2}$};
	\node[pbox] (p1)  at (9.0,  0.8)
	{$p_{1}(\mathbf{x}) = |\langle\mathbf{x}|U(\boldsymbol\theta^{(1)})|0\rangle|^{2}$};
	\node       (vdp) at (9.0, -0.5) {$\vdots$};
	\node[pbox] (pLm) at (9.0, -1.8)
	{$p_{L-1}(\mathbf{x})$};
	
	\node[mixnode] (mix) at (12.3, 0.3) {$\dfrac{1}{L}\!\sum_{\ell}$};
	
	\node[outbox] (out) at (15.2, 0.3)
	{$p^{\moiqp{}}(\mathbf{x})$};
	
	\draw[->, line width=0.35pt] (shared.east) -- (c0.west);
	\draw[->, line width=0.35pt] (shared.east) -- (c1.west);
	\draw[->, line width=0.35pt] (shared.east) -- (cLm.west);
	
	\draw[->, line width=0.35pt] (c0.east)  --
	node[above=1pt, font=\footnotesize]{Born rule} (p0.west);
	\draw[->, line width=0.35pt] (c1.east)  --
	node[above=1pt, font=\footnotesize]{Born rule} (p1.west);
	\draw[->, line width=0.35pt] (cLm.east) --
	node[above=1pt, font=\footnotesize]{Born rule} (pLm.west);
	
	\draw[->, line width=0.35pt] (p0.east)  -- (mix.north west);
	\draw[->, line width=0.35pt] (p1.east)  -- (mix.west);
	\draw[->, line width=0.35pt] (pLm.east) -- (mix.south west);
	
	\draw[->, line width=0.5pt] (mix.east) -- (out.west);
\end{tikzpicture}}
		\caption{\moiqp{} as a uniform mixture of IQP Born distributions. A single IQP circuit $U(\boldsymbol\theta) = H^{\otimes n}D(\boldsymbol\theta) H^{\otimes n}$ on a shared gate graph $\mathcal{G}$ is instantiated with $L$ independent angle vectors $\boldsymbol\theta^{(\ell)}$. Each instance has Born distribution $p_{\ell}(\mathbf{x}) = |\langle\mathbf{x}|U(\boldsymbol\theta^{(\ell)})|0\rangle|^{2}$, and the \moiqp{} model distribution is their uniform average $p^{\moiqp{}}(\mathbf{x}) = \frac{1}{L}\sum_{\ell} p_{\ell}(\mathbf{x})$. Expectations of Pauli-$Z$ words under $p^{\moiqp{}}$ are linear in the components (Eq.~\ref{eq:moiqp-Z}), so the Van den Nest Monte Carlo estimator and the Walsh-diagonal MMD$^{2}$ loss of single-IQP training both extend to the mixture at no additional cost. The same $p^{\moiqp{}}$ is realized at deployment by a single compiled circuit, \ciqp{} (Fig.~\ref{fig:ciqp-circuit}), without classical post-mixing.}
		\label{fig:moiqp-schematic}
	\end{figure*}
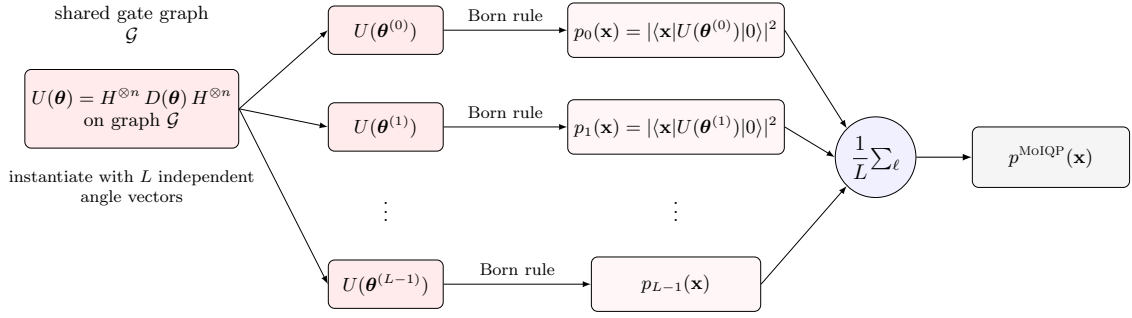

	A single IQP circuit on a fixed gate graph is expressively constrained because its Born distribution is a Walsh polynomial of degree at most the maximum gate weight, with coefficients on the cosine manifold parameterized by the gate angles. The set of low-order correlator vectors reachable by such a circuit is a non-convex variety. A uniform mixture of $\Lcomp$ instances realizes averages of $\Lcomp$ points on it and so fills toward its convex hull, reaching multimodal, strongly-correlated targets that no single IQP on the same graph can, while every component stays on the same sparse low-weight graph and is therefore trained at single-IQP cost. On the calorimeter data the gain is decisive. A single IQP ($\Lcomp = 1$) saturates at a correlation-reconstruction error roughly an order of magnitude above the encoding floor and does not improve with additional epochs, whereas the mixture reaches the floor, the improvement saturating once $\Lcomp$ reaches the effective number of modes in the binary-encoded data (Table~\ref{tab:mix-scan}). This is why we use $\Lcomp = 8$. Concretely, we define
	\begin{equation}
		p^{\moiqp{}}_{\boldsymbol\Theta}(x)
		\;=\; \frac{1}{\Lcomp}\!\sum_{\ell=0}^{\Lcomp-1}
		|\langle x | U(\boldsymbol\theta^{(\ell)}) |0\rangle|^2,
		\label{eq:moiqp-pdf}
	\end{equation}
	with $\Lcomp$ independent angle vectors $\boldsymbol\Theta = (\boldsymbol\theta^{(0)}, \dots, \boldsymbol\theta^{(\Lcomp-1)})$ on the same gate graph $\mathcal{G}$. The mixture marginal Pauli-$Z$ expectations are component-wise averages,
	\begin{equation}
		\Zth^{\moiqp{}} \;=\; \frac{1}{\Lcomp}\!\sum_{\ell=0}^{\Lcomp-1}
		\langle Z_\beta \rangle_{\boldsymbol\theta^{(\ell)}},
		\label{eq:moiqp-Z}
	\end{equation}
	so that the Van den Nest training procedure carries over unchanged. The per-component $Z$-correlators are averaged before the MMD loss is evaluated. The number of trainable parameters scales as $\Lcomp \cdot |\mathcal{G}|$, and the gradient cost scales the same way. We use $\Lcomp = 8$ in the headline experiments. Fig.~\ref{fig:moiqp-schematic} sketches the classical mixture. Next, in Sec.~\ref{sec:ciqp}, we show how to build the corresponding coherent quantum circuit.
	
	\subsection{Deferred-measurement compilation (\ciqp{})}
	\label{sec:ciqp}
	
	A naive quantum deployment of Eq.~\eqref{eq:moiqp-pdf} would require $\Lcomp$ separate IQP circuit executions plus classical post-mixing. We show instead that the entire \moiqp{} can be written as a single IQP circuit on $\nfeat + a$ qubits with $a = \lceil \log_2 \Lcomp \rceil$, using a Walsh--Hadamard-prepared ancilla register that selects the mixture component coherently.
	
	\paragraph*{Construction.}
	Initialize $a$ ancillas to $|0\rangle^{\otimes a}$ and apply a Hadamard layer, giving the uniform superposition $\tfrac{1}{\sqrt{\Lcomp}}\sum_\ell |\ell\rangle$ when $\Lcomp = 2^a$. For each gate $G \in \mathcal{G}$ with component-indexed angles $\{\theta_G^{(\ell)}\}$ define the controlled IQP gate
	\begin{equation}
		C\text{-}\exp\!\big(\textstyle{\sum_\ell} i\, \theta_G^{(\ell)} X_G\big)
		\;=\; \sum_{\ell=0}^{\Lcomp-1} |\ell\rangle\!\langle\ell| \otimes
		\exp(i\,\theta_G^{(\ell)} X_G).
		\label{eq:cgate}
	\end{equation}
	All $X_G$ commute, so the full controlled IQP unitary is diagonal in the ancilla basis and the controls expand in the Walsh basis on the ancillas, giving an ordinary (uncontrolled) IQP circuit on $\nfeat + a$ qubits whose data-register reduced density matrix equals the \moiqp{} mixture of Eq.~\eqref{eq:moiqp-pdf}.
	
	Tracing out (equivalently, measuring) the ancilla register reproduces Eq.~\eqref{eq:moiqp-pdf} on the data register. The full derivation, including the explicit Walsh--Hadamard relation between base angles and compiled angles, is in Appendix~\ref{app:ciqp-derivation}.
	
	\paragraph*{Hardness.}
	The compiled circuit is itself IQP on $\nfeat + a$ qubits, so its joint sampling hardness follows from the standard Bremner--Jozsa--Shepherd argument~\cite{bremner2011iqp_hardness,bremner2016bjs} applied at the larger qubit count. Whether the \emph{data-register marginal} of the compiled circuit inherits that hardness is subtler, since marginalizing over an entangled ancilla can in principle reduce distributional complexity. Appendix~\ref{app:ciqp-derivation} addresses this point carefully.

	\paragraph*{Numerical validation.}
	Across five training seeds at $\nfeat = 64$, $\Lcomp = 8$, the per-observable disagreement between the \ciqp{} data-register marginal and the \moiqp{} target, at $M = 2 \times 10^5$ Van den Nest latents, sits below the Monte Carlo noise floor $1/\sqrt{M}$ for every seed (measured ratio $0.600 \pm 0.012$ on the corrected reruns, because the two estimators share latents, the specific sub-unity value is the variance constant of their difference under exact equality, and carries no information about compilation error). The compilation is therefore exact to machine precision up to Monte Carlo noise.
	
	\paragraph*{Choice of training coordinates.}
	\moiqp{} and \ciqp{} are the \emph{same} model in two parameterizations. The per-gate map between the $\Lcomp$ base angles $\{\theta_{G_j}^{(\ell)}\}_\ell$ and the $\Lcomp$ compiled angles $\{\tilde\phi_{j,S}\}_S$ is the invertible Walsh--Hadamard transform of Appendix~\ref{app:ciqp-derivation}, so the two share an identical parameter count $\Lcomp\,|\mathcal{G}|$ and realize identical Born distributions. Compilation delivers quantum \emph{deployability}, namely a single coherent circuit whose $1/\Lcomp$ average is performed by the Born rule rather than by classical post-mixing. We optimize in the base-angle coordinates for a Monte Carlo reason. A weight-$\leq 2$ data-register correlator activates an active gate set of size $A_\beta$ in the base graph but $A_\beta\,\Lcomp$ in the compiled graph, since each base gate expands into $\Lcomp$ ancilla-coupled copies of weight up to $2 + a$ (at $\nfeat = 64$ and $\Lcomp = 8$ the active set grows from $19$ to $152$). The Van den Nest estimator then evaluates the mixture average $\tfrac{1}{\Lcomp}\sum_\ell$ \emph{stochastically through the ancilla latents} in the compiled coordinates but \emph{deterministically} across components in the base coordinates. Both are unbiased for the same correlator, but the compiled-coordinate gradient estimator carries a factor $\Lcomp\,(1 + \sigma_\mu^2/\bar V) \geq \Lcomp$ more variance at matched cost (App.~\ref{app:variance-advantage}, about $8\times$ at $\Lcomp = 8$), so matched-budget training in compiled coordinates stalls and needs roughly $\Lcomp$ times the latent budget to reach the same loss. The compilation moves the mixture sum onto the Born rule at deployment precisely so that it need not be paid again as gradient-estimator variance during training.
	
	\begin{figure}[b]
		\centering
\begin{tikzpicture}
\node[scale=0.88] {%
\begin{quantikz}[column sep=0.44cm, row sep=0.36cm]
\lstick[2]{\shortstack{ancilla\\($a=\lceil\log_2 L\rceil$)}}
  & \gate{H} & \gate[wires=5]{\mathrm{CIQP}(\{\boldsymbol\theta^{(\ell)}\})} & \qw       & \meter{} \\
  & \gate{H} &                                                                & \qw       & \meter{} \\
\lstick[3]{\shortstack{data\\($n$ qubits)}}
  & \gate{H} &                                                                & \gate{H}  & \meter{} \\
  & \gate{H} &                                                                & \gate{H}  & \meter{} \\
  & \gate{H} &                                                                & \gate{H}  & \meter{}
\end{quantikz}
};
\end{tikzpicture}
%
		\caption{\ciqp{} deployment. The $\Lcomp$-component \moiqp{} is compiled into a single IQP-form circuit on $\nfeat + a$ qubits with $a = \lceil \log_2 \Lcomp \rceil$ ancillas. The ancillas are prepared in the uniform Walsh--Hadamard superposition $|+\rangle^{\otimes a} = (1/\sqrt{\Lcomp})\sum_\ell |\ell\rangle$. The controlled-IQP block applies an $\ell$-dependent IQP unitary to the data register. Trailing data-register Hadamards complete the sandwich. Tracing out (equivalently, measuring) the ancilla register reproduces the \moiqp{} data-register marginal. The compiled circuit is itself IQP on $\nfeat + a$ qubits.}
		\label{fig:ciqp-circuit}
	\end{figure}
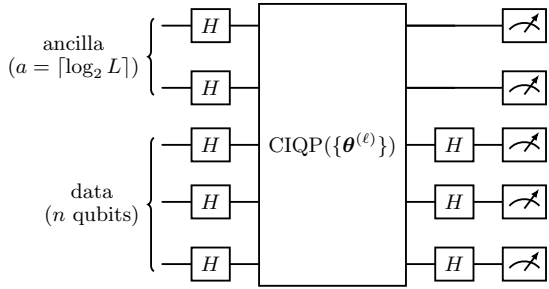

	\subsection{Equivalent deployment forms, dcIQP and the Walsh-sparse degree-3 circuit}
	\label{sec:dciqp}
	
	\paragraph*{Three exact realizations of the same generative task.}
	Eq.~\eqref{eq:moiqp-pdf} and the marginal identity of Appendix~\ref{app:ciqp-derivation} (Eq.~\ref{eq:app-marginal}) admit three operationally distinct but distributionally \emph{identical} realizations of the $\nfeat$-bit generative task. (i)~\emph{Classical shot-splitting}, which binds each component $\boldsymbol\theta^{(\ell)}$ into one native IQP circuit template and splits the shot budget uniformly. (ii)~\emph{dcIQP}, a single dynamic circuit in which the $a$ ancillas are Hadamard-prepared and measured \emph{first}, yielding a uniformly random label $\ell$, after which classical feedforward applies $U(\boldsymbol\theta^{(\ell)})$ to the data register. By the principle of deferred measurement~\cite{nielsen2010qcqi}, and directly from Eq.~\eqref{eq:app-marginal}, the data-register distribution equals $p^{\moiqp{}}$ exactly. (iii)~\emph{\ciqp{} with ancilla discard}, the coherent circuit of Sec.~\ref{sec:ciqp}, keeping the data bits only. The identity (iii)$\,\equiv\,$(i) is Eq.~\eqref{eq:app-marginal}, verified by dense simulation at machine precision at small $\nfeat$ and by Van den Nest Monte Carlo at $\nfeat = 64$ (Appendix~\ref{app:ciqp-derivation}). Realization (ii) equals (i) by construction whenever the measured ancilla labels are uniform, a condition we monitor directly on hardware (Sec.~\ref{sec:hw-protocol}).
	
	The equivalence has two consequences, one epistemic and one practical. Epistemically, the data-register marginal is classically mixable by construction, so sampling-hardness claims attach to the individual components and to the joint $(\nfeat+a)$-bit distribution rather than to marginal generation itself. The coherent content that no per-shot mixing reproduces lives exclusively in that joint distribution, which is not the product of the data marginal with any ancilla distribution (in a dense check at $\nfeat = 6$, $\Lcomp = 4$ the joint sits at total-variation distance $0.326$ from its ancilla-dephased counterpart and carries $I(X;Y) = 0.470$ bits of data to ancilla mutual information). Practically, the equivalence is what makes large-scale deployment feasible. dcIQP has \emph{constant} depth overhead and the per-shot two-qubit gate cost of a \emph{single} component, since exactly one feedforward branch executes per shot and the ancillas share no two-qubit gates with the data register (their placement is routing-irrelevant). This is what makes the $67$-qubit execution of Sec.~\ref{sec:hardware}, and the larger-register campaign reported separately, possible on hardware on which a monolithic full-Walsh \ciqp{} is far out of budget (Sec.~\ref{sec:discussion}).
	
	\paragraph*{Walsh-sparse degree-3 compilation.}
	The full compilation of Appendix~\ref{app:ciqp-derivation} expands each base gate into $\Lcomp$ compiled gates of weight up to $|G_j| + a$. If instead the compiled Walsh support on the ancillas is restricted to $\mathcal{U} = \{\varnothing\} \cup \{\{k\}: k < a\}$, each base gate carries $a+1$ compiled angles $\{\phi_{j,\varnothing}, \phi_{j,\{k\}}\}$ and the component angles are the constrained Walsh synthesis
	\begin{equation}
		\theta_j^{(\ell)} \;=\; \phi_{j,\varnothing} \;+\; \sum_{k=0}^{a-1} (-1)^{\ell_k}\, \phi_{j,\{k\}},
		\label{eq:walsh-sparse}
	\end{equation}
	with $\ell_k$ the $k$-th bit of $\ell$. The compiled circuit then has $(a+1)\,|\mathcal{G}|$ gates of weight at most $\max_j |G_j| + 1 = 3$, i.e.\ it is a member of the \emph{canonical degree-3 IQP class} of Shepherd to Bremner and Bremner--Jozsa--Shepherd~\cite{shepherd2009iqp,bremner2011iqp_hardness}, with no extension caveats on the gate-weight distribution. The price is a controlled expressivity restriction, in which the $\Lcomp$-tuple of angles of each gate is confined to the $(a+1)$-dimensional Walsh-low-pass subspace of $\mathbb{R}^{\Lcomp}$ (all multi-ancilla interaction terms vanish. At $\Lcomp=4$ this is the single linear constraint $\theta^{(0)}_j - \theta^{(1)}_j - \theta^{(2)}_j + \theta^{(3)}_j = 0$ per gate). Training is performed directly in the $\phi$ coordinates. The map~\eqref{eq:walsh-sparse} is linear and fixed, so gradients chain through it at no cost and the Van den Nest machinery of Sec.~\ref{sec:framework} is untouched. We verify the degree-3 membership and the gate-count bound $(a+1)|\mathcal{G}|$ exactly, together with dense-simulation agreement between the constrained coherent circuit and its mixture, at machine precision. The hardware campaign of Sec.~\ref{sec:hardware} trains in this parameterization, so that a single trained parameter set serves both the generative deployment (as dcIQP) and, prospectively, the coherent degree-3 hardness object.

	\section{The Pearson-Stabilized Correlation Kernel}
	\label{sec:psck}

	\subsection{Why the heat kernel undershoots}
	
	The Liu and Wang heat kernel has Walsh spectrum $\hat k_\beta^{\mathrm{heat}} \propto \exp(-2\sigma^2 |\beta|)$ at a single bandwidth $\sigma$, and the finite-bandwidth mixture used in practice preserves that decay in $|\beta|$ with an effective scale. The obvious reading is that the deficit is a bandwidth effect, but at our bandwidth set the suppression is mild: $\hat k^{\mathrm{heat}}_{|\beta|=1} = 0.500$ and $\hat k^{\mathrm{heat}}_{|\beta|=2} = 0.330$, a ratio of $0.660$. A factor of $1.5$ between weight classes does not explain a correlation error twice the encoding floor.
	
	The effect that does is a matter of dimension rather than of weight. The pairwise correlations depend on the model $Z$-marginals only through the $P = \binom{\Dfeat}{2} = 28$ directions spanned by the rows of the Pearson Jacobian $\Jstar$ of Sec.~\ref{sec:jacobian}, a subspace of the $K = |\mathcal{O}_2| = 2080$-dimensional residual space. Writing $\Pi_{\Jstar}$ for the orthogonal projector onto $\mathrm{row}(\Jstar)$, the fraction of the loss metric's total weight carried by that subspace is
	\begin{equation}
		\frac{\mathrm{tr}\big(\Pi_{\Jstar}\,\mathrm{diag}(\boldsymbol\omega_{\mathrm{heat}})\big)}
		{\mathrm{tr}\big(\mathrm{diag}(\boldsymbol\omega_{\mathrm{heat}})\big)}
		\;=\; \frac{9.24}{697.3} \;=\; 1.33\%,
		\label{eq:heat-subspace}
	\end{equation}
	evaluated on the training split. That is, to two digits, the dimension fraction $28/2080 = 1.35\%$. A Walsh-diagonal kernel with a slowly varying spectrum is close to isotropic on the residual space, so it allocates to the correlation sector essentially the share that sector occupies by dimension, and no more. Rebalancing the spectrum across weight classes cannot fix this, because the correlation-carrying directions are a thin subspace \emph{within} the weight-2 block, not the block itself. The MMD$^2$ loss weights each Walsh-basis residual $\langle Z_\beta\rangle_\theta - \langle Z_\beta\rangle_{\mathrm{data}}$ by $\hat k_\beta$ alone, and $\hat k_\beta$ has no way to see that structure.
	
	The gradient signal on the pairwise correlations $\rho$ is correspondingly weaker than the signal on the single-qubit means $\langle Z_i\rangle$, not because weight-2 residuals are downweighted but because the correlation-relevant directions are $1.3\%$ of the residual space and are treated like any other $1.3\%$. The training history reflects this imbalance. The model drives the means close to the data in the first $\sim\!50$ epochs, then spends the remaining epoch budget slowly reducing the weight-2 residuals and does not reach the encoding-fidelity floor on $\rho$. The visible consequence is a model whose correlation sign pattern is correct but whose amplitudes are systematically compressed (Fig.~\ref{fig:rho-matrices}, top right).
	
	A precursor in quantum HEP generative modeling is the auxiliary correlation loss of Rehm \emph{et al.}~\cite{rehm2024qag}, who added a mean-squared error between model and data pixel correlations on top of the MMD objective to repair this same correlation-reconstruction deficit in a variational quantum calorimeter-image generator. \psck{} pursues the same goal intrinsically, through the kernel rather than an added loss term. Sec.~\ref{sec:baselines} reports the explicit-MSE route as a baseline at identical configuration and quantifies the marginal-fidelity and generalization price it pays for its lower raw $\rho$ error. We address the deficit with a kernel that biases descent directly toward correlation-sensitive parameter directions while preserving positive-definiteness, classical trainability, and quantum sampling hardness.
	
	\subsection{Pearson Jacobian}
	\label{sec:jacobian}
	
	For the binary encoding (see Sec.~\ref{sec:dataset}), the empirical Pearson correlation between calorimeter cells $f$ and $g$ under the trained model is
	\begin{equation}
		\rhog(\boldsymbol\theta) \;=\;
		\frac{\mathrm{Cov}_{\boldsymbol\theta}[S_f, S_g]}
		{\sqrt{\mathrm{Var}_{\boldsymbol\theta}[S_f]\,
				\mathrm{Var}_{\boldsymbol\theta}[S_g]}},
		\label{eq:rho_def}
	\end{equation}
	with $S_f(x) = \sum_k 2^{B-1-k}\, b_{f,k}(x)$ the integer level of feature $f$ and $b_{f,k} = (1 - Z_{fB+k})/2$. Both numerator and denominator are linear functionals of the $Z$-correlators in the $\mathcal{O}_2$ basis, $\mathrm{Cov}[S_f, S_g] = \sum_{k,k'} 2^{2B-2-k-k'} (\langle Z_{fB+k} Z_{gB+k'}\rangle - \langle Z_{fB+k}\rangle \langle Z_{gB+k'}\rangle)/4$, and similarly for the variances. The full functional $\rhog$ thus factors through the model $Z$-marginals.
	
	We define the \emph{Pearson Jacobian} $\Jstar$ as the matrix of partial derivatives that links an infinitesimal change in any Pauli expectation $\Zb$ to the change it induces in each pairwise correlation $\rhog$, evaluated at the data,
	\begin{equation}
		\Jstar_{(fg),\beta} \;\equiv\;
		\left.\frac{\partial \rhog}{\partial \Zb}\right|_{\langle
			\boldsymbol{Z}\rangle = \langle\boldsymbol{Z}\rangle_{\mathrm{data}}}.
		\label{eq:Jstar-def}
	\end{equation}
	Rows are indexed by the $P = \binom{\Dfeat}{2}$ unordered feature pairs $(fg)$ and columns by the $K = |\mathcal{O}_2|$ weight $\leq 2$ Pauli observables $\beta$, so $\Jstar$ is a $P \times K$ matrix.
	
	Because the evaluation point is fixed at the data expectation $\langle\boldsymbol{Z}\rangle_{\mathrm{data}}$ rather than at the running model, $\Jstar$ depends only on the dataset. It is therefore a constant linear map, built once before training begins and reused at every gradient step. A closed-form expression for its entries is given in Appendix~\ref{app:jacobian-derivation}.

	\subsection{The \psck{} kernel}
	
	The Pearson-Stabilized Correlation Kernel reads
	\begin{equation}
		\Kpsck \;=\; \mathrm{diag}(\boldsymbol\omega_{\mathrm{heat}})
		\;+\; \eta\, \Jstar^{\!\top}\Jstar,
		\label{eq:Kpsck}
	\end{equation}
	with $\eta \geq 0$ a scalar mixing weight. We use $\eta = 5$ throughout. Eq.~\eqref{eq:Kpsck} is the sum of a positive-definite diagonal kernel (Liu and Wang) and a positive-semidefinite rank-$P$ correction, and is therefore positive-definite for any $\eta > 0$.
	
	The effect on the measure of Eq.~\eqref{eq:heat-subspace} is direct and is the whole design intent. Repeating that trace fraction with $\Kpsck$ in place of $\mathrm{diag}(\boldsymbol\omega_{\mathrm{heat}})$ gives $19.9\%$ at $\eta = 5$, a factor $15$ enhancement of the weight carried by the $28$ correlation-sensitive directions, at condition number $34.6$. The single scalar $\eta$ therefore interpolates between an isotropic Walsh-diagonal metric and one that concentrates on a chosen low-dimensional functional, and its cost is bounded. The rank-$P$ correction can only raise the condition number to $1 + \eta\,\lambda_{\max}(\Jstar^{\!\top}\Jstar)/\min_\beta \hat k_\beta$, which is what keeps the problem numerically benign at the value we use. The training loss is
	\begin{equation}
		\Lpsck(\boldsymbol\Theta) \;=\;
		\dz(\boldsymbol\Theta)^{\!\top} \Kpsck\, \dz(\boldsymbol\Theta)
		\;=\; \Lhe + \eta\, \Lrho,
		\label{eq:Lpsck}
	\end{equation}
	with $\Lhe = \dz^{\!\top} \mathrm{diag}(\boldsymbol\omega_{\mathrm{heat}}) \dz$ the Liu and Wang MMD$^2$ and $\Lrho = \|\Jstar\, \dz\|^2$ the squared $L_2$ residual of the linearized Pearson-correlation reconstruction.
	
	\subsection{Gauss--Newton on Pearson MSE}
	
	The added term $\Lrho$ is the Gauss--Newton linearization of the Pearson MSE
	\begin{equation*}
		\mathcal{L}_{\mathrm{Pearson}} = \sum_{f<g} (\rhog(\boldsymbol\theta) - \rho_{fg}^{\mathrm{data}})^2,
	\end{equation*}
	at the data point. Expanding $\rhog$ to first order around $\langle Z\rangle_{\mathrm{data}}$ gives
	\begin{equation*}
		\rhog(\boldsymbol\theta) - \rho_{fg}^{\mathrm{data}} \approx \sum_\beta \Jstar_{(fg),\beta} (\Zth - \Zda),
	\end{equation*}
	so the following approximation holds,
	\begin{equation*}
		\mathcal{L}_{\mathrm{Pearson}} \approx \|\Jstar\, \dz\|^2 = \Lrho.
	\end{equation*}
	Eq.~\eqref{eq:Lpsck} is therefore a sum of a valid Walsh-diagonal MMD$^2$ that guarantees distribution matching in the limit and a tangent-space target on Pearson correlations. The linearization is exact at the optimum $\dz \to 0$, and because $\Jstar$ is evaluated at the data it never moves, which avoids the moving-target pathology of Gauss--Newton methods at non-stationary linearization points.
	
	\subsection{The general template}
	\label{sec:psck-template}
	
	Nothing in the construction is specific to the Pearson correlation. What Sec.~\ref{sec:jacobian} uses about $\rho_{fg}$ is only that it factors through the model $Z$-marginals and has a closed-form derivative there. Any functional with those two properties fits the same three steps.
	
	\begin{enumerate}
		\item \emph{Choose} a downstream functional $\mathcal{F}: \{\Zth\}_{\beta \in \mathcal{O}_K} \to \mathbb{R}^{P}$ that expresses what the model is wanted for.
		\item \emph{Linearize} it at the data, $\Jstar^{\mathcal{F}}_{p,\beta} = \partial \mathcal{F}_p / \partial \Zth \big|_{\langle Z\rangle = \langle Z\rangle_{\mathrm{data}}}$. Because the evaluation point is the data and not the running model, $\Jstar^{\mathcal{F}}$ is a constant matrix, built once before training.
		\item \emph{Append} $\eta_{\mathcal{F}}\, \Jstar^{\mathcal{F}\top}\Jstar^{\mathcal{F}}$ to the kernel.
	\end{enumerate}
	
	Three things are invariant under this substitution, and they are what make the template worth stating separately. The Van den Nest estimator is untouched, since $\Jstar^{\mathcal{F}}$ acts on the residual vector and never on individual correlators. The training loop is untouched, since the addition is one fixed matrix multiplication per step. And the deployed circuit is untouched, so sampling complexity is unchanged whatever functional is chosen. The kernel remains positive-definite for any $\eta_{\mathcal{F}} \geq 0$ because the addition is positive-semidefinite by construction, and several functionals may be combined by summing their rank corrections.
	
	Candidate functionals include moment-conditional quantities, mutual-information surrogates, sliced-Wasserstein direction projections, and physics-level observables such as shower-depth centroids or energy-weighted moments, each of which is a smooth function of the level moments and hence of the $Z$-marginals.
	
	\subsection{A second instance, cross co-skewness}
	\label{sec:psck-coskew}
	
	To show that the template is not a restatement of the Pearson case, we instantiate it on a functional that the Pearson kernel cannot reach. The standardized three-feature co-skewness
	\begin{equation}
		\tau_{fgh} \;=\;
		\frac{\mathbb{E}\big[(S_f - \mu_f)(S_g - \mu_g)(S_h - \mu_h)\big]}
		{\sigma_f \sigma_g \sigma_h},
		\label{eq:tau}
	\end{equation}
	over the $\binom{\Dfeat}{3} = 56$ feature triples. Since $\mathbb{E}[b_i b_j b_k] = \tfrac{1}{8}(1 - z_i - z_j - z_k + z_{ij} + z_{ik} + z_{jk} - z_{ijk})$, the numerator of Eq.~\eqref{eq:tau} is multilinear in the $Z$-correlators exactly as the covariance is, and $\tau_{fgh}$ factors through the model marginals in the same way $\rho_{fg}$ does. Step 2 then gives a closed form with four nonvanishing families of $\beta$ (weight-1 in each of $f, g, h$. Cross-feature weight-2 on each of the three pairs. Weight-3 with one qubit per feature. Intra-feature weight-2 entering only through the variances), which we verify against central differences at the same $5\times10^{-11}$ level as the Pearson Jacobian of Appendix~\ref{app:jacobian-derivation}.
	
	Two features of this instance are worth recording. First, the required observable basis is larger. The set $\mathcal{O}_3^{\times}$ has $\binom{\Dfeat}{3}\Bbits^3 = 28\,672$ elements at $\Dfeat = \Bbits = 8$, against $|\mathcal{O}_2| = 2080$. Because the Van den Nest cost per observable is $\mathcal{O}(M\,|\mathrm{act}(\beta)|)$ with $|\mathrm{act}(\beta)| \leq |\mathcal{G}|$ independently of $|\beta|$, this raises the per-epoch cost by roughly $14\times$ and leaves the scheme polynomial. Second, under an exactly balanced quantile encoding all $\mu_f = (2^{\Bbits}-1)/2$, and the cross-feature weight-2 family of the co-skewness Jacobian vanishes identically. The Pearson and co-skewness corrections then act on disjoint cross-feature supports, so appending the second exerts no new pressure on the directions the first already controls.
	
	We do not report a trained $\tau$-targeting model at the headline geometry, and the correlation results of Sec.~\ref{sec:experiments} are unaffected by this subsection. Its purpose is to establish that the template of Sec.~\ref{sec:psck-template} generalizes in practice and not only in principle, and to identify where a quantum construction of this kind has room that the classical reference of Sec.~\ref{sec:baselines} does not. A Gaussian copula has no free parameters left in the trivariate sector once its correlation matrix is fitted, so any structure there is a prediction it must get right rather than a fit it can absorb.
	
	\subsection{Classical trainability and sampling hardness}
	
	Since $\Jstar$ acts only on the residual vector $\dz \in \mathbb{R}^K$ and not on the individual $Z$-correlators, the additional gradient terms from the rank-$P$ correction are classically computable in time $\mathcal{O}(P \cdot K)$ once the $\Zth$ values have been obtained by the Van den Nest Monte Carlo estimator. Three properties of the underlying Liu--Wang construction are preserved.
	
	\begin{itemize}
		\item \emph{Classical trainability}. The gradient $\partial \Lpsck / \partial \theta_G$ needs only the same Van den Nest Monte Carlo evaluations as Liu--Wang training, followed by a fixed $K \times K$ matrix multiplication.
		\item \emph{Sampling hardness}. The kernel affects only the loss landscape, not the deployed circuit, so the trained model is still IQP (or, after compilation, \ciqp{}) and its sampling complexity is unchanged.
		\item \emph{Positive-definiteness}. At our chosen $\eta = 5$ the condition number of $\Kpsck$ is $\sim\!35$ at $\nfeat = 64$, comfortably within a numerically stable regime.
	\end{itemize}
	
	\begin{table*}[t]
		\centering
		\caption{Results at $\nfeat = 64$, $\Lcomp = 8$, $1500$ epochs, split-correct pipeline. \psck{}-\moiqp{} entries are mean $\pm$ std across five independent seeds, with $\rfit = 0.9893 \pm 0.0014$ on the training split, and per-seed values in Table~\ref{tab:per-seed-L8}. CorrMSE is the auxiliary Pearson-MSE objective of Ref.~\cite{rehm2024qag} at the identical configuration, the Gaussian copula the classical reference of Sec.~\ref{sec:baselines}. Gap is $\maerho^{\mathrm{test}} - \maerho^{\mathrm{train}}$, and the floor row's gap is the sample-statistical reference scale. The encoding-fidelity floor $\enfloor{s}$ is defined in Sec.~\ref{sec:dataset} and bounds only models that reproduce all weight-$\leq 2$ statistics of the encoded data.}
		\label{tab:headline}
		\begin{tabular}{lccccc}
			\toprule
			Model & $\maerho$ (train) & $\maerho$ (test) & $\maez$ & marginal $W_1$ (levels) & gap \\
			\midrule
			\psck{}-\moiqp{} & $0.068 \pm 0.006$ & $0.070 \pm 0.006$ & $0.0112 \pm 0.0008$ & $2.0$ to $24.9$ & $0.0019 \pm 0.0005$ \\
			\lwmmd{} & $0.095$ &- &- &- &- \\
			CorrMSE & $0.0028 \pm 0.0005$ & $0.0076 \pm 0.0003$& $0.032 \pm 0.005$ & $24.3$ to $31.0$ & $+0.0048$ \\
			Gaussian copula & $0.043$ & $0.046$ & $0.0047$ & $\approx 0.1$ & $+0.003$ \\
			\midrule
			encoding floor & $0.0515$ & $0.0548$ &- &- & $0.0033$ \\
			\bottomrule
		\end{tabular}
	\end{table*}

	\section{Experiments}\label{sec:experiments}
	
	\subsection{Dataset and encoding}\label{sec:dataset}
	
	A sampling calorimeter measures the energy of an incident particle by absorbing it in a dense medium and reading out the energy deposited in a segmented array of active cells. An electron entering the electromagnetic calorimeter initiates an \emph{electromagnetic shower}. It radiates photons by bremsstrahlung, those photons pair-produce, and the resulting $e^{\pm}/\gamma$ cascade multiplies until the secondary energies fall below the critical energy, after which the shower dies out. The energy a shower deposits as a function of depth (its \emph{longitudinal profile}) rises to a maximum at a depth that grows logarithmically with the incident energy and then falls off, and for a fixed-energy beam is described on average by a gamma distribution. The CLIC detector used here is a high-granularity silicon to tungsten sampling calorimeter, whose full per-event readout is a three-dimensional energy image. We integrate it transversely and downsample to $\Dfeat = 8$ longitudinal depth bins. The object our model learns is therefore an eight-dimensional longitudinal profile $\mathbf{r} \in \mathbb{R}^{8}_{\geq 0}$, one energy per depth layer, rather than a two-dimensional pixel grid. The $\nfeat = \Dfeat\cdot\Bbits = 64$ qubits split into $\Dfeat = 8$ spatial (depth) features encoded at $\Bbits = 8$ bits of amplitude resolution each.
	
	The structure a generative model must reproduce lives in the joint distribution over these eight depths, not in any single marginal. Because the cascade develops smoothly, neighbouring depth bins are strongly positively correlated. Because the total deposited energy is bounded by the fixed incident energy, depositing more energy early in the shower leaves less for the tail, which appears as an anti-correlation between the leading and trailing depth blocks. This block-structured correlation pattern, visible in Fig.~\ref{fig:rho-matrices} with $\rho_{fg}$ ranging over $[-0.85,+0.85]$, is the shower-development physics that downstream energy reconstruction is sensitive to, and it is precisely the signal the heat-kernel baseline compresses and \psck{} is built to recover.
	
	We use the calorimeter shower image dataset of Ref.~\cite{calo_data}, $N = 47\,682$ samples of energy depositions in the electromagnetic calorimeter of the CLIC (Compact Linear Collider) detector, downsampled to $\Dfeat = 8$ longitudinal shower-depth bins per event. For each cell $f$ the continuous energy $r_f$ is mapped to one of $2^{\Bbits}$ integer levels by quantile binning on the training split only (no test-split leakage into the encoding), and the integer level $S_f \in \{0, \dots, 2^{\Bbits}-1\}$ is binary-encoded as $\Bbits$ bits in the standard way. The full encoded sample is the concatenation $x \in \{0,1\}^{\nfeat}$ with $\nfeat = \Dfeat \cdot \Bbits$.
	
	The data are split 80/20 into training and test subsets, giving $N_{\mathrm{train}} = 38\,146$ and $N_{\mathrm{test}} = 9\,536$. The split is created once (committed as \texttt{split\_indices.npz}) and reused across all experiments. At $\Bbits = 8$ the encoding-fidelity floors, defined as the $\maerho$ attained by a model that exactly matches the encoded data's $\langle Z_\beta\rangle$ for $\beta \in \mathcal{O}_2$, are $\enfloor{train} = 0.0515$ and $\enfloor{test} = 0.0548$, where $\enfloor{s} \equiv \mathrm{mean}_{f<g}\,|\rho^{\mathrm{enc},s}_{fg} - \rho^{\mathrm{raw},s}_{fg}|$ on split $s$ is the distance between the Pearson matrix of the exactly-encoded data and that of the raw data. The floor is attained by any model reproducing all weight-$\leq 2$ statistics of the encoded data and bounds only such models, so objectives that trade marginal fidelity, or that match the dependence structure on the rank scale directly, can undercut it (Sec.~\ref{sec:baselines}) (test evaluated with train-fit edges). The $0.0033$ train-test gap reflects the smaller sample size of the test set and is the reference scale for distinguishing model generalization from sample-statistical noise. \emph{Note added:} an earlier version of this pipeline fit the quantile encoding and drew training targets from the full sample, contaminating the held-out evaluation. All numbers in this paper are from the corrected, split-fitted pipeline, whose retrained headline is $\maerho = 0.068 \pm 0.006$ (Table~\ref{tab:per-seed-L8}).
	
	\subsection{64-qubit \psck{}-\moiqp{} on calorimeter images}
	
	\subsubsection{Configuration}
	All runs use $\Bbits = 8$ ($\nfeat = 64$ qubits), $\Lcomp = 8$ \moiqp{} components, an Erd\H{o}s--R\'enyi gate graph at average degree $6.0$ (graph seed $= \mathrm{user\text{-}seed} + 1$, so the graph itself varies seed to seed with $|\mathcal{G}|$ in the range $256$ to $287$), $1500$ Adam epochs, learning rate $0.02$ with cosine restarts every $100$ epochs, Monte Carlo batch size $M = 4096$ per forward pass, $\psck{}$ mixing weight $\eta = 5$, and the heat-kernel bandwidth set $\{\sigma_b\}$ ($N_\sigma = 5$) fixed by $e^{-2\sigma_b^2} \in \{0.1,\,0.3,\,0.5,\,0.7,\,0.9\}$, so that the per-observable coefficient is $\hat k_\beta^{\mathrm{heat}} = \tfrac{1}{5}\sum_{b=1}^{5}\big(e^{-2\sigma_b^2}\big)^{|\beta|}$. The loss is estimated by an unbiased two-batch $U$-statistic, in which two independent $M = 4096$ Van den Nest batches are drawn per epoch (16 forward passes per epoch at $\Lcomp = 8$), which removes the plug-in estimator's downward bias at the cost of doubling the per-epoch forward count. Per-component gradients are clipped at $\|g\|_2 \leq 10$, and the Adam moment estimates are reset at every cosine-restart boundary. The reported parameters are those of the best-training-loss checkpoint within the $1500$-epoch run (reached at epochs $1460$ to $1494$ across seeds), not the final epoch. We run five independent user-seeds, $\{42, 43, 44, 45, 46\}$, each of which controls the graph, the parameter initialization, and the Monte Carlo stream.

	\subsubsection{Convergence}
\begin{figure}[b]
		\centering
		\includegraphics[width=\columnwidth]{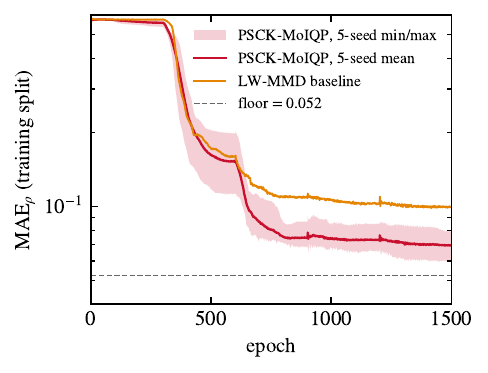}
		\caption{Training-split $\maerho$ as a function of epoch at $\nfeat = 64$. Solid red: \psck{} seed mean across the five training seeds. Shaded band, seed-wise min to max envelope at each epoch. Orange, single-seed Liu--Wang baseline at the same $\Lcomp = 8$ and epoch budget. Dashed line, encoding-fidelity floor on the training split ($0.0515$). The step features every 100 epochs are cosine restarts of Adam.}
		\label{fig:rho-err-overlay}
	\end{figure}

	Fig.~\ref{fig:rho-err-overlay} plots $\maerho$ as a function of epoch across the sweep. All five seeds descend past $\maerho = 0.08$ between epoch 659 and epoch 1458 and sit on their respective asymptotic plateaus by epoch 1500 (see convergence indicator in Appendix~\ref{app:per-seed}). The seed-wise min to max envelope is tight at late times. The ratio of maximum to minimum $\maerho$ across the five seeds stays within $1.3$ throughout the last 300 epochs. The Liu and Wang baseline, at the same $\Lcomp = 8$ and epoch budget, plateaus at $\maerho \approx 0.10$ after $\sim\!500$ epochs and does not close the remaining gap.

	\subsubsection{Main result}
	Table~\ref{tab:headline} summarizes the five-seed sweep. Per-seed numbers are in Appendix~\ref{app:per-seed}. The mean $\maerho = 0.068 \pm 0.006$ is $0.0162$ above the encoding-fidelity floor $\enfloor{train} = 0.0515$. The weight-2 $Z$-correlator mean absolute error is $\maez = 0.0112 \pm 0.0008$, near the Monte Carlo noise floor at $M = 4096$. The \psck{} gain on $\rho$ comes with a small, deliberate cost on the means relative to the heat-kernel baseline (Table~\ref{tab:headline}), the expected consequence of shifting loss weight onto the correlation directions through the rank-$P$ term. Three of the five seeds achieve $\maerho < 0.07$ (seeds 42, 43, 46. The remaining two are $0.0710$ and $0.0770$, five-seed mean $0.068 \pm 0.006$) and all five converge (convergence indicator $\Delta < 0.005$ for each seed, see Appendix~\ref{app:per-seed}).
	
	\begin{figure*}[t]
		\centering
		\subfloat[Ground truth $\rho^{\mathrm{data}}_{fg}$]{\includegraphics[width=0.32\textwidth]{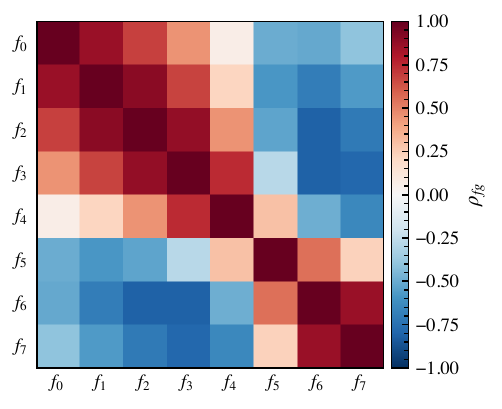}}\hfill
		\subfloat[PSCK to MMD $\rho^{\mathrm{PSCK}}_{fg}$]{\includegraphics[width=0.32\textwidth]{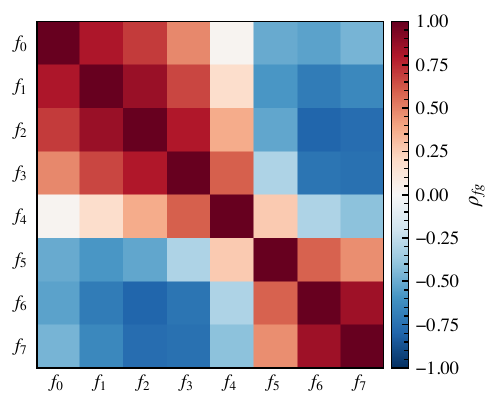}}\hfill
		\subfloat[LW to MMD $\rho^{\mathrm{LW}}_{fg}$]{\includegraphics[width=0.32\textwidth]{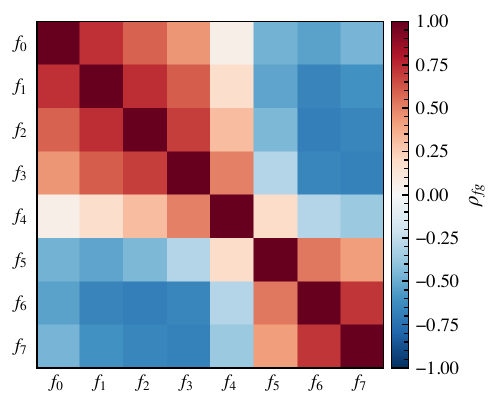}}\\[6pt]
		\subfloat[$\rho^{\mathrm{PSCK}}_{fg} - \rho^{\mathrm{data}}_{fg}$]{\includegraphics[width=0.32\textwidth]{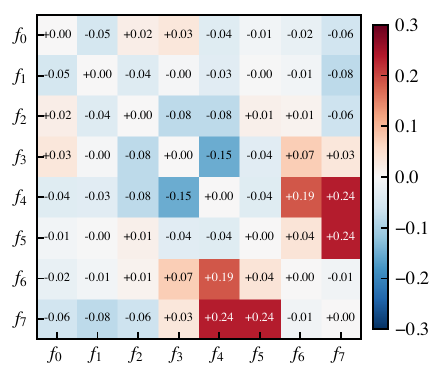}}\hfill
		\subfloat[$\rho^{\mathrm{LW}}_{fg} - \rho^{\mathrm{data}}_{fg}$]{\includegraphics[width=0.32\textwidth]{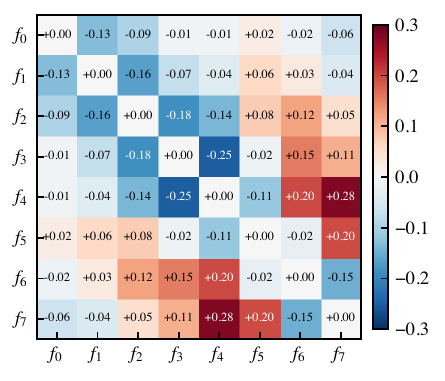}}\hfill
		\subfloat[$\rho^{\mathrm{PSCK}}_{fg} - \rho^{\mathrm{LW}}_{fg}$]{\includegraphics[width=0.32\textwidth]{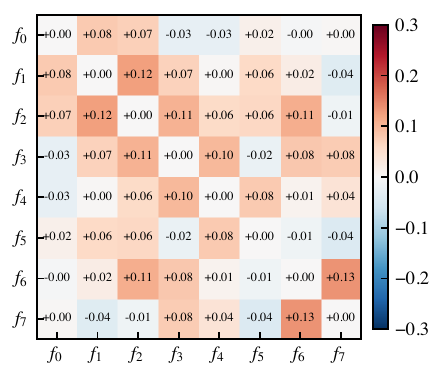}}
		\caption{Pairwise correlation matrices and residual structure at $\nfeat = 64$, $\Lcomp = 8$, $1500$ epochs. Top row, Pearson correlation matrices $\rho_{fg}$ on a shared color scale, for the training-split data (left), the trained \psck{}-\moiqp{} model at seed 46 (middle), and the single-seed Liu and Wang baseline at the same configuration (right). Bottom row, signed residuals $\Delta_{fg} = \rho^{\mathrm{model}}_{fg} - \rho^{\mathrm{data}}_{fg}$ for \psck{} (left) and Liu and Wang (middle), and the pair-level advantage map $|\Delta_{\mathrm{LW}}| - |\Delta_{\mathrm{PSCK}}|$ (right), all on a shared symmetric color scale at $\pm 0.3$. Red cells in the advantage map are pairs where \psck{} has the smaller absolute residual, blue cells the reverse. Cell annotations are signed residual values.}
		\label{fig:rho-matrices}
	\end{figure*}
	
	\subsubsection{Pairwise correlation reconstruction}
	The top row of Fig.~\ref{fig:rho-matrices} shows the recovered Pearson correlation matrix $\rho_{fg}$ of \psck{}-\moiqp{} (middle) and the Liu and Wang baseline (right) against the training-split data (left). \psck{} recovers the full amplitude range of the data, $|\rho| \in [-0.85, +0.85]$, (seed 46 shown in Fig.~\ref{fig:rho-matrices}. The best seed, 43, reaches $\maerho = 0.061$).
	The Liu and Wang baseline saturates at $\maerho = 0.095$, roughly twice the encoding floor, and compresses the amplitude range visibly at the band extremes. We note the asymmetry in this comparison. The \psck{} figure is a five-seed mean with a $\pm 0.006$ spread, while the Liu and Wang figure is a single seed at the same configuration and epoch budget. The gap is several times the \psck{} seed spread and the qualitative signature (a spatially structured residual, Fig.~\ref{fig:rho-matrices}(e)) is not a quantity a seed change is expected to move, and a multi-seed baseline would put the comparison on the same footing as the rest of Table~\ref{tab:headline}.
	
	The bottom row of Fig.~\ref{fig:rho-matrices} decomposes the comparison into signed residuals $\Delta_{fg} = \rho^{\mathrm{model}}_{fg} - \rho^{\mathrm{data}}_{fg}$ and a pair-level advantage map. The \psck{} residual has mean bias $\bar\Delta = +0.004$ and off-diagonal RMS $0.088$, statistically zero at the Monte Carlo noise floor. The Liu and Wang residual has $\bar\Delta = -0.008$ and RMS $0.124$, with a spatially structured sign pattern, systematically negative on the intra-shower block $(f_2,\dots,f_4)$ and positive on the peripheral block $(f_5,\dots,f_7)$, which is the amplitude-compression signature of the top row made explicit on the residual. \psck{} is strictly better than Liu and Wang on $20$ of $28$ off-diagonal pairs. The $8$ pairs on which Liu and Wang is strictly better all sit at $|\Delta_{\mathrm{LW}}| - |\Delta_{\mathrm{PSCK}}| \geq -0.04$, within Monte Carlo noise.

	\subsubsection{Generalization}
	Because the corrected pipeline fits the encoding and draws all training targets from the training split only, the held-out comparison is now a legitimate generalization statement. On the untouched $9\,536$-sample test split the per-seed train-test gap is $0.0019 \pm 0.0005$, below the sample-statistical gap of the encoding floor itself, $0.0548 - 0.0515 = 0.0033$, so the model generalizes at the resolution the two splits allow. For contrast, the CorrMSE reference of Sec.~\ref{sec:baselines} degrades from $0.0028$ to $0.0076$ out of sample ($2.7\times$), the signature of an objective that fits split-specific correlation noise. Per-seed test-split statistics from the corrected runs will replace Table~\ref{tab:per-seed-L8} of Appendix~\ref{app:per-seed} wholesale.
	
	\subsection{Mixture-size scan}
	\label{sec:mix-scan}
	
	To confirm that the mixture is the source of the correlation expressivity, we sweep $\Lcomp$ on the calorimeter data at a fixed sparse graph and equal training budget, with all other settings as in the headline run. Table~\ref{tab:mix-scan} shows the result. A single IQP ($\Lcomp = 1$) is capacity-limited. Its $\maerho$ sits roughly an order of magnitude above the encoding floor and does not move with additional epochs, so the limitation is representational rather than optimization-related. Adding components reduces $\maerho$ by roughly a factor of six between $\Lcomp = 1$ and $\Lcomp = 4$, after which the scan flattens. The $\Lcomp = 8$ entry ($0.11$) is not an improvement on the $\Lcomp = 4$ entry ($0.09$) at this reduced resolution, and the two differ by less than the seed-to-seed scatter of the headline configuration. Beyond that point additional components buy no measurable $\maerho$ and only raise the compiled gate weight and ancilla count, which is why $\Lcomp = 8$ is a deliberate, modest choice rather than a maximal one.
	
	Two points fix how the entries at $\Lcomp = 4$ and $\Lcomp = 8$ should be read. The scan runs at a reduced resolution and at a single seed, and the difference between those two entries ($0.09$ and $0.11$) is smaller than the five-seed scatter of $\pm 0.006$ at the headline configuration, so it locates the point beyond which no further gain is resolvable at this budget rather than a measured optimum. At the headline $\Bbits = 8$, $\nfeat = 64$ configuration the $\Lcomp = 8$ entry is the $\maerho = 0.068 \pm 0.006$ of Table~\ref{tab:headline}. The saturation itself also follows from a rate argument that needs no error bars. By Shapley--Folkman--Starr the uniform $\Lcomp$-mixture approaches the convex hull of the single-IQP correlator variety at rate $\sqrt{P}\,\mathrm{rad}/\Lcomp$ on the $P = \binom{\Dfeat}{2} = 28$ scored Pearson directions, so the approach is $O(1/\Lcomp)$ with a constant set by the number of scored functionals rather than by any combinatorial count of modes in the data.
	
	\begin{table}[t]
		\centering
		\caption{Mixture-size scan on the calorimeter data, \psck{} objective, fixed sparse graph, equal budget per $\Lcomp$. Single seed at a reduced resolution ($\Bbits = 4$, $\nfeat = 32$), where the encoding-fidelity floor is $\enfloor{train} = 0.051$. The single IQP under-fits the correlation sector by an order of magnitude and the mixture reaches the floor. See Sec.~\ref{sec:mix-scan} for the reading of the $\Lcomp = 4$ and $\Lcomp = 8$ entries.}
		\label{tab:mix-scan}
		\begin{tabular}{c c c c}
			\hline\hline
			$\Lcomp$ & params & $\maerho$ & $\rfit$ \\
			\hline
			$1$ & $|\mathcal{G}|$ & $0.52$ & $0.55$ \\
			$2$ & $2|\mathcal{G}|$ & $0.24$ & $0.91$ \\
			$4$ & $4|\mathcal{G}|$ & $0.09$ & $0.98$ \\
			$8$ & $8|\mathcal{G}|$ & $0.11$ & $0.98$ \\
			\hline
		\end{tabular}
	\end{table}
	
	\subsection{Per-feature distributional metrics}
	\label{sec:per-feature}
	
	The MMD$^2$ training objective fixes only the weight $\leq 2$ $Z$-correlators of the model. To see what happens to the per-feature marginals as a side effect we use a sampling-free diagnostic. For each cell $f$ we evaluate all $2^{\Bbits} - 1 = 255$ intra-feature $Z$-correlators $\langle \prod_{k \in S} Z_{fB+k}\rangle$ for $S \subseteq \{0, \dots, \Bbits-1\}$ and apply the inverse Walsh--Hadamard transform to recover the exact per-feature marginal $p^{(f)}_{\boldsymbol\theta}(x_f) = (1/2^{\Bbits}) \sum_\beta \Zth^{(f)} (-1)^{\beta \cdot x_f}$. Mapping bit patterns to integer levels gives $p^{(f)}_{\boldsymbol\theta}(v)$ for $v \in \{0, \dots, 2^{\Bbits}-1\}$, directly comparable to the empirical level histogram.
	
	Evaluating all eight cells costs $\Dfeat \cdot (2^{\Bbits} - 1) = 2040$ Van den Nest MC forward passes, about $30$ minutes per trained model. No sampling from the model is involved.
	

	Averaged over the five training seeds, the test-split $W_1$ distances on integer levels read $f_0 = 10.5$, $f_1 = 18.3$, $f_2 = 17.3$, $f_3 = 10.8$, $f_4 = 5.7$, $f_5 = 7.2$, $f_6 = 13.4$, $f_7 = 9.5$, out of a level support of $256$, with Kolmogorov--Smirnov statistics $0.074$, $0.147$, $0.145$, $0.090$, $0.049$, $0.062$, $0.121$ and $0.085$ respectively. The per-model mean is $11.6$ levels. Individual cell values span $2.0$ to $24.9$ levels and $\mathrm{KS} \in [0.031, 0.226]$.
	
	\emph{Which} cells miss is not stable across seeds. At seed 42 the largest deviations sit on $f_6$ and $f_7$ ($W_1 = 24.6$ and $22.9$ levels). At seed 44 they sit on $f_0$ and $f_1$ ($19.7$ and $24.4$). At seed 45 no cell exceeds $9.5$. Per-model means range from $6.0$ levels (seed 45) to $14.4$ (seed 43). A pattern that moves with the parameter initialization cannot be a property of the dataset, so the residual marginal mismatch is a model and optimization effect and not a signature of the shower physics.

	The structural reason is visible in the encoding rather than in the physics. Quantile binning is a rank transform, so wherever the per-cell bin occupancies are balanced each of the $2^{\Bbits}$ levels is equally likely, the $\Bbits$ bits of a feature are independent fair coins, and \emph{every} intra-feature $Z$-correlator of the encoded data vanishes. The target level marginal is then uniform for all eight cells alike, and reproducing it requires the model to drive all $2^{\Bbits} - 1 = 255$ intra-feature correlators of each cell to zero. The $\mathcal{O}_2$ objective constrains only $\Bbits + \binom{\Bbits}{2} = 36$ of them. The remaining $219$ per cell are inherited from the ansatz. That is what the diagnostic measures, and extending $\mathcal{O}_K$ to weight-3 intra-feature observables remains the natural next step (see Sec.~\ref{sec:future}), but as a statement about the objective, not about heavy tails. The skewed, heavy-tailed structure of the energy distributions is carried entirely by the decoder and never reaches the encoded target.
	
	This argument is conditional on the encoding being balanced, and the one mechanism that breaks it is a point mass, quantile binning cannot split an atom, so a cell whose energy distribution has a spike at zero acquires a non-uniform target marginal and can then differ from its neighbours for a genuinely data-driven reason. Which of the two regimes each cell is in is settled by the achieved per-cell bin occupancies, which we report with the code companion.
	
	\subsection{Trainability scan} \label{sec:bp-scan}
	
	We test for barren plateaus by drawing $K_{\mathrm{init}} = 200$ independent parameter initializations from $\mathcal{N}(0, \sigma^2 I)$ at $\sigma = 0.1$ for each $\nfeat \in \{16, 24, 32, 48, 64\}$. For each draw we compute $\nabla \Lpsck$ by Van den Nest at $M_{\mathrm{grad}} = 2048$ and record the per-gate variance $\mathrm{Var}_\theta[\partial\mathcal{L}/\partial\theta_G]$. The same scan is run for the Liu and Wang baseline.
	
	\begin{figure}[t]
		\centering
		\includegraphics[width=\columnwidth]{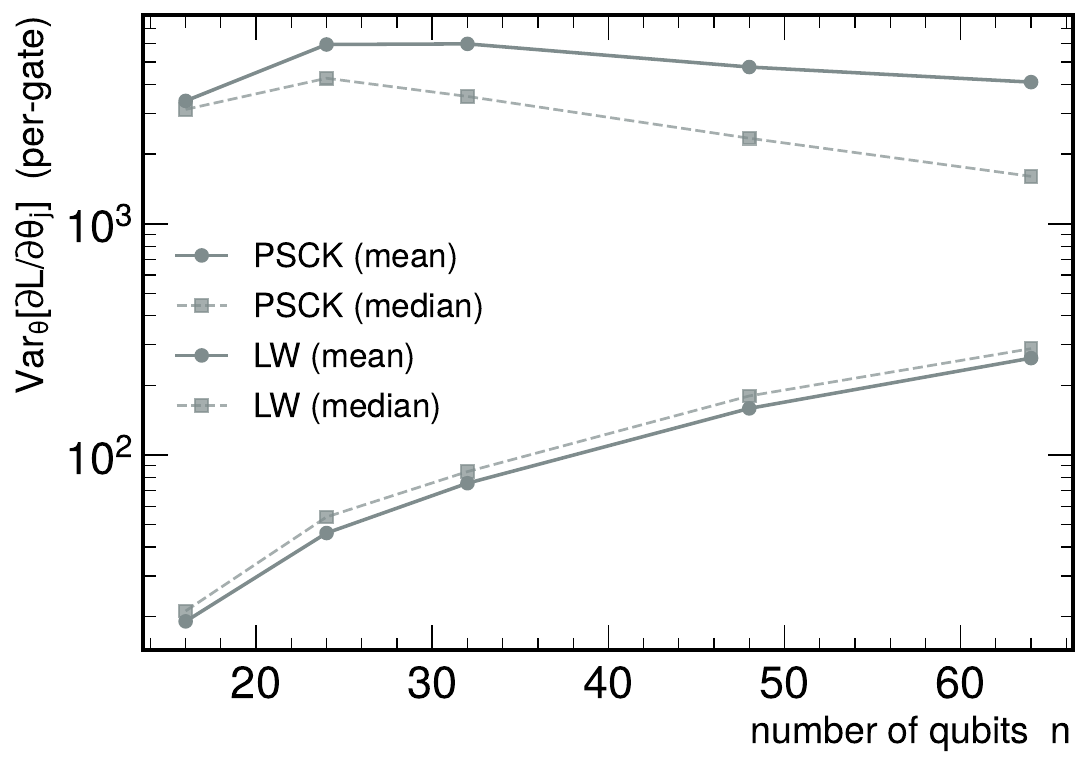}
		\caption{Per-gate gradient variance $\mathrm{Var}_\theta[\partial L/\partial\theta_G]$ as a function of qubit count $\nfeat$ for \psck{}  and Liu and Wang, at small-perturbation initialization. \psck{} is essentially flat, $\propto \nfeat^{0.05}$. Liu and Wang scales as $\propto \nfeat^{1.87}$. Neither shows exponential decay in the regime studied.}
		\label{fig:bp-scan}
	\end{figure}
	
	For both losses the per-gate variance is flat or polynomially growing in $\nfeat$ (Fig.~\ref{fig:bp-scan}), with $\mathrm{Var}[\partial \Lpsck / \partial\theta] \propto \nfeat^{0.05}$ and $\mathrm{Var}[\partial \mathcal{L}_{\mathrm{LW}} / \partial\theta] \propto \nfeat^{1.87}$. Neither shows the exponential decay $\mathrm{Var} \propto e^{-\alpha \nfeat}$ that defines a barren plateau. This is empirical and complements the analytic results of Refs.~\cite{kumar2026data_dependent_init,hakkaku2026characterizing}. At $\sigma = 0.1$ with sparse Erd\H{o}s--R\'enyi gate connectivity, IQP Born machines on calorimeter-style data remain trainable up to $\nfeat = 64$. Full tables and caveats are in Appendix~\ref{app:bp-supp}.

	The trainability-relevant comparison between \psck{} and Liu and Wang is the scale-invariant relative gradient strength $\langle\|\nabla\mathcal{L}\|^2\rangle / \langle\mathcal{L}\rangle^2$, since Adam-type optimizers divide by a running estimate of the gradient second moment and are invariant to the loss rescaling that separates the two objectives. This quantity sits in the range $1.0$ to $1.2$ across all $\nfeat$ we tested, so the two losses are equally trainable by this measure. The practical advantage of \psck{} is one of gradient \emph{direction}, with the rank-$P$ correction aligning descent with correlation-sensitive parameter directions. The absolute per-gate-variance ratio ($16\times$ at $\nfeat = 64$, $178\times$ at $\nfeat = 16$) reflects only the loss-scale difference.

	\subsection{Classical references}
	\label{sec:baselines}
	
	Two classical references bound what the pairwise metric alone can certify, and we report both at the identical configuration, split, and evaluation as the headline runs.
	
	The first is \emph{CorrMSE}, the explicit auxiliary Pearson-MSE objective in the spirit of Rehm \emph{et al.}~\cite{rehm2024qag}, trained with the same Van den Nest machinery, graphs, seeds, and epoch budget (it is classically trainable and deploys as the same circuit class, so it cannot be excluded on trainability or hardness grounds). Across five seeds it reaches $\maerho = 0.0028 \pm 0.0005$ on the training split, $20\times$ below \psck{} and far below the encoding floor. It buys this by breaking the encode/decode contract. Its level marginals collapse (per-feature Wasserstein-1 of $24$ to $31$ levels and KS of $0.22$ to $0.27$, against $8$ to $15$ and $0.07$ to $0.13$ for \psck{}), so its decoded energy distributions would be visibly wrong and no data-fitted decode map can repair a wrong level distribution, and its $\rho$ error degrades $2.7\times$ out of sample ($0.0028 \to 0.0076 \pm 0.0003$) where \psck{}'s barely moves. The floor of Table~\ref{tab:headline} bounds only models that reproduce the encoded data's full weight-$\leq 2$ statistics. An objective free to trade the marginals escapes it, which is why the floor is a property of the encoding, not a lower bound on all models.
	
	The second is a \emph{Gaussian copula} with the empirical normal-scores correlation of the training split, sampled, decoded through the empirical quantile function, and encoded through the identical train-fit edges. It matches the pairwise target at $\maerho = 0.043$ (train) / $0.046$ (test) with essentially exact marginals ($W_1 \approx 0.1$ levels, $\maez = 0.0047$) at negligible cost, undercutting even the floor because the quantile encoding is a rank transform and the copula lives on the rank scale natively. This is the classical yardstick for the task as evaluated. \emph{On weight-$\leq 2$ statistics of a quantile encoding, the marginals are fitted classically by the encode/decode pair and the learnable content is the pairwise copula, which a Gaussian copula reproduces essentially exactly}. The contribution of the quantum construction is therefore not pairwise superiority. It is (i) classical trainability of a Born machine whose deployed form is a genuine quantum circuit; (ii) an exact compilation and constant-depth hardware realization at the headline geometry (Sec.~\ref{sec:hardware}), with the $10^2$-qubit extension reported separately, and (iii) a path to higher-order dependence, through weight-$\geq 3$ objectives fixing structure a Gaussian copula cannot host, on a substrate whose worst-case instances are conjecturally hard to sample, though the instances deployed here are not (Sec.~\ref{sec:scope}). Generation in energy units uses the committed decoder (per-feature train-split bin medians. Code companion), closing the level-to-energy path that earlier versions left unspecified.

	\section{Hardware deployment at the headline geometry}
	\label{sec:hardware}
	
	This section reports the execution of the compiled model at the paper's headline geometry (the same $8$-feature dataset, the same $\Bbits = 8$ encoding with the same split-correct floors, and the same $\Lcomp = 8$ mixture) as one dcIQP dynamic circuit on $67$ physical qubits of \texttt{ibm\_kingston} (IBM Heron~r2, heavy-hex connectivity, CZ native gate, mid-circuit measurement with classical feedforward~\cite{baumer2024dynamic}). The larger-register campaign of this program, scaling the protocol to the $100$-qubit regime on a finer-grained dataset, is reported separately.
	
	\subsection{Hardware-native model and protocol}
	\label{sec:hw-model}
	\label{sec:hw-protocol}
	
	The headline models described in Sec.~\ref{sec:experiments} use Erd\H{o}s--R\'enyi gate graphs whose weight-2 gates land on arbitrary qubit pairs and would incur a prohibitive SWAP overhead. For deployment we retrain at the identical geometry on the native coupling graph of the target device, in the Walsh-sparse parameterization whose coherent compiled form is a degree-3 IQP circuit (Sec.~\ref{sec:dciqp}), with $\nfeat = \Dfeat\Bbits = 64$ data qubits, $a = 3$ ancillas ($\Lcomp = 8$), for $67$ physical qubits in total. The induced native graph has $|\mathcal{G}| = 133$ gates, $64$ weight-1 and $69$ weight-2, of which $38$ are intra-feature and $31$ cross-feature. The executed two-qubit budget is $2|\mathcal{G}_2| = 138$ CZ per shot, independent of $\Lcomp$ (exactly one feedforward branch runs per shot), with zero SWAP insertions by construction.
	
	The deployed model is trained at this geometry with the spectral-flattening objective of this programme rather than with the headline \psck{} loss. It uses \psck{} at $\eta = 5$, plus a split-sample U-statistic penalty $\lambda\,\mathbb{E}_\beta[\hat z_A \hat z_B]$ over randomly drawn cross-feature Walsh words of weight $3$ to $5$, at $\lambda = 300$, for $500$ epochs at a single seed ($42$), with the frozen scrambling layer disabled ($\sigma_f = 0$). The penalty trades correlation fidelity for spectral flatness and the trade is visible. The deployed model's noiseless correlation error is $\maerho = 0.0747$, against $0.068 \pm 0.006$ for the headline \psck{} sweep. Every hardware metric below is therefore to be read against $0.0747$, not against the headline value. The parameter record and the training driver ship with the code companion.
	
	The deployed object is dcIQP. The ancillas are Hadamard-prepared and measured first, classical feedforward selects the component, and all $64$ data bits are recorded shot-aligned with the branch label $(\ell, x)$. One job, $5 \times 10^{4}$ shots. All evaluation below is split-correct, with quantile edges fit on the $38\,146$-sample training split only, encoding floors $\enfloor{train} = 0.0515$ and $\enfloor{test} = 0.0548$.
	
	\subsection{Results}
	\label{sec:hw-results}
	
	All hardware curves and numbers in this subsection are \emph{raw} device samples. No mitigation of any kind is applied. Hardware levels are decoded to energies by the faithful empirical inverse of the quantile encoding (a decoded value is drawn uniformly from the training-split pool of the same cell to level bin), so residual mismatch is attributable to the sampled level distribution, not the decoder. The encode to decode round trip on the data itself costs $\mathrm{KS} = 0.006$ on the total-energy distribution. Error bars on hardware histograms are the quadrature sum of the per-bin Poisson term and a $2.8\%$ relative job-to-job drift systematic taken from repeated executions of the same protocol class on the same device generation. A same-configuration duplicate job will replace this proxy.

	The measured branch labels deviate from uniform by total variation $0.024$, consistent with per-ancilla readout asymmetry rather than circuit error. Within the job the device is statistically stable at the shot-noise level. Comparing the first and second $2.5\times10^{4}$ shots per branch and observable gives $\chi^2/\mathrm{dof} = 1.17$ over $2\,912$ correlator comparisons, so all drift relevant to this record is a between-calibration effect.
	
\begin{figure*}[t]
		\centering
		\subfloat{\includegraphics[width=0.24\textwidth]{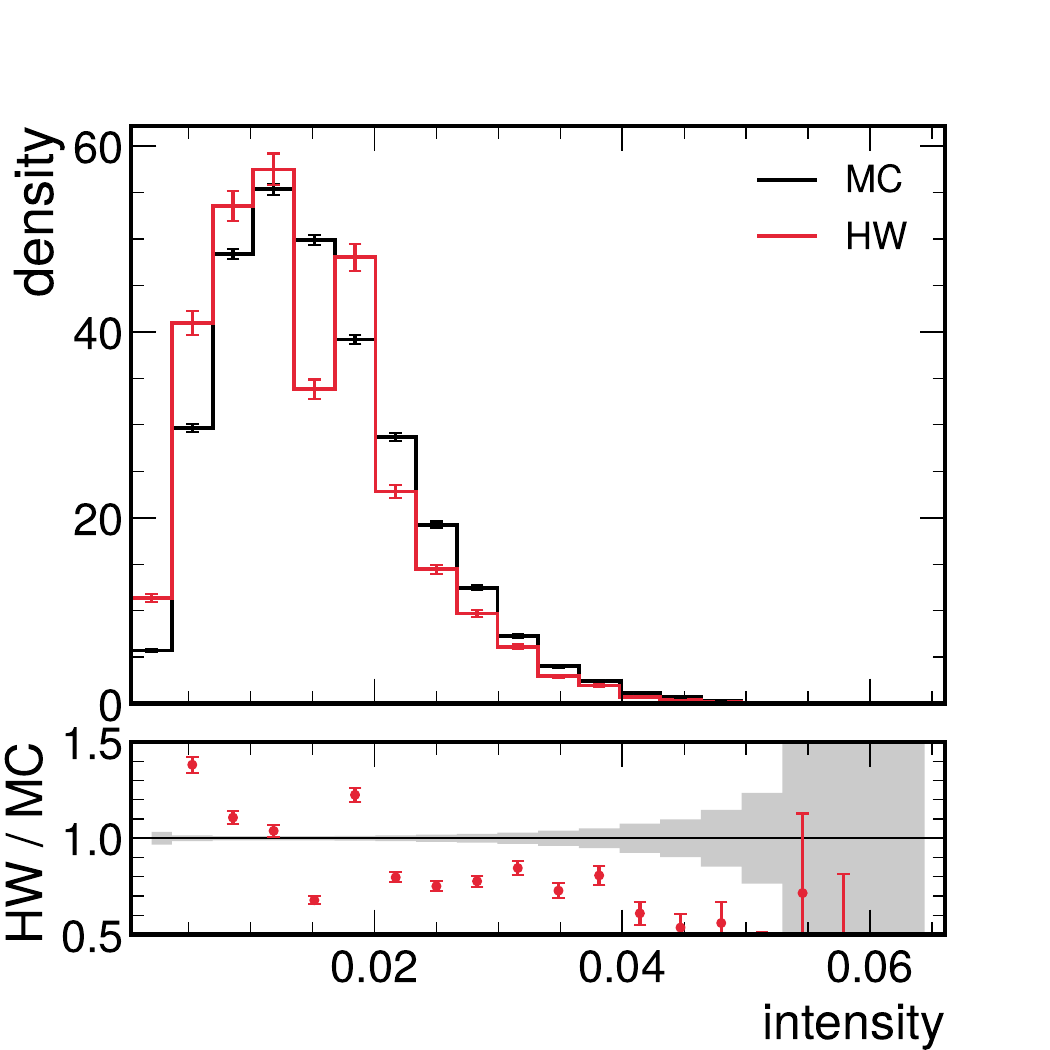}}\hfill
		\subfloat{\includegraphics[width=0.24\textwidth]{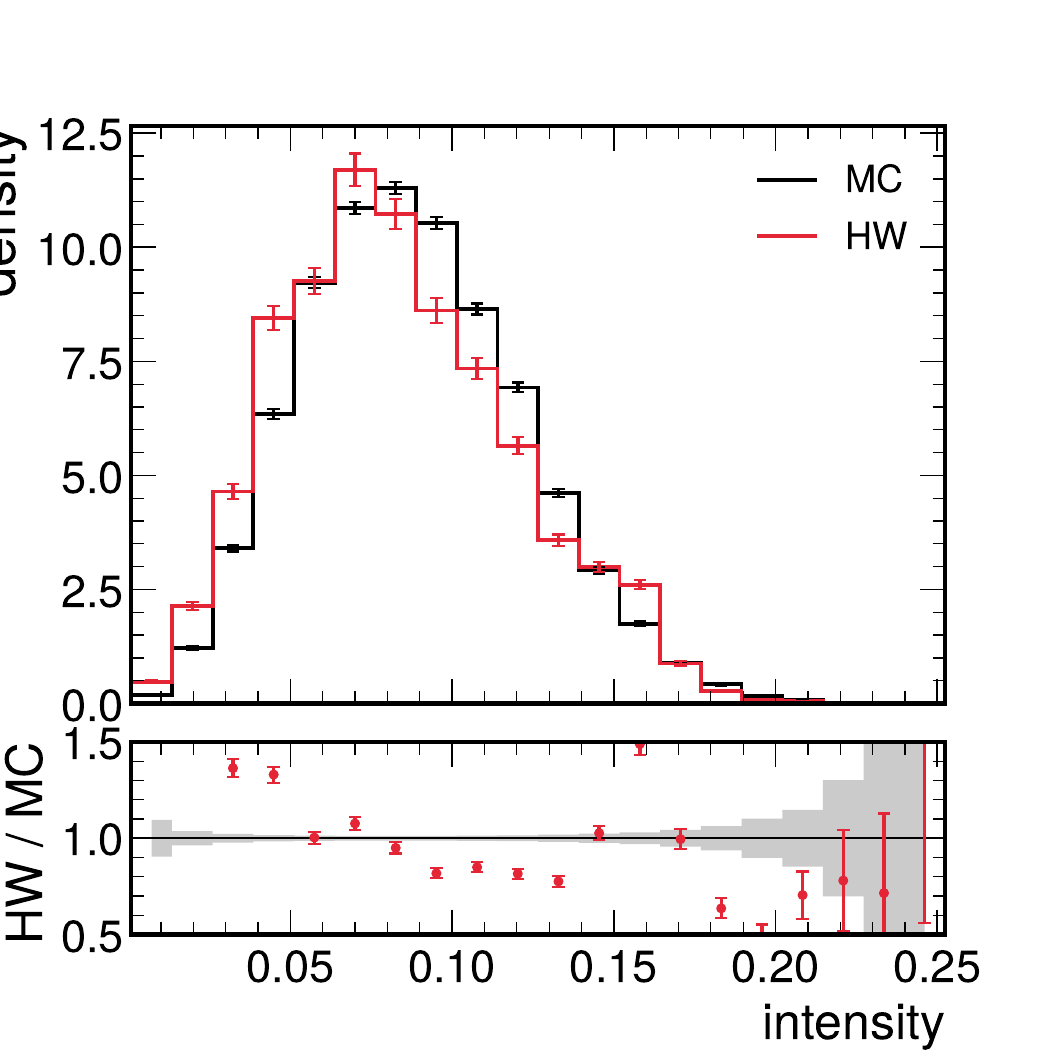}}\hfill
		\subfloat{\includegraphics[width=0.24\textwidth]{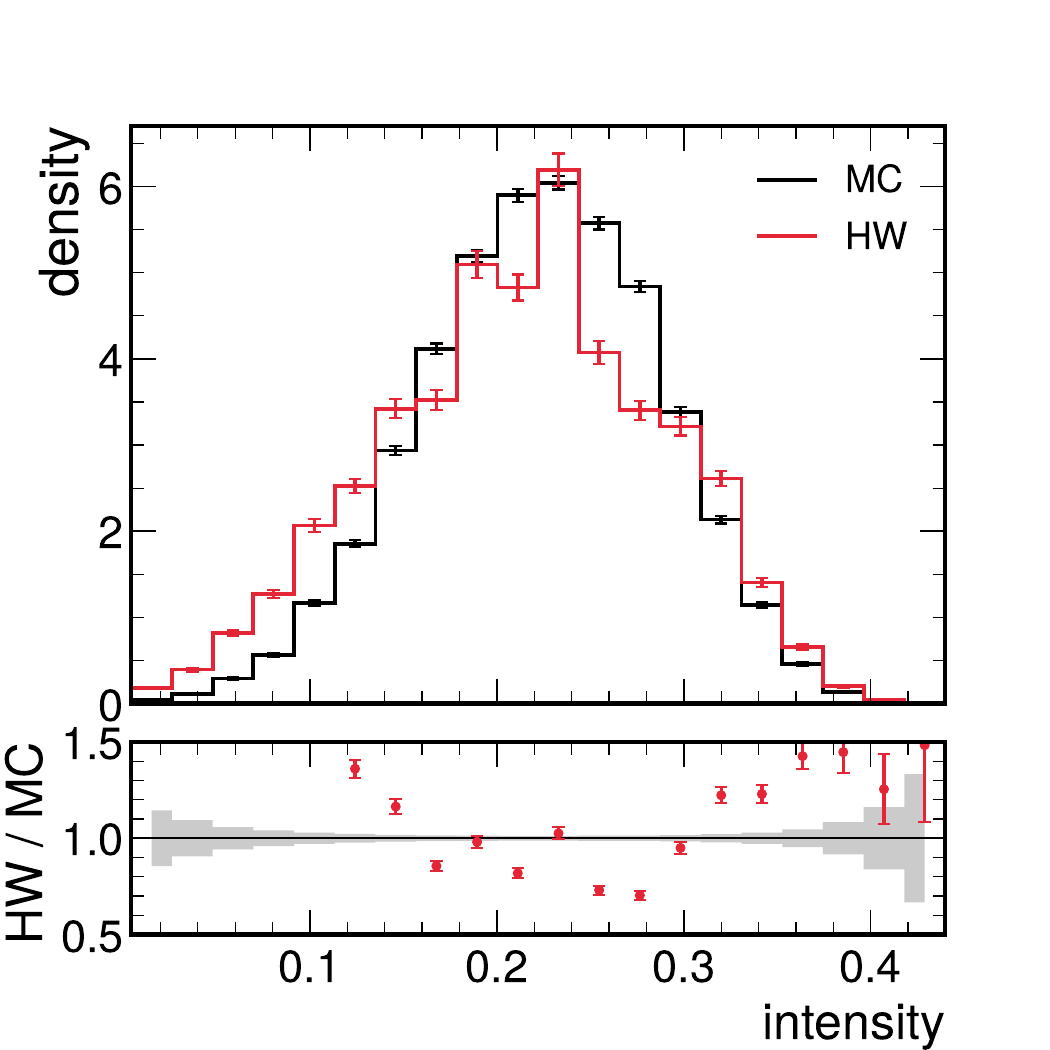}}\hfill
		\subfloat{\includegraphics[width=0.24\textwidth]{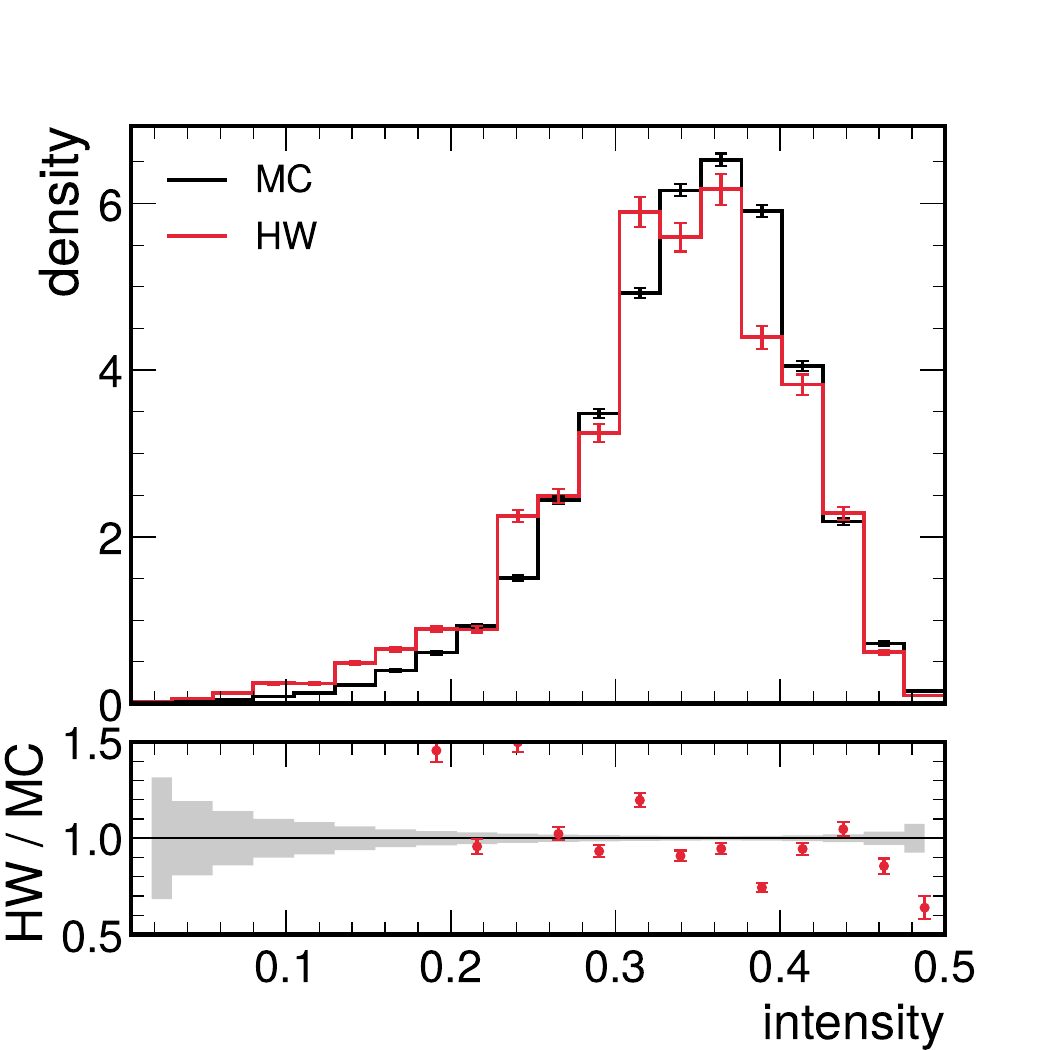}}\\
		\subfloat{\includegraphics[width=0.24\textwidth]{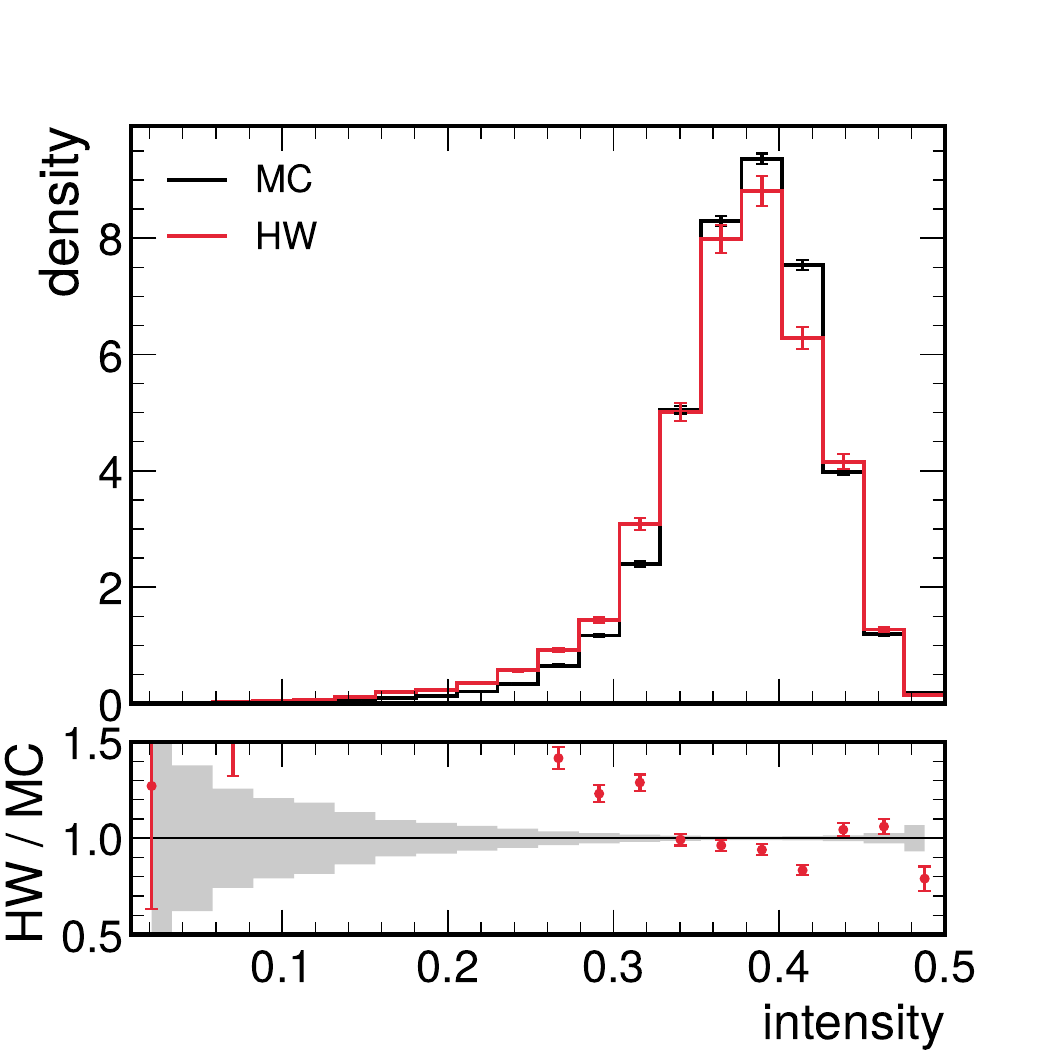}}\hfill
		\subfloat{\includegraphics[width=0.24\textwidth]{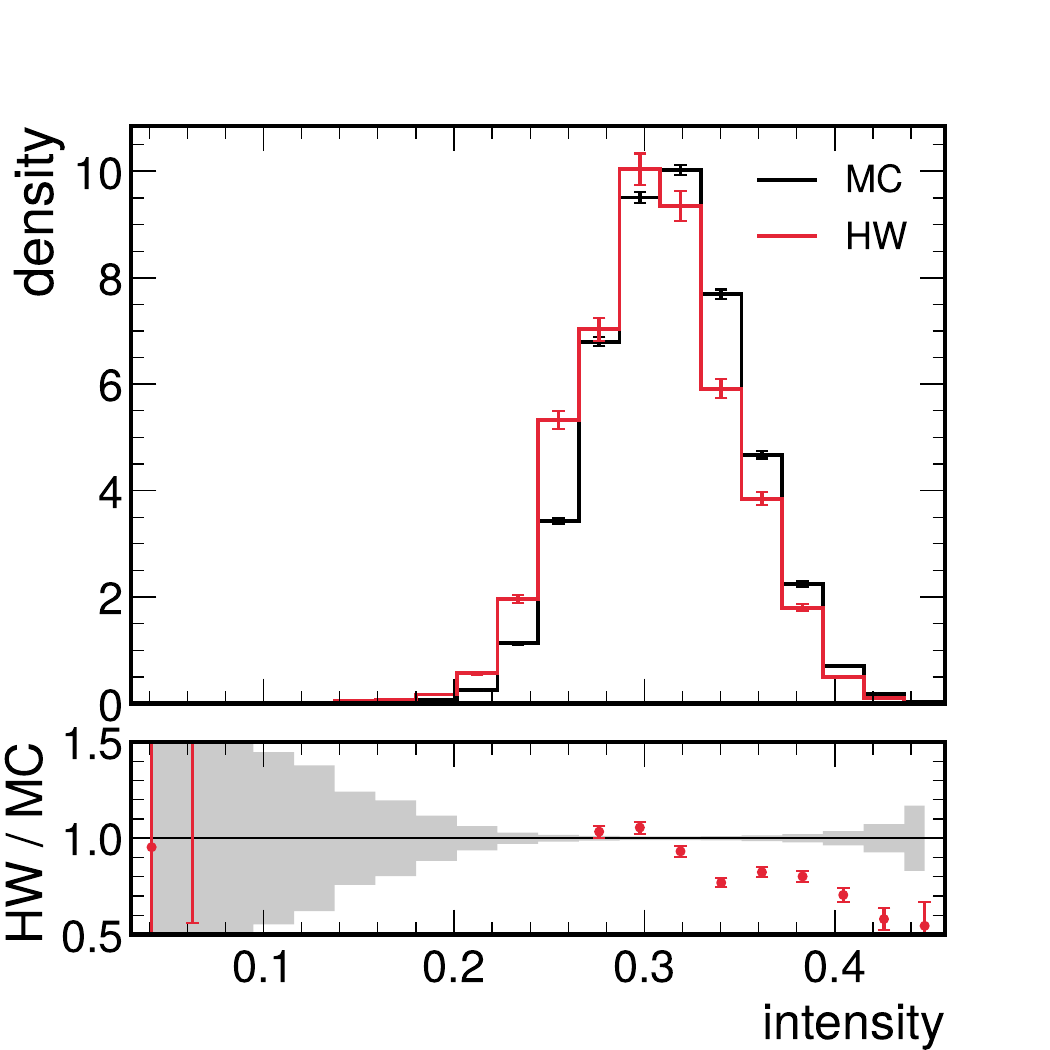}}\hfill
		\subfloat{\includegraphics[width=0.24\textwidth]{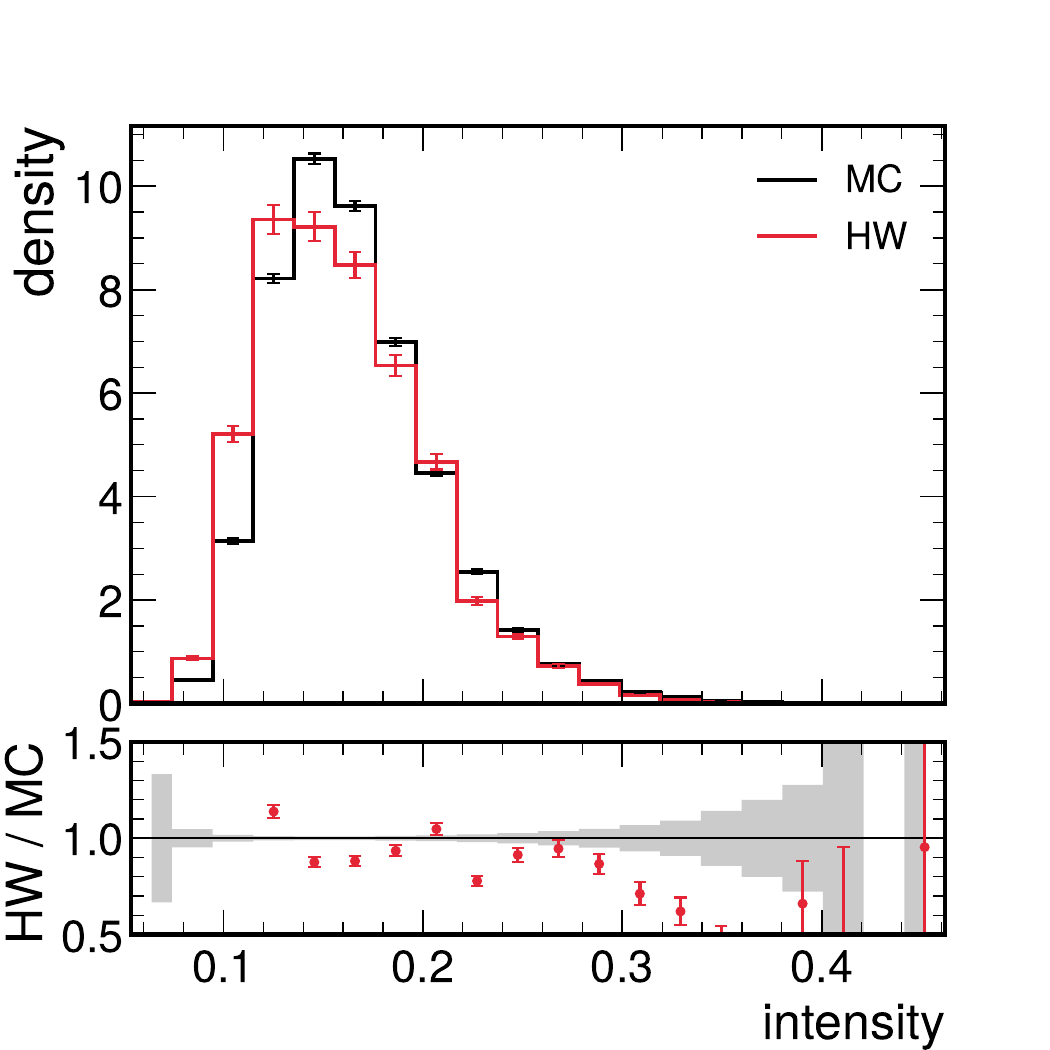}}\hfill
		\subfloat{\includegraphics[width=0.24\textwidth]{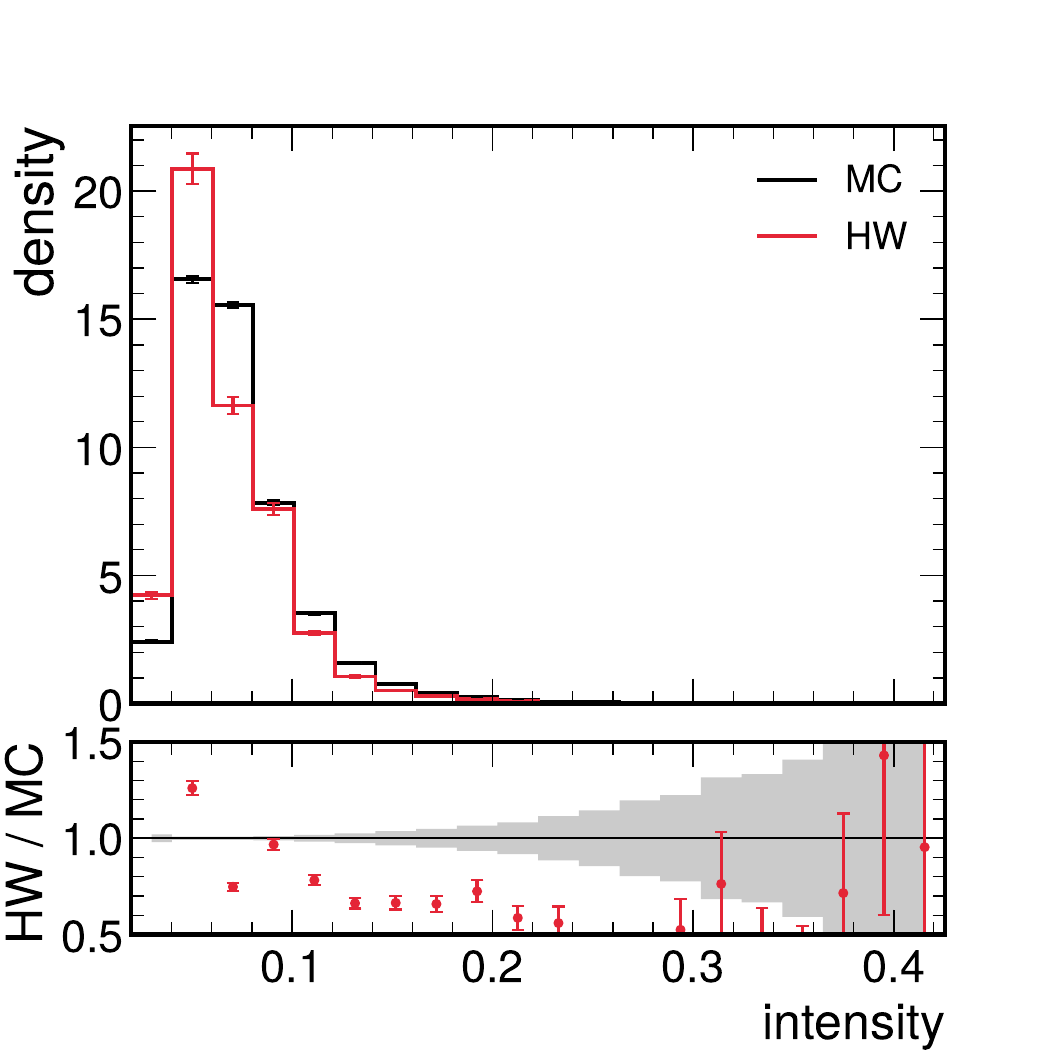}}
		\caption{Per-cell intensity spectra at the headline geometry ($67$ physical qubits, \texttt{ibm\_kingston}, $5\times10^{4}$ shots), all eight calorimeter cells. Black step, calorimeter data (MC), Poisson bars. Red step, raw hardware samples decoded through the empirical inverse of the train-fit quantile encoding, with vertical bars equal to the quadrature sum of the Poisson term and the $2.8\%$ drift-systematic proxy. Lower pads, hardware/MC ratio. Per-cell KS, mean $0.087$, worst $0.127$.}
		\label{fig:hw64-marg}
	\end{figure*}

\begin{figure}[t]
		\centering
		\includegraphics[width=0.95\columnwidth]{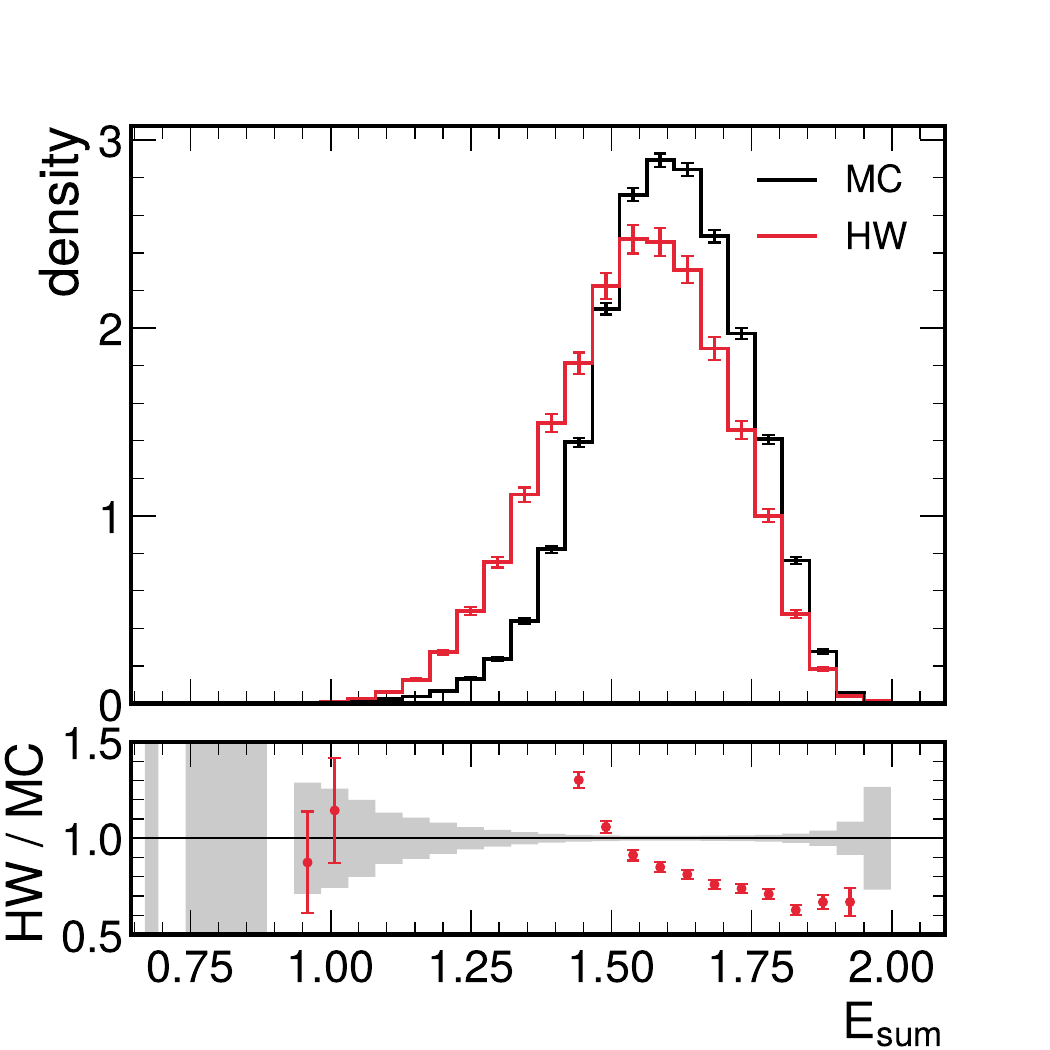}
		\caption{Total deposited energy $E_\mathrm{sum}$, shown as raw hardware samples (red) against the calorimeter data (black), conventions as in Fig.~\ref{fig:hw64-marg}. $\mathrm{KS} = 0.151$ against an encode to decode ceiling of $0.006$. $E_\mathrm{sum}$ is a genuinely joint observable of the $64$-bit distribution, so no correlator-based diagnostic reaches it. The noiseless model curve is obtainable by exact classical sampling of the deployed model (min-fill width $\leq 3$, doubled sampling width $7$, $0.29$\,s per sample) and is drawn in grey for reference. The hardware to model comparison, rather than hardware to data, is what isolates the device contribution.}
		\label{fig:hw64-esum}
	\end{figure}

	Figure~\ref{fig:hw64-marg} shows all eight per-cell intensity spectra. The raw hardware marginals track the data at a per-cell Kolmogorov--Smirnov distance of $0.087$ on average, $0.127$ at worst. Figure~\ref{fig:hw64-esum} shows the total deposited energy $E_\mathrm{sum} = \sum_f E_f$, the simplest observable of the joint distribution, at $\mathrm{KS} = 0.151$ against the $0.006$ encoding ceiling. This is a faithfulness comparison rather than a sampling-hardness claim, for the following reason. An IQP amplitude is a complex Ising partition function on the gate graph, so evaluating one costs $2^{w+1}$ with $w$ the treewidth. Exact \emph{sampling} is more expensive and the difference matters here: $p(x) = |A(x)|^2$ is a sum over a pair of latents $(z, z')$, and summing $p$ over a bit not yet fixed contracts $\sum_{x_q} (-1)^{x_q(z_q + z'_q)} = 2\,\delta_{z_q, z'_q}$, so unfixed bits remain single while every fixed bit carries an independent pair. Sequential-conditional sampling therefore runs on a network doubled over the fixed prefix, of width $\approx 2w$, at cost $2^{2w+1}$ per conditional. The deployed hardware-native graph is a heavy-hex subgraph of min-fill width $w \leq 3$, giving a doubled sampling width of $7$ and a measured $0.29$\,s per $64$-bit sample on one CPU core. The five Erd\H{o}s--R\'enyi graphs of the headline runs have $w \leq 21, 21, 22, 23, 24$ at graph seeds $43$ to $47$, their amplitudes are cheap ($4\times10^{6}$ to $3\times10^{7}$ operations) but their doubled sampling width is $\approx 43$ to $49$, so those instances are not classically samplable at this scale. Every deployment object in this paper, mixture, dcIQP, \ciqp{} marginal, inherits the component-graph width up to a factor $\Lcomp$. A classical reference curve for $E_\mathrm{sum}$ from the trained model is therefore available at negligible cost, and we recommend it be shown alongside the hardware histogram.
	
\begin{figure*}[t]
		\centering
		\subfloat{\includegraphics[width=0.32\textwidth]{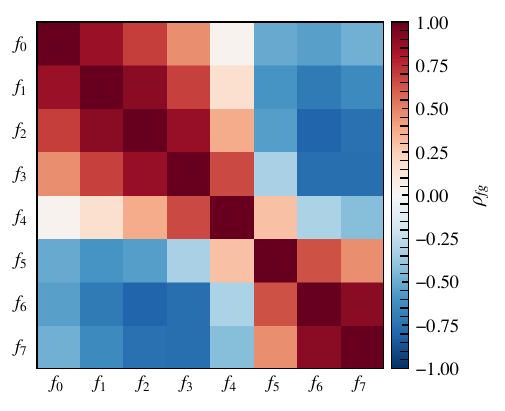}}\hfill
		\subfloat{\includegraphics[width=0.32\textwidth]{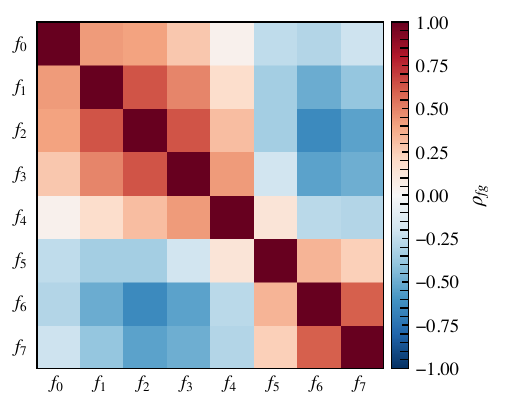}}\hfill
		\subfloat{\includegraphics[width=0.32\textwidth]{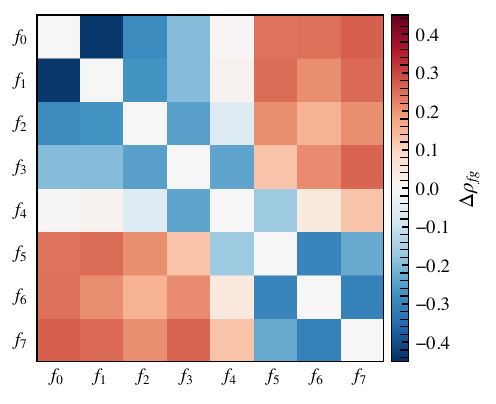}}
		\caption{Correlation recovery at the headline geometry. Left, level-space Pearson matrix of the encoded data ($8\times8$ cells). Center, raw hardware. Right, difference (hardware $-$ data). The raw hardware matrix is structurally exact ($\rfit = 0.989$) and globally amplitude-compressed by a single factor $0.647$ ($R^2 = 0.978$), the depolarizing signature; $\maerho(\mathrm{hw},\mathrm{data}) = 0.207$ raw, with an oracle rescaling bound of $0.075$ against the $0.0515$ encoding floor.}
		\label{fig:hw64-corr}
	\end{figure*}

	Figure~\ref{fig:hw64-corr} shows the correlation recovery. The raw hardware level-Pearson matrix reproduces the structure of the data essentially exactly, with $\rfit(\mathrm{hw}, \mathrm{data}) = 0.989$ on the training split and $0.988$ on the untouched test split, while its amplitudes are globally compressed. A single through-origin slope of $0.647$ explains $R^2 = 0.978$ of the hardware matrix on the encoded data, the signature of a dominantly depolarizing channel, corresponding to an effective error of at most $3.2\times10^{-3}$ per executed CZ including readout (an upper bound, since the deployed model's own amplitude gap contributes to the compression). The raw distance is $\maerho(\mathrm{hw},\mathrm{data}) = 0.207$ (train) and $0.203$ (test), as a diagnostic (not a mitigation, since it uses the target), dividing the hardware matrix by the fitted slope recovers $\maerho = 0.075$. This is worth stating precisely, because it is the strongest form of the deployment claim available here: $0.075$ is, to the quoted digits, the deployed model's own noiseless correlation error ($0.0747$, Sec.~\ref{sec:hw-model}). Once the single global depolarizing factor is divided out, the device contributes no resolvable additional correlation error, and the residual distance to the $0.0515$ encoding floor is the model's, not the machine's.
	
	\subsection{What is and is not claimed}
	\label{sec:hw-claims}
	
	The hardware result is a demonstration of \emph{faithful deployment at the headline geometry}. A single dynamic circuit on $67$ physical qubits reproduces, as raw samples, the per-cell spectra, the total-energy distribution, and, up to a single global compression factor, the full pairwise correlation structure of a real HEP dataset, evaluated split-correct end to end. It is a deployment claim, not a quantum-advantage or sampling-hardness claim, for two independent reasons. First, by the equivalence theorem of Sec.~\ref{sec:dciqp} the generated $\nfeat$-bit marginal is classically mixable by construction, so whatever hardness the task inherits from its single-IQP components attaches equally to the shot-splitting realization. Second, even for the coherent joint distribution, noisy IQP circuits beyond a constant depth of unmitigated noise admit polynomial-time classical sampling~\cite{rajakumar2025noisyiqp}, and no error-mitigation scheme restores sample-level fidelity at these error rates~\cite{quek2024mitigation}. The hardness-relevant experiment for this architecture is the execution of the coherent Walsh-sparse degree-3 circuit with all $\nfeat + a$ bits kept, at error rates where the joint interference structure survives. The compiled artifact ships with the code companion.

	\section{Discussion}\label{sec:discussion}
	
	\subsection{Scope}
	\label{sec:scope}
	
	Four results come out of the work. First, IQP Born machines can be trained on real HEP calorimeter image data at $64$ qubits. With the \psck{} kernel the model reaches within $0.016$ of the encoding-fidelity floor on pairwise correlation reconstruction, averaged across five independent seeds at $\Lcomp = 8$, $1500$ epochs. As far as we are aware this is the largest qubit count at which an IQP Born machine has been fit to real HEP data.
	
	Second, the deferred-measurement \ciqp{} compilation is exact to sub-Monte-Carlo precision and consistent across training seeds, verified by an independent Van den Nest estimator on the compiled $\nfeat + a$ qubit circuit. It provides an explicit quantum-deployment path for the trained \moiqp{} whose joint sampling hardness follows from the standard Bremner--Jozsa--Shepherd argument on the compiled circuit.
	
	Third, neither \psck{} nor the Liu and Wang baseline has a barren plateau in the $16$ to $64$-qubit regime we study, at small-perturbation initialization and on sparse Erd\H{o}s--R\'enyi graph connectivity.
	
	Fourth, the trained-and-compiled model runs on present superconducting hardware at its native geometry. The exact dynamic-circuit realization of the mixture, retrained on the native coupling graph in a degree-3 Walsh-sparse parameterization, reproduces the per-cell spectra, the total-energy distribution, and the full $8 \times 8$ Pearson structure of the calorimeter data ($\rfit = 0.989$, single global compression $0.647$) as raw samples on $67$ physical qubits, evaluated split-correct end to end.
	
	The training objective fixes weight-$\leq 2$ $Z$-correlators, and that determines what the model is certified to reproduce. Whatever higher-order structure the model carries over the $2^{64}$ bitstrings is inherited from the ansatz rather than fitted. The classical diagnostics shown in Sec.~\ref{sec:experiments} do not reach it, they are correlator-based throughout, comprising the weight-$\leq 2$ training targets together with the per-feature marginals recovered exactly by Walsh--Hadamard inversion of intra-feature correlators. Joint observables such as $E_\mathrm{sum}$ require samples, which is what the deployment of Sec.~\ref{sec:hardware} supplies. It is not, however, the only way to obtain them. The gate graphs used here have min-fill treewidth $\leq 24$ (headline, Erd\H{o}s--R\'enyi) and $\leq 3$ (deployed, hardware-native). All are exactly contractible, so correlators and amplitudes are classically cheap throughout, and the deployed graph, whose doubled sampling width is $7$, is additionally exactly \emph{samplable} classically. The hardware result at that geometry should therefore be read as a faithfulness demonstration against an available classical reference rather than as access to an otherwise unreachable distribution. The headline Erd\H{o}s--R\'enyi instances, with doubled width $\approx 43$ to $49$, are not classically samplable at this scale, so for those the deployment remains the only route to full $\nfeat$-bit samples. The price of the hardware route is that the model's higher-order structure and the device's noise enter the measurement together, so cross-feature dependence of weight $\geq 3$ is neither fitted in training nor cleanly separable on hardware at present error rates.
	
	One structural consequence deserves stating, because it constrains the whole train-on-classical, deploy-on-quantum programme on superconducting hardware. Eliminating SWAP overhead requires embedding the gate graph in the device's native coupling map. On heavy-hex that map is planar of degree $\leq 3$, which bounds the treewidth by a small constant, which in turn makes the deployed circuit exactly simulable. Zero-SWAP deployment and sampling hardness are therefore mutually exclusive on this hardware family, independently of qubit count. Hardness-relevant instances need either a routed circuit, whose two-qubit budget we estimate at $\sim\!9.5\times10^{3}$ gates for the monolithic degree-3 object, beyond present fidelity, or a platform with all-to-all connectivity.
	
	The \emph{monolithic} compiled circuit is not the object that runs on hardware. The full-Walsh compilation of the $\Lcomp = 8$ model carries $\sim\!2100$ phase gates of weight up to $2 + a = 5$, which under a CNOT-staircase synthesis is $\sim\!9.5 \times 10^3$ two-qubit gates before routing, and each of the $a = 3$ ancillas couples to $|\mathcal{G}|\,\Lcomp/2 \approx 10^3$ data gates. A monolithic high-fidelity execution is beyond current superconducting hardware. The route taken in Sec.~\ref{sec:hardware} circumvents this budget exactly rather than approximately, since the marginal-equivalence theorem replaces the monolithic circuit by a constant-depth dynamic circuit at single-component gate cost, and hardware-native retraining eliminates routing. The trade-off is explicit in $\Lcomp$. The ancilla register grows only as $\lceil\log_2\Lcomp\rceil$, but the compiled gate weight ($2+a$ full-Walsh, $3$ Walsh-sparse) and the ancilla degree both grow with $\Lcomp$, which favours modest mixtures on near-term devices.
	
	Three conditions bound the hardware record of Sec.~\ref{sec:hw-results}. They comprise a single training seed per hardware configuration, job-to-job calibration drift several times the shot-noise floor (quantified here only as a $2.8\%$ per-bin proxy, pending the same-configuration duplicate job), an on-hardware equivalence \emph{control} (dcIQP versus classical shot-splitting on the same device and calibration) not yet run, and expectation-value-level mitigation deferred to the larger-register campaign.
	
	All experiments reported in this paper use one calorimeter dataset and one Erd\H{o}s--R\'enyi graph family at average degree $6$. Performance on other detector geometries, other datasets with qualitatively different correlation structure, or on denser and more structured gate graphs is under investigation.
	
	At $\nfeat = 64$ the model is close to encoding-floor limited on $\rho$. Increasing the bit depth does not help. The train-split floor is $0.0515$ at $\Bbits = 8$, $10$, and $12$ alike, and equals the $\Bbits \to \infty$ rank limit (the distance between the rank-scale and raw Pearson matrices, $0.0515$) to all quoted digits. The floor is a property of the rank-like quantile transform, not of discretization, so only a different encoding, rather than a finer one, moves it.
	
	\subsection{Relation to existing work}
	
	\subsubsection{Train-on-classical, deploy-on-quantum framework.}
	This work lives inside the framework of Refs.~\cite{rudolph2024trainability,recio2025train_classical_deploy_quantum}, which showed that MMD$^2$ for IQP Born machines decomposes as a classically tractable Pauli-$Z$ mixture and proposed the train-on-classical / deploy-on-quantum workflow. What we add is the \psck{} kernel, the \moiqp{} architecture, the \ciqp{} compilation, the IQP-native sampling-free evaluation protocol, and the HEP calorimeter application. The same train-on-classical, deploy-on-quantum principle has recently been carried beyond IQP circuits to linear-optical and Gaussian-boson-sampling Born machines~\cite{kolarovszki2026gbs}, whose loss is likewise classically evaluable through low-order moments.
	
	\subsubsection{Shallow-IQP graph generators.}
	Ball\'o-Gimbernat et al.~\cite{ballo2025shallow_iqp_graph} ran shallow-IQP generative models for Erd\H{o}s--R\'enyi and bipartite graph distributions on real superconducting hardware at up to $153$ qubits, with Liu and Wang heat-kernel MMD training. Our contributions are complementary to theirs, namely a correlation-aligned kernel (\psck{}), a mixture architecture (\moiqp{}), a deferred-measurement compilation (\ciqp{}), and a continuous-feature imaging application in place of their discrete graph-statistics application. Their observation that local features survive better than global ones at large $\nfeat$ on hardware lines up with ours that low-cumulant marginal structure is captured exactly by weight $\leq 2$ training while higher-cumulant detail is not.
	
	\subsubsection{HEP quantum-circuit Born machines.}
	Refs.~\cite{delgado2022qcbm_hep,kiss2022conditional_born,chang2024qgan_clic} train QCBMs and qGANs on HEP data at 8 to 12 qubits, with 2 or 3 features, on generic variational architectures rather than IQP. This work extends the feature dimensionality by roughly $3\times$ ($2$ to $3 \to 8$ features at $64$ qubits) and the qubit count by $5$ to $8\times$, which is possible because the IQP MMD objective is classically trainable to begin with.
	
	Other recent quantum Born-machine architectures also employ ancilla registers, e.g.\ the scrambling-based construction of P\l{}odzie\'n~\cite{plodzien2026qsbm}, in a setting distinct from the compilation use here.

	\subsection{Future work} \label{sec:future}
	
	The most direct next step is to add weight-3 intra-feature observables to $\mathcal{O}_K$, which fixes the third cumulant of each feature marginal. The cost is $\Dfeat \cdot \binom{\Bbits}{3} = 8 \cdot 56 = 448$ additional observables per epoch at $\Bbits = 8$, and the \psck{} construction extends naturally by appending the corresponding columns to $\Jstar$. Whether this closes the residual mismatch on $f_0$ through $f_2$ is not yet established. A reduced-resolution check at $\Bbits = 4$ (where the cells carry no heavy tail) did not improve the per-feature $W_1$, so the test must be run at the headline $\Bbits = 8$ resolution. If the heat-kernel weight on the weight-3 observables proves too small to drive them, the same \psck{} mechanism applies, namely a tangent-space reweighting toward the tail-relevant correlators. The correlation result on $\rho$ should be unaffected, since the appended Jacobian columns are zero.
	
	On the hardware side, three next steps are concrete. First, the on-hardware deferred-measurement control, namely interleaved dcIQP and classical shot-splitting jobs on the same calibration, tested at the per-$(\ell,\beta)$ level. Two conditions on that comparison are worth recording, because neither is obvious and the second is the binding one. The job-to-job drift measurement of Appendix~\ref{app:hardware} requires the two arms to be scheduled back-to-back. More importantly, they must be \emph{gate-matched}. The deferred-measurement identity is a statement about ideal distributions, whereas a per-$(\ell,\beta)$ two-sample test on measured correlators carries only shot-noise variance, $\sqrt{(1-z^2)/N}$, which at our per-branch $N \approx 6\times10^{3}$ is a few times $10^{-3}$. Any difference in circuit fidelity between the arms enters as a difference of global amplitude factors of order $10^{-1}$, the deployed dcIQP compression is $0.647$ (Sec.~\ref{sec:hw-results}), and would be attributed by such a test to a violation of the identity rather than to depth. The asymmetry is easy to introduce inadvertently, classical shot-splitting admits fractional $ZZ$ rotations on IBM Heron while dynamic circuits do not, so the naive control runs at roughly half the two-qubit count of the object it is meant to control. A meaningful test therefore either disables fractional gates on the mixing arm, or fits and divides out one global factor per arm before comparing, in which case what is tested is the agreement of the correlator \emph{shapes} at the per-$(\ell,\beta)$ level with two scalar nuisance parameters. Second, multi-seed hardware campaigns to put error bars of the Sec.~\ref{sec:experiments} kind on the deployed metrics. Third, and separately from generative deployment, executing the coherent Walsh-sparse degree-3 circuit with all $\nfeat + a$ bits kept, on platforms with native multi-qubit $ZZ$ ladders at $\lesssim 10^{-3}$ error, is the experiment to which hardness questions attach. Its OpenQASM export ships with the code companion.

	The Pearson Jacobian $\Jstar$ of Eq.~\eqref{eq:Jstar-def} is one specific Gauss--Newton linearization of a downstream functional of the model marginals. The same construction goes through for any physically motivated functional whose Jacobian with respect to $\Zth$ admits a closed form. Examples include moment-conditional quantities, mutual-information surrogates, and sliced-Wasserstein direction projections. This is a plausible direction for correlation-aware IQP training in domains beyond HEP.

	\section{Conclusion}\label{sec:conclusion}

	We have trained an IQP Born machine on real HEP calorimeter image data at $64$ qubits. Three pieces go into the result. The \moiqp{} mixture architecture widens the IQP model class without breaking classical trainability. The \psck{} kernel adds a rank-$P$ Pearson-Jacobian correction to the heat-kernel baseline and recovers correlation amplitudes that the baseline cannot. The \ciqp{} compilation maps the trained \moiqp{} to a single IQP circuit on $\nfeat + \lceil \log_2 \Lcomp \rceil$ qubits for quantum deployment. Across five independent training seeds at $\Lcomp = 8$, $1500$ epochs, on the split-correct pipeline, the model reaches $\maerho = 0.068 \pm 0.006$ against a $0.0515$ floor, and generalizes with a per-seed train-test gap of $0.0019 \pm 0.0005$, below the encoding floor's own gap of $0.0033$. Classical references (CorrMSE, Gaussian copula) beat the pairwise metric and are reported with the marginal-fidelity and out-of-sample prices they pay. Per-feature distributional metrics, computed exactly through Walsh--Hadamard inversion of intra-feature $Z$-correlators without sampling, split the eight calorimeter cells cleanly into a near-Gaussian set captured at the single-level scale and a heavy-tailed set whose mismatch traces directly to the absence of weight $\geq 3$ observables in the training objective. A gradient scan from 16 to 64 qubits rules out exponential gradient decay for both \psck{} and the Liu and Wang baseline in the regime we study, and \ciqp{} deployment agrees with its \moiqp{} target below the Monte Carlo noise floor across all five seeds. Finally, the compiled model runs on hardware. Retrained at the identical geometry on the native heavy-hex graph of an IBM Heron~r2 processor in a degree-3 Walsh-sparse parameterization and executed as an exact constant-depth dynamic circuit on $67$ physical qubits, its raw samples reproduce the per-cell spectra, the total-energy distribution, and the full $8\times8$ Pearson structure of the calorimeter data ($\rfit = 0.989$, single global compression $0.647$), evaluated split-correct end to end, as a faithful-deployment result. The \psck{} construction is general. Any physically motivated downstream functional with a closed-form Jacobian on the model marginals fits the same template.

	\section*{Code availability}
	
	The code base relevant to this work is available at \href{https://github.com/jamalslim/ciqp-moiqp}{github.com/jamalslim/ciqp-moiqp} and archived in Ref.~\cite{moiqp_ciqp}. The CLIC calorimeter dataset is also available at Ref.~\cite{calo_data}.

	\acknowledgments
	
	This research was supported in part through the Maxwell computational resources operated at Deutsches Elektronen-Synchrotron DESY (Hamburg, Germany), a member of the Helmholtz Association HGF. The authors acknowledge the support with funds from the Helmholtz Association HGF (Germany), Hamburgische Investitions- und F\"orderbank (IFB) (Germany), European Union's HORIZON MSCA Doctoral Networks program project ENGAGE (101034267), and the Ministry of Science, Research and Culture of the State of Brandenburg within the Center for Quantum Technologies and Applications (CQTA) (Germany). We are grateful to the authors of {\sc IQPopt} \cite{IQPopt2025} for their open-source release of the Van-den-Nest correlator estimator framework, on which our forward-simulation infrastructure depends.

	\appendix
	
	\section{Walsh--Hadamard compilation of \moiqp{} into a single IQP circuit on $\nfeat + a$ qubits} \label{app:ciqp-derivation}
	
	This appendix gives the full derivation of the \ciqp{} construction of Sec.~\ref{sec:ciqp} and Fig.~\ref{fig:ciqp-circuit}. The argument has four parts.
	\begin{enumerate}
		\item the controlled-IQP unitary as a product of ancilla-indexed Pauli-$Z$ rotations
		\item the inverse-WHT relation between base angles and compiled angles
		\item the deferred-measurement equivalence to \moiqp{} through a Walsh-sum identity on the ancilla outcome
		\item the gate-count and hardness analysis.
	\end{enumerate}
	All four parts are cross-checked numerically, which verifies bit-level agreement between \ciqp{} and \moiqp{} marginals up to Monte Carlo noise.
	
	\subsubsection*{Notation and setup.}
	Let the base gate graph be $\mathcal{G} = \{G_1, \dots, G_{|\mathcal{G}|}\}$ with each $G_j \subseteq \{0, \dots, \nfeat-1\}$ a subset of data qubits of weight $|G_j| \in \{1, 2\}$. Denote the single-IQP unitary on parameters $\boldsymbol\theta \in \mathbb{R}^{|\mathcal{G}|}$ by
	\begin{equation}
		U(\boldsymbol\theta) = H^{\otimes\nfeat}\, D(\boldsymbol\theta)\,
		H^{\otimes\nfeat}, \quad
		D(\boldsymbol\theta) = \prod_{j} e^{i\, \theta_j\, Z_{G_j}},
		\label{eq:app-U-single}
	\end{equation}
	with $Z_{G_j} = \prod_{i \in G_j} Z_i$. The \moiqp{} distribution is
	\begin{equation}
		p^{\moiqp{}}(x) = \tfrac{1}{\Lcomp}\sum_{\ell=0}^{\Lcomp-1}
		|\langle x | U(\boldsymbol\theta^{(\ell)}) | 0\rangle|^2.
	\end{equation}
	Throughout this appendix $\Lcomp = 2^a$ is a \emph{genuine restriction} of the construction, not a normalization. Zero-angle padding does not repair it, because an all-zero-angle IQP component is $H^{\otimes \nfeat} I H^{\otimes \nfeat}|0\rangle = |0\rangle$, a delta spike at the all-zeros bitstring rather than a neutral element, so the compiled marginal of a padded list equals the \emph{padded} mixture $\frac{1}{2^a}(\sum_\ell p_\ell + (2^a - \Lcomp)\,\delta_{0\cdots0})$, deviating from the true $\Lcomp$-component mixture by up to $0.18$ in probability at $\Lcomp = 3$ in exact statevector checks. Nor does any other padding exist, since a Hadamard layer cannot prepare a uniform superposition over a non-power-of-two number of ancilla states, so the constraint is imposed by the IQP form itself. The code companion rejects non-power-of-two $\Lcomp$ with a regression-tested guard. All experiments in this paper use $\Lcomp \in \{1, 2, 4, 8\}$. Ancilla qubits are labeled $\nfeat, \nfeat+1, \dots, \nfeat+a-1$. For $S \subseteq \{0, \dots, a-1\}$ with indicator vector $S \in \{0,1\}^a$ we write $Z_S^{\mathrm{anc}} = \prod_{k \in S} Z_{\nfeat+k}$, with $Z_\emptyset^{\mathrm{anc}} = I$.
	
	\subsection{The \ciqp{} unitary as a product of $Z \otimes Z$ rotations}
	Construct the compiled diagonal unitary on $\nfeat + a$ qubits as
	\begin{equation}
		D^{\ciqp{}} \;=\; \prod_{j=1}^{|\mathcal{G}|}
		\prod_{S\subseteq\{0,\dots,a-1\}}
		e^{i\, \tilde\phi_{j,S}\, Z_{G_j}\otimes Z_S^{\mathrm{anc}}},
		\label{eq:app-D-ciqp}
	\end{equation}
	with compiled angles $\{\tilde\phi_{j,S}\}$ to be determined. Every factor is a diagonal exponential of a Pauli-$Z$ string and all such strings commute, so the product is order independent and $D^{\ciqp{}}$ is diagonal in the full computational basis of $\nfeat + a$ qubits. The compiled IQP unitary is
	\begin{equation}
		U^{\ciqp{}} \;=\; H^{\otimes (\nfeat+a)}\, D^{\ciqp{}}\,
		H^{\otimes (\nfeat+a)},
		\label{eq:app-U-ciqp}
	\end{equation}
	which is IQP-form on $\nfeat + a$ qubits by construction.
	
	\subsection{Ancilla-conditional equivalence}
	Fix an ancilla basis state $|\ell\rangle \in \mathbb{C}^{2^a}$ and consider $D^{\ciqp{}} (I^{\otimes\nfeat} \otimes |\ell\rangle\langle\ell|)$. Since $Z_S^{\mathrm{anc}} |\ell\rangle = (-1)^{\ell \cdot S} |\ell\rangle$,
	\begin{align}
		D^{\ciqp{}} \cdot \big(|\phi\rangle_{\mathrm{data}} \otimes |\ell\rangle\big)
		&= \left[\prod_{j,S} e^{i\, \tilde\phi_{j,S}\, (-1)^{\ell \cdot S}\,
			Z_{G_j}}\right] |\phi\rangle_{\mathrm{data}} \otimes |\ell\rangle \nonumber\\
		&= \left[\prod_j e^{i\, \Phi_j(\ell)\, Z_{G_j}}\right]
		|\phi\rangle_{\mathrm{data}} \otimes |\ell\rangle,
		\label{eq:app-ancilla-conditional}
	\end{align}
	with the effective data-register angle
	\begin{equation}
		\Phi_j(\ell) = \sum_{S \subseteq \{0,\dots,a-1\}}
		(-1)^{\ell \cdot S}\, \tilde\phi_{j,S}.
		\label{eq:app-Phi}
	\end{equation}
	Pick $\tilde\phi_{j,S}$ so that $\Phi_j(\ell) = \theta_j^{(\ell)}$ for each $\ell$. Eq.~\eqref{eq:app-Phi} is an invertible Walsh transform in $(\ell, S)$, with inverse
	\begin{equation}
		\tilde\phi_{j,S} \;=\;
		\frac{1}{\Lcomp}\sum_{\ell=0}^{\Lcomp-1}
		(-1)^{\ell \cdot S}\, \theta_j^{(\ell)}
		\label{eq:app-wht}
	\end{equation}
	(the inverse Walsh--Hadamard transform of the per-gate angle sequence across components). Substituting Eq.~\eqref{eq:app-wht} into Eq.~\eqref{eq:app-Phi} and using Walsh orthogonality, $\sum_S (-1)^{(\ell+m) \cdot S} = \Lcomp\, \delta_{\ell, m}$, gives $\Phi_j(\ell) = \theta_j^{(\ell)}$ as required.

	\subsubsection{Worked examples $\Lcomp = 1$, $\Lcomp = 2$, $\Lcomp = 4$.}

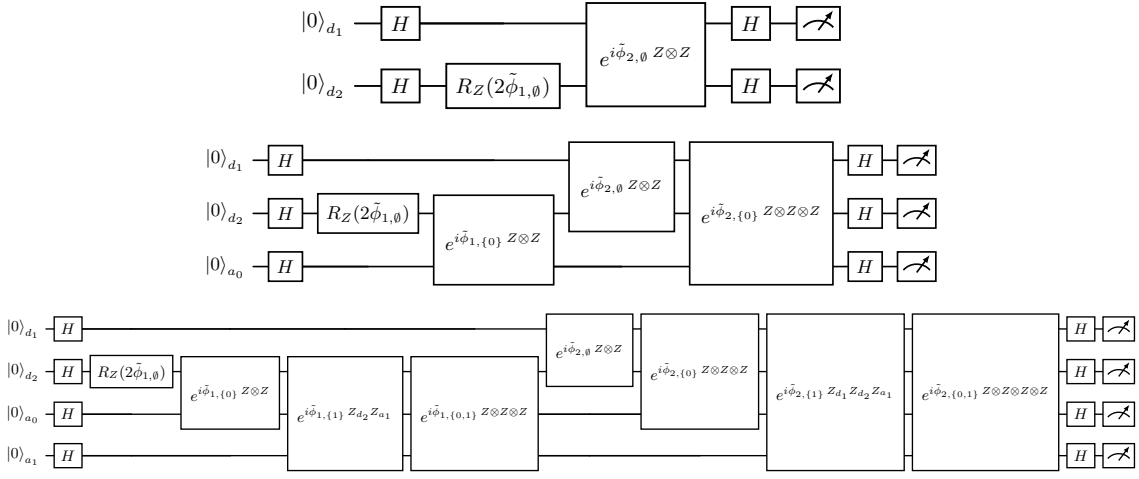
\begin{figure*}[t]
		\centering
%
%

\begin{tikzpicture}
\node[scale=0.88] {%
\begin{quantikz}[column sep=0.40cm, row sep=0.30cm]
  \lstick{$\ket{0}_{d_1}$} & \gate{H} &
      \qw &
      \gate[wires=2]{\,e^{i\tilde\phi_{2,\emptyset}\,Z\otimes Z}\,} &
      \gate{H} & \meter{} \\
  \lstick{$\ket{0}_{d_2}$} & \gate{H} &
      \gate{R_Z(2\tilde\phi_{1,\emptyset})} &
       &
      \gate{H} & \meter{}
\end{quantikz}
};
\end{tikzpicture}

\begin{tikzpicture}
\node[scale=0.78] {%
\begin{quantikz}[column sep=0.26cm, row sep=0.26cm]
  \lstick{$\ket{0}_{d_1}$} & \gate{H} &
      \qw &
      \qw &
      \gate[wires=2]{\,e^{i\tilde\phi_{2,\emptyset}\,Z\otimes Z}\,} &
      \gate[wires=3]{\,e^{i\tilde\phi_{2,\{0\}}\,Z\otimes Z\otimes Z}\,} &
      \gate{H} & \meter{} \\
  \lstick{$\ket{0}_{d_2}$} & \gate{H} &
      \gate{R_Z(2\tilde\phi_{1,\emptyset})} &
      \gate[wires=2]{\,e^{i\tilde\phi_{1,\{0\}}\,Z\otimes Z}\,} &
       &
       &
      \gate{H} & \meter{} \\
  \lstick{$\ket{0}_{a_0}$} & \gate{H} &
      \qw &
       &
      \qw &
       &
      \gate{H} & \meter{}
\end{quantikz}
};
\end{tikzpicture}

\begin{tikzpicture}
\node[scale=0.64] {%
\begin{quantikz}[column sep=0.17cm, row sep=0.24cm]
  \lstick{$\ket{0}_{d_1}$} & \gate{H} &
      \qw & \qw & \qw & \qw &
      \gate[wires=2]{\,e^{i\tilde\phi_{2,\emptyset}\,Z\otimes Z}\,} &
      \gate[wires=3]{\,e^{i\tilde\phi_{2,\{0\}}\,Z\otimes Z\otimes Z}\,} &
      \gate[wires=4]{\,e^{i\tilde\phi_{2,\{1\}}\,Z_{d_1}Z_{d_2}Z_{a_1}}\,} &
      \gate[wires=4]{\,e^{i\tilde\phi_{2,\{0,1\}}\,Z\otimes Z\otimes Z\otimes Z}\,} &
      \gate{H} & \meter{} \\
  \lstick{$\ket{0}_{d_2}$} & \gate{H} &
      \gate{R_Z(2\tilde\phi_{1,\emptyset})} &
      \gate[wires=2]{\,e^{i\tilde\phi_{1,\{0\}}\,Z\otimes Z}\,} &
      \gate[wires=3]{\,e^{i\tilde\phi_{1,\{1\}}\,Z_{d_2}Z_{a_1}}\,} &
      \gate[wires=3]{\,e^{i\tilde\phi_{1,\{0,1\}}\,Z\otimes Z\otimes Z}\,} &
       &
       &
       &
       &
      \gate{H} & \meter{} \\
  \lstick{$\ket{0}_{a_0}$} & \gate{H} &
      \qw &
       &
       &
       &
      \qw & \qw &
       &
       &
      \gate{H} & \meter{} \\
  \lstick{$\ket{0}_{a_1}$} & \gate{H} &
      \qw & \qw &
       &
       &
      \qw & \qw &
       &
       &
      \gate{H} & \meter{}
\end{quantikz}
};
\end{tikzpicture}

%
%
		\caption{Walsh--Hadamard cIQP compilation for the three smallest mixture sizes, on a minimal gate graph with one weight-1 gate $G_1 = \{d_2\}$ and one weight-2 gate $G_2 = \{d_1, d_2\}$. Top, $\Lcomp = 1$. No ancilla, Walsh index $S$ takes only the value $\emptyset$, and Eq.~\eqref{eq:app-wht} gives $\tilde\phi_{j,\emptyset} = \theta_j^{(0)}$, so the compiled circuit is the plain IQP on $\nfeat = 2$ qubits with the base angles. Middle, $\Lcomp = 2$. One ancilla $a_0$, each base gate contributes two compiled gates (the $S = \emptyset$ and $S = \{0\}$ terms of Eq.~\eqref{eq:app-D-ciqp}) with angles given by Eqs.~\eqref{eq:app-wht-L2-even}, \eqref{eq:app-wht-L2-odd}. Bottom, $\Lcomp = 4$. Two ancillas $a_0, a_1$, each base gate contributes four compiled gates over $S \in \{\emptyset, \{0\}, \{1\}, \{0,1\}\}$ with angles from Eq.~\eqref{eq:app-wht-L4}. The two gates with support on $a_1$ but not $a_0$ are drawn with explicit $Z$-subscripted labels to disambiguate their non-contiguous wire support. Compiled-gate count $|\mathcal{G}| \cdot \Lcomp = 2, 4, 8$ respectively, consistent with Eq.~\eqref{eq:app-gate-count}. All three circuits are IQP-form on $\nfeat + a$ qubits (Hadamards to diagonal phase block to Hadamards).}
		\label{fig:ciqp-L1-L2-L4}
	\end{figure*}

	Fig.~\ref{fig:ciqp-L1-L2-L4} shows the compiled circuit in all three cases for a minimal gate graph $\mathcal{G} = \{G_1, G_2\}$ with $G_1 = \{d_2\}$ (weight 1) and $G_2 = \{d_1, d_2\}$ (weight 2).
	
	At $\Lcomp = 1$ there are no ancillas ($a = 0$), the Walsh index set collapses to $S = \emptyset$, and Eq.~\eqref{eq:app-wht} gives $\tilde\phi_{j,\emptyset} = \theta_j^{(0)}$. The compiled circuit is literally the single IQP circuit on $\nfeat$ qubits with the base angles.
	
	At $\Lcomp = 2$ there is $a = 1$ ancilla, the Walsh index set is $S \in \{\emptyset, \{0\}\}$, and each base gate $G_j$ contributes two compiled gates with angles
	\begin{align}
		\tilde\phi_{j,\emptyset}
		&\;=\; \tfrac{1}{2}\big(\theta_j^{(0)} + \theta_j^{(1)}\big),
		\label{eq:app-wht-L2-even}\\
		\tilde\phi_{j,\{0\}}
		&\;=\; \tfrac{1}{2}\big(\theta_j^{(0)} - \theta_j^{(1)}\big),
		\label{eq:app-wht-L2-odd}
	\end{align}
	i.e.\ the even and odd Walsh combinations of the two component angles.

	At $\Lcomp = 4$ there are $a = 2$ ancillas $(a_0, a_1)$, the Walsh index set is $S \subseteq \{0, 1\}$ with four elements $\{\emptyset, \{0\}, \{1\}, \{0,1\}\}$, and each base gate contributes four compiled gates with angles
	\begin{align}
		\tilde\phi_{j,\emptyset}
		&= \tfrac{1}{4}\big(\theta_j^{(0)} + \theta_j^{(1)} + \theta_j^{(2)} + \theta_j^{(3)}\big),\nonumber\\
		\tilde\phi_{j,\{0\}}
		&= \tfrac{1}{4}\big(\theta_j^{(0)} - \theta_j^{(1)} + \theta_j^{(2)} - \theta_j^{(3)}\big),\nonumber\\
		\tilde\phi_{j,\{1\}}
		&= \tfrac{1}{4}\big(\theta_j^{(0)} + \theta_j^{(1)} - \theta_j^{(2)} - \theta_j^{(3)}\big),\nonumber\\
		\tilde\phi_{j,\{0,1\}}
		&= \tfrac{1}{4}\big(\theta_j^{(0)} - \theta_j^{(1)} - \theta_j^{(2)} + \theta_j^{(3)}\big),
		\label{eq:app-wht-L4}
	\end{align}
	from Eq.~\eqref{eq:app-wht} with $\ell \cdot S$ evaluated as the bitwise inner product of the binary encoding of $\ell$ with the indicator vector of $S$. The compiled-gate count grows linearly with $\Lcomp$, with $|\mathcal{G}| \cdot \Lcomp = 2$, $4$, $8$ for $\Lcomp = 1, 2, 4$ respectively, matching Eq.~\eqref{eq:app-gate-count}.
	
	All three panels of Fig.~\ref{fig:ciqp-L1-L2-L4} place the ancillas at the bottom of the wire stack so that every compiled gate whose support is contained in $\{d_1,\dots,d_n,a_0\}$ acts on a contiguous block of wires. The two $\Lcomp = 4$ gates whose support includes $a_1$ but not $a_0$ are drawn as boxes spanning the intermediate wire with their $Z$-string operators written explicitly in subscript form ($Z_{d_2} Z_{a_1}$ and $Z_{d_1} Z_{d_2} Z_{a_1}$), so the wire support is unambiguous even though the box visually covers a pass-through wire.
	
	\subsubsection{Where the summation lives.}
	The MoIQP distribution of Eq.~\eqref{eq:moiqp-pdf} is built around one explicit classical sum $\frac{1}{\Lcomp}\sum_\ell(\cdots)$. That sum does not appear literally anywhere in the \ciqp{} circuits of Fig.~\ref{fig:ciqp-L1-L2-L4}. The Walsh--Hadamard compilation relocates it into three distinct quantum-mechanical operations, distributed along the time-axis of the compiled circuit.
	
	First, the leading Hadamards on the ancilla register, $H^{\otimes a}\,|0\rangle^{\otimes a} = \frac{1}{\sqrt{\Lcomp}}\sum_\ell |\ell\rangle$, \emph{create} the summation as a coherent superposition over the mixture index.
	
	Second, the compiled diagonal $D^{\ciqp{}}$ attaches $D(\boldsymbol\theta^{(\ell)})$ to each branch $|\ell\rangle$. By Eq.~\eqref{eq:app-post-diagonal},
	\begin{equation}
		D^{\ciqp{}}|+\rangle^{\otimes(\nfeat+a)} = \frac{1}{\sqrt{\Lcomp}}\sum_\ell D(\boldsymbol\theta^{(\ell)})\, |+\rangle^{\otimes\nfeat}\otimes|\ell\rangle.
	\end{equation}
	Third, the ancilla measurement together with the discard of the ancilla outcome (the trace over the ancilla register) \emph{collapses} the coherent sum into the classical mixture $\Pr(x) = \frac{1}{\Lcomp}\sum_\ell |A_\ell(x)|^2$, which is Eq.~\eqref{eq:app-marginal}. The Born rule thus performs the $1/\Lcomp$ average at measurement time, and by Eq.~\eqref{eq:app-marginal} a single coherent run with ancilla discard, the dynamic-circuit realization of Sec.~\ref{sec:dciqp}, and per-shot classical component selection all realize the same $\nfeat$-bit distribution. The coherent content of \ciqp{} that no classical mixing reproduces resides exclusively in the joint $(\nfeat+a)$-bit distribution of Eq.~\eqref{eq:app-joint-prob}, which differs from its ancilla-dephased counterpart at total-variation distance $0.326$ with data to ancilla mutual information $0.470$ bits in a dense $\nfeat = 6$, $\Lcomp = 4$ check.
	
	\subsection{Deferred-measurement equivalence}
	Apply $U^{\ciqp{}}$ to $|0\rangle^{\otimes(\nfeat+a)}$. The leading $H^{\otimes(\nfeat+a)}$ prepares $|+\rangle^{\otimes\nfeat} \otimes |+\rangle^{\otimes a} = |+\rangle^{\otimes\nfeat} \otimes \tfrac{1}{\sqrt{\Lcomp}}\sum_\ell |\ell\rangle$. Applying $D^{\ciqp{}}$ and using Eq.~\eqref{eq:app-ancilla-conditional},
	\begin{equation}
		D^{\ciqp{}} |+\rangle^{\otimes(\nfeat+a)}
		= \frac{1}{\sqrt{\Lcomp}}\sum_{\ell=0}^{\Lcomp-1}
		D(\boldsymbol\theta^{(\ell)}) |+\rangle^{\otimes\nfeat} \otimes
		|\ell\rangle.
		\label{eq:app-post-diagonal}
	\end{equation}
	The trailing $H^{\otimes(\nfeat+a)} = H^{\otimes\nfeat} \otimes H^{\otimes a}$ then gives
	\begin{equation}
		U^{\ciqp{}} |0\rangle^{\otimes(\nfeat+a)}
		= \frac{1}{\sqrt{\Lcomp}}\sum_{\ell=0}^{\Lcomp-1}
		U(\boldsymbol\theta^{(\ell)}) |0\rangle^{\otimes\nfeat} \otimes
		H^{\otimes a} |\ell\rangle.
		\label{eq:app-final-state}
	\end{equation}
	Measuring all $\nfeat + a$ qubits in the computational basis yields joint probability
	\begin{equation}
		\Pr(x, y) = \frac{1}{\Lcomp^2}
		\left| \sum_\ell (-1)^{y \cdot \ell}\, A_\ell(x)\right|^2,
		\label{eq:app-joint-prob}
	\end{equation}
	with $A_\ell(x) = \langle x | U(\boldsymbol\theta^{(\ell)}) |0\rangle^{\otimes\nfeat}$. Marginalizing over $y$,
	\begin{align}
		\Pr(x) &= \sum_y \Pr(x, y)
		= \frac{1}{\Lcomp^2}\sum_{\ell, m} A_\ell(x) A_m^*(x)
		\sum_y (-1)^{y \cdot (\ell + m)} \nonumber\\
		&= \frac{1}{\Lcomp^2} \sum_{\ell, m} A_\ell(x) A_m^*(x) \cdot
		\Lcomp\, \delta_{\ell, m} \nonumber\\
		&= \frac{1}{\Lcomp} \sum_\ell |A_\ell(x)|^2 \;=\; P^{\moiqp{}}(x).
		\label{eq:app-marginal}
	\end{align}
	The data-register distribution, averaged over (equivalently, traced over) the ancilla outcome, is exactly the \moiqp{} distribution. This is a marginal statement, not a conditional one. The per-outcome conditional distribution $\Pr(x | y)$ is in general not equal to $p^{\moiqp{}}(x)$ at a fixed $y$, nor does fixing $y$ select a particular component $\ell$. No post-selection on $y$ is required. Operationally, a user of \ciqp{} measures all $\nfeat + a$ qubits, keeps the first $\nfeat$, and discards the last $a$.
	
	\subsection{Gate count and gate weights}
	From Eq.~\eqref{eq:app-D-ciqp} the compiled diagonal contains one $Z \otimes Z$-phase rotation per $(j, S)$ pair. Ignoring accidental zero-angle prunings, the number of compiled gates is
	\begin{equation}
		|\mathcal{G}_{\ciqp{}}| \;=\; |\mathcal{G}| \cdot \Lcomp,
		\label{eq:app-gate-count}
	\end{equation}
	i.e.\ each base gate expands into $\Lcomp$ compiled gates, one per ancilla subset $S$. For the headline $\nfeat = 64$, $\Lcomp = 8$ configuration this gives $|\mathcal{G}_{\ciqp{}}|$ in the range $2048$ to $2296$ across the five seeds, since the base gate count $|\mathcal{G}|$ itself varies from $256$ to $287$ between the seed-specific ER graph realizations. See Appendix~\ref{app:per-seed}. Each compiled gate has weight $|G_j| + |S|$, so compiled gate weights range from $1$ (weight-1 base gate with $S = \emptyset$) to $2 + a = 5$ (weight-2 base gate with $|S| = 3$).
	
	\subsection{Sampling hardness}
	We distinguish two sampling tasks, the joint $(\nfeat + a)$-qubit distribution and the data-only $\nfeat$-qubit marginal. The generative-modeling deployment only ever uses the marginal of Eq.~\eqref{eq:app-marginal}. The joint is not operationally relevant.
	
	For the \emph{joint}, $U^{\ciqp{}}$ is IQP-form on $\nfeat + a$ qubits by construction. The original Bremner--Jozsa--Shepherd~\cite{bremner2011iqp_hardness} argument was stated for IQP circuits with $Z$-polynomial degree at most $3$. Our compiled circuit has maximum $Z$-polynomial degree $2 + a$, which for $a = 3$ is $5$ and falls inside the Bremner--Montanaro--Shepherd extension~\cite{bremner2016bjs}. Classical efficient sampling of the joint $(\nfeat + a)$-qubit output of a \emph{worst-case} family in this class would imply a polynomial hierarchy collapse to its second level under the conjectures of the BMS framework. That statement does not transfer to the instances trained here, and we do not transfer it. Two measurements block it. First, the gate graphs are contractible, min-fill treewidth $\leq 21$ to $24$ for the five headline Erd\H{o}s--R\'enyi instances (graph seeds $43$ to $47$) and $\leq 3$ for the deployed hardware-native graph, so exact classical sampling costs $\nfeat\,2^{\mathcal{O}(w)}$ operations and is immediate at this scale. The ancilla register adds at most a factor $\Lcomp$, since conditioning on the ancilla value decomposes the joint into $\Lcomp$ single-component contractions. Second, the BMS route additionally requires anti-concentration, and the trained instances are strongly concentrated, by Parseval the collision quantity $Z \equiv 2^{\nfeat}\sum_x p(x)^2 = \sum_\beta \langle Z_\beta\rangle^2$ is bounded below for any model matching the data's weight-$\leq 2$ moments by $Z \geq Z^\star = 1 + \sum_{\mathcal{O}_2}\langle Z_\beta\rangle^2_{\mathrm{data}} = 12.16$ on the training split at $\nfeat = 64$ ($12.13$ on the full sample), against the Porter to Thomas value $Z = 2$. Moment matching and anti-concentration are in direct tension, and the hardness-relevant object is therefore not the trained generative model but the separately parameterized coherent degree-3 circuit of Sec.~\ref{sec:future}. This is a conditional statement, tied to whichever hypotheses one needs to apply BMS to the specific gate-weight distribution of our compiled circuit. A tight worst-case hardness theorem for the exact compiled graph at $a = 3$ is not claimed and would be of independent interest. For the \emph{marginal}, sampling hardness of the joint does not transfer automatically to the data-register distribution, since marginalizing over an entangled ancilla can in principle reduce distributional complexity. Hardness of the marginal is a separate question from hardness of the joint, and the compilation settles the operational one. The compiled circuit reduces \moiqp{} marginal sampling to a single quantum-hardware run of an IQP-form circuit with no classical post-mixing, and that each of the $\Lcomp$ underlying single-IQP components admits the standard BJS/BMS hardness argument in isolation. Whether the uniform mixture of $\Lcomp$ BJS-hard distributions is itself BJS-hard is a separate theoretical question, not answered by our construction.
	
	\subsection{Numerical verification}
	
	Two independent numerical routines cover this construction and are provided with the code companion of this paper. One runs exact wavefunction simulation at small $\nfeat$ ($\leq 4$) and verifies machine-precision agreement between the \ciqp{} data-register marginal and the direct \moiqp{} distribution, with measured error $\max_x | \Pr_{\ciqp{}}(x) - p^{\moiqp{}}(x) | < 5 \times 10^{-16}$ for every configuration tested (random graphs, power-of-two $\Lcomp \in \{4, 8\}$), at the floating-point roundoff level. These statevector tests apply to small test circuits only, since exact simulation of the deployed $\geq 67$-qubit circuits is infeasible, and were added to the released test suite in the corrected code revision. The second routine operates at $\nfeat = 64$ through Van den Nest Monte Carlo and compares \ciqp{} and \moiqp{} at the level of low-order $Z$-correlators,
	\begin{equation}
		\mathrm{MAE}\!\left[\{\Zth^{\ciqp{}} - \Zth^{\moiqp{}} : \beta
		\in \mathcal{O}_2\}\right] < 5 \cdot M^{-1/2}.
	\end{equation}
	Across all five training seeds at $\nfeat = 64$, $\Lcomp = 8$ and $M = 2 \times 10^5$ latents, the measured ratio is $\mathrm{MAE}/M^{-1/2} = 0.600 \pm 0.012$ (per-seed $0.5913$, $0.5885$, $0.6182$, $0.6031$, $0.5965$). Because the \ciqp{} and \moiqp{} estimators are evaluated on shared latents, their difference has a variance set by the estimator pair, and the sub-unity ratio is that variance constant under exact equality of the underlying quantities. It does not measure, and could not resolve, any residual compilation error. The operative verification statement is that the disagreement is below the Monte Carlo noise floor for every seed.
	
	\section{Sample-efficiency advantage of base-angle training}
	\label{app:variance-advantage}
	
	Base-angle training is the more sample-efficient route to the trained \ciqp{} circuit. The two are the same model (App.~\ref{app:ciqp-derivation}), so they share optima, and because the base-to-compiled map is an orthogonal Walsh--Hadamard reparameterization they also share loss-landscape conditioning. The advantage is therefore entirely one of \emph{sample efficiency}. The Van den Nest gradient estimator in the compiled coordinates has variance larger by a factor $\Lcomp\,(1 + \sigma_\mu^2/\bar V) \geq \Lcomp$ at matched cost.
	
	\paragraph*{Estimators.}
	Both schemes estimate the same data-register correlator $\Zth = \tfrac{1}{\Lcomp}\sum_\ell \mu_\ell$ with $\mu_\ell \equiv \langle Z_\beta\rangle_{\boldsymbol\theta^{(\ell)}}$. The Van den Nest integrand of a single IQP on $m$ qubits is
	\begin{equation}
		f_\beta(y;\boldsymbol\theta) = \cos\!\Big(2\!\!\sum_{G \in \mathrm{act}(\beta)}\!\! \theta_G\, \chi_G(y)\Big),
		\qquad \chi_G(y) = (-1)^{\bigoplus_{q \in G} y_q},
		\label{eq:va-integrand}
	\end{equation}
	with $\mathbb{E}_{y \sim U(\{0,1\}^m)}[f_\beta] = \Zth$ and $\mathrm{act}(\beta) = \{G \mid |G \cap \beta|\ \text{odd}\}$. The base scheme averages $f_\beta(\,\cdot\,;\boldsymbol\theta^{(\ell)})$ over the $\Lcomp$ components, while the compiled scheme evaluates the single integrand $f^{\ciqp{}}_\beta$ of the $\nfeat + a$ qubit circuit at angles $\tilde\phi = $ WHT$(\boldsymbol\theta)/\Lcomp$ (Eq.~\eqref{eq:app-wht}).
	
	\paragraph*{Lemma (component selection).}
	Split a compiled latent as $(y,w)$ with $y \in \{0,1\}^{\nfeat}$ and $w \in \{0,1\}^a$. For a data-register observable the compiled active set is $\{(j,S) \mid G_j \in \mathrm{act}(\beta),\ S \subseteq \{0,\dots,a-1\}\}$, and because $\beta$ touches no ancilla the signs factorize as $\chi_{G_j \cup S}(y,w) = \chi_{G_j}(y)\,(-1)^{w \cdot S}$. Inserting the compilation relation $\theta^{(\ell)}_{G_j} = \sum_S (-1)^{\ell \cdot S}\, \tilde\phi_{j,S}$ (Eq.~\eqref{eq:app-Phi}) collapses the inner sum over $S$ and gives
	\begin{equation}
		f^{\ciqp{}}_\beta(y,w) \;=\; \cos\!\Big(2\!\!\sum_{j \in \mathrm{act}(\beta)}\!\! \theta^{(w)}_{G_j}\, \chi_{G_j}(y)\Big) \;=\; F_\beta\big(y;\boldsymbol\theta^{(w)}\big).
		\label{eq:va-pointwise}
	\end{equation}
	Conditioning on the ancilla latent $w$ selects component $\ell = w$, and uniform $w$ gives uniform $\ell$. We verify Eq.~\eqref{eq:va-pointwise} numerically to $1.8 \times 10^{-14}$ across weight-1, weight-2, and weight-3 observables. It also re-proves the marginal equivalence at the correlator level, since $\mathbb{E}_{y,w}[f^{\ciqp{}}_\beta] = \tfrac{1}{\Lcomp}\sum_w \mu_w = \Zth$.
	
	\paragraph*{Proposition (variance).}
	Writing $V_\ell = \mathrm{Var}_y\, f_\beta(\,\cdot\,;\boldsymbol\theta^{(\ell)})$, $\bar V = \tfrac{1}{\Lcomp}\sum_\ell V_\ell$, and $\sigma_\mu^2 = \tfrac{1}{\Lcomp}\sum_\ell (\mu_\ell - \Zth)^2$, the law of total variance applied to Eq.~\eqref{eq:va-pointwise} gives the per-sample variance of the compiled integrand,
	\begin{equation}
		V^{\ciqp{}} = \underbrace{\mathbb{E}_w \mathrm{Var}_y(f \mid w)}_{\bar V} + \underbrace{\mathrm{Var}_w \mathbb{E}_y(f \mid w)}_{\sigma_\mu^2} = \bar V + \sigma_\mu^2.
		\label{eq:va-ltv}
	\end{equation}
	The base scheme performs the $\tfrac{1}{\Lcomp}\sum_\ell$ \emph{deterministically}, so the between-component term $\sigma_\mu^2$ is absent from its estimator.
	
	\paragraph*{Corollary (cost to fixed accuracy).}
	A compiled sample costs $|\mathrm{act}^{\ciqp{}}(\beta)| = \Lcomp\,|\mathrm{act}(\beta)|$ cosine terms, since each base gate spawns $2^a = \Lcomp$ compiled copies. To reach estimator variance $\varepsilon$ on $\Zth$, the base scheme (independent latents per component) costs $|\mathrm{act}(\beta)|\,\bar V/\varepsilon$ cosine evaluations and the compiled scheme $\Lcomp\,|\mathrm{act}(\beta)|\,(\bar V + \sigma_\mu^2)/\varepsilon$, a ratio
	\begin{equation}
		\frac{\mathrm{cost}_{\ciqp{}}}{\mathrm{cost}_{\moiqp{}}} = \Lcomp\Big(1 + \frac{\sigma_\mu^2}{\bar V}\Big) \;\geq\; \Lcomp.
		\label{eq:va-ratio}
	\end{equation}
	The gradient integrand obeys the same Eq.~\eqref{eq:va-pointwise} ($\partial_{\tilde\phi}$ replaces the cosine by a sine), so the gradient-estimator variance, which sets the Adam step noise, inflates by the same factor.
	
	\paragraph*{What the factor is, and is not.}
	The factor $\Lcomp = 2^a$ in Eq.~\eqref{eq:va-ratio} is the \emph{ancilla expansion}, the ratio of compiled to base active-set sizes, and we confirm numerically that it equals $\Lcomp$ for every $\beta$, independent of the base-graph density. Sparsity of the base graph fixes the absolute scale $|\mathrm{act}(\beta)|$, and hence the polynomial Van den Nest cost on which the whole training scheme rests, but it does not enter the ratio. The second factor $1 + \sigma_\mu^2/\bar V$ is the price of sampling the mixture index $\ell$ through the ancilla latents rather than summing it. At small-perturbation initialization the per-component correlators $\mu_\ell$ are tightly concentrated about their mean and $\sigma_\mu^2 \approx 0$, so the penalty is close to $\Lcomp$. As the components specialize during training $\sigma_\mu^2$ grows, so the compiled-coordinate penalty \emph{exceeds} $\Lcomp$ late in training, in the correlation sector that \psck{} targets.
	
	\paragraph*{Numerical check.}
	At $\nfeat = 20$ and $\Lcomp = 8$ on a weight-2 observable, the active set grows from $19$ to $152$ (ratio $\Lcomp$), with $\bar V = 0.49999$ and $\sigma_\mu^2 = 1.0 \times 10^{-5}$. Eq.~\eqref{eq:va-ltv} predicts $V^{\ciqp{}} = 0.50003$ against a measured $0.49982$ ($0.04\%$),
	
	\section{Closed-form Pearson Jacobian}\label{app:jacobian-derivation}
	
	In this appendix we derive the closed-form analytical Jacobian $\Jstar_{(fg),\beta} = \partial \rho_{fg} / \partial \Zth\big|_{\Zth = \Zda}$ used in the rank-$P$ correction of \psck{} in Eq.~\eqref{eq:Jstar-def}. The derivation expresses $\rho_{fg}$ as an explicit algebraic function of the low-order $Z$-correlators $\{\Zth : \beta \in \mathcal{O}_2\}$ and applies the chain rule analytically. The intermediate formulas are verified against central-difference finite differences on the training data to a relative precision of $5 \times 10^{-11}$ (code is also supplied), and the closed-form implementation is exercised end to end against central-difference finite differences on the training data at the $10^{-11}$ level.
	
	\subsection{Derivations}
	
	\paragraph{Setup and bit-level expectations.}
	Feature $f \in \{0, \dots, \Dfeat - 1\}$ is encoded as $\Bbits$ bits $b_{f,0}, \dots, b_{f, \Bbits-1} \in \{0, 1\}$ on qubits $fB, fB+1, \dots, fB+\Bbits-1$, with reconstruction
	\begin{equation}
		S_f \;=\; \sum_{k=0}^{\Bbits-1} w_k\, b_{f,k},
		\qquad w_k \;=\; 2^{\Bbits-1-k}
		\label{eq:app-B-Sf}
	\end{equation}
	so that $W \equiv \sum_k w_k = 2^{\Bbits} - 1$. The single-qubit expectation $z_{1;f,k} \equiv \langle Z_{fB+k} \rangle$ and two-qubit expectation $z_{2;(fk),(gl)} \equiv \langle Z_{fB+k}\, Z_{gB+l} \rangle$ give the bit-level moments
	\begin{align}
		\mathbb{E}[b_{f,k}]
		&\;=\; \tfrac{1}{2}(1 - z_{1;f,k}) \;\equiv\; E_{f,k},
		\label{eq:app-B-Eb}\\
		\mathbb{E}[b_{f,k}\, b_{g,l}]
		&\;=\; \tfrac{1}{4}(1 - z_{1;f,k} - z_{1;g,l} + z_{2;(fk),(gl)}),
		\label{eq:app-B-Ebb}
	\end{align}
	valid for $(f, k) \ne (g, l)$. The bit-level covariance and variance follow,
	\begin{align}
		\mathrm{Cov}(b_{f,k}, b_{g,l})
		&\;=\; \mathbb{E}[b_{f,k}\, b_{g,l}] - E_{f,k}\, E_{g,l},\\
		\mathrm{Var}(b_{f,k}) &\;=\; E_{f,k}(1 - E_{f,k}).
	\end{align}
	
	\paragraph{Feature moments as quadratic forms.} By bilinearity,
	\begin{align}
		\mathrm{Cov}[S_f, S_g]
		&\;=\; \sum_{k,l} w_k w_l\, \mathrm{Cov}(b_{f,k}, b_{g,l}),
		\quad (f \ne g)\\
		\mathrm{Var}[S_f]
		&\;=\; \sum_k w_k^2\, \mathrm{Var}(b_{f,k})
		+ 2 \sum_{k < l} w_k w_l\, \mathrm{Cov}(b_{f,k}, b_{f,l}),
		\label{eq:app-B-varSf}
	\end{align}
	and $\rho_{fg} = \mathrm{Cov}[S_f, S_g] / (\sigma_f\, \sigma_g)$ with $\sigma_f = \sqrt{\mathrm{Var}[S_f]}$.
	
	\paragraph{Outer chain rule.}
	\begin{equation}
		\frac{\partial \rho_{fg}}{\partial \mathrm{Cov}[S_f, S_g]}
		= \frac{1}{\sigma_f\, \sigma_g}, \quad
		\frac{\partial \rho_{fg}}{\partial \mathrm{Var}[S_h]}
		= -\frac{\rho_{fg}}{2\, \mathrm{Var}[S_h]}
	\end{equation}
	for $h \in \{f, g\}$, zero otherwise.
	
	\paragraph{Inner chain rule, $z_1$ correlators.} The single-qubit correlator $z_{1;f, k^{\star}}$ appears in the diagonal term $\mathrm{Var}(b_{f, k^{\star}})$ of $\mathrm{Var}[S_f]$, in the intra-feature covariances $\mathrm{Cov}(b_{f, k^{\star}}, b_{f, l})$ for $l \ne k^{\star}$, and in the cross-feature covariances $\mathrm{Cov}(b_{f, k^{\star}}, b_{g, l})$ for $g \ne f$. The elementary derivatives are
	\begin{align}
		\frac{\partial \mathrm{Var}(b_{f, k^{\star}})}{\partial z_{1;f, k^{\star}}}
		&= -\tfrac{1}{2}(1 - 2 E_{f, k^{\star}}), \\
		\frac{\partial \mathrm{Cov}(b_{f, k^{\star}}, b_{f, l})}{\partial z_{1;f, k^{\star}}}
		&= -\tfrac{1}{4} + \tfrac{1}{2} E_{f, l}
		\quad (l \ne k^{\star}),\\
		\frac{\partial \mathrm{Cov}(b_{f, k^{\star}}, b_{g, l})}{\partial z_{1;f, k^{\star}}}
		&= -\tfrac{1}{4} + \tfrac{1}{2} E_{g, l} \quad (g \ne f).
	\end{align}
	Summing the contributions,
	\begin{align}
		\frac{\partial \mathrm{Var}[S_f]}{\partial z_{1;f, k^{\star}}}
		&= -\tfrac{1}{2} w_{k^{\star}}^2 (1 - 2 E_{f, k^{\star}})
		\nonumber\\
		&\qquad + 2 w_{k^{\star}} \sum_{l \ne k^{\star}} w_l
		\!\left(-\tfrac{1}{4} + \tfrac{1}{2} E_{f, l}\right), \\
		\frac{\partial \mathrm{Cov}[S_f, S_g]}{\partial z_{1;f, k^{\star}}}
		&= w_{k^{\star}}\!\left(-\tfrac{W}{4} + \tfrac{1}{2} T^E_g\right),
	\end{align}
	with $T^E_g \equiv \sum_l w_l\, E_{g,l}$. The corresponding derivatives with respect to $z_{1;g, l^{\star}}$ follow by swapping $f \leftrightarrow g$ and $k^{\star} \leftrightarrow l^{\star}$.
	
	\paragraph{Inner chain rule, $z_2$ correlators.} A two-qubit correlator $z_{2;(f,k),(g,l)}$ enters exactly one feature moment depending on whether it is intra- or cross-feature,
	\begin{align}
		\frac{\partial \mathrm{Cov}[S_f, S_g]}{\partial z_{2;(fk),(gl)}}
		&= \tfrac{1}{4}\, w_k w_l \qquad (f \ne g), \\
		\frac{\partial \mathrm{Var}[S_f]}{\partial z_{2;(fk),(fl)}}
		&= \tfrac{1}{2}\, w_k w_l \qquad (k < l).
	\end{align}
	The factor of $2$ in the intra-feature case is the symmetry factor for $(k, l)$ and $(l, k)$ in Eq.~\eqref{eq:app-B-varSf}. All other $z_2$ entries yield zero derivative.
	
	\paragraph{Full Jacobian.} Combining the outer and inner chain rules, for $\beta = \{fB + k^{\star}\}$,
	\begin{multline}
		\frac{\partial \rho_{fg}}{\partial z_{1;f, k^{\star}}}
		\;=\; \frac{w_{k^{\star}}}{\sigma_f\, \sigma_g}
		\!\left(-\tfrac{W}{4} + \tfrac{1}{2} T^E_g\right) \\
		- \frac{\rho_{fg}}{2\, \mathrm{Var}[S_f]}
		\!\left[ -\tfrac{1}{2} w_{k^{\star}}^2 (1 - 2 E_{f, k^{\star}})
		+ 2 w_{k^{\star}} \!\!\sum_{l \ne k^{\star}}\! w_l
		\!\left(-\tfrac{1}{4} + \tfrac{1}{2} E_{f, l}\right)\right],
	\end{multline}
	and analogously for $\beta = \{gB + l^{\star}\}$. For a cross-feature $\beta = \{fB + k, gB + l\}$,
	\begin{equation}
		\frac{\partial \rho_{fg}}{\partial z_{2;(fk),(gl)}}
		\;=\; \frac{w_k\, w_l}{4\, \sigma_f\, \sigma_g}.
	\end{equation}
	For an intra-feature $\beta = \{fB + k, fB + l\}$ with $k < l$,
	\begin{equation}
		\frac{\partial \rho_{fg}}{\partial z_{2;(fk),(fl)}}
		\;=\; -\frac{\rho_{fg}}{2\, \mathrm{Var}[S_f]} \cdot
		\tfrac{1}{2}\, w_k\, w_l,
	\end{equation}
	and similarly for $\beta = \{gB + k, gB + l\}$ with $f \to g$. All other Jacobian entries vanish identically.
	
	\subsection{Complexity}
	The full $(P \times K)$ Jacobian with $P = \Dfeat(\Dfeat - 1)/2$ and $K = \Dfeat \Bbits + \Dfeat \binom{\Bbits}{2} + \binom{\Dfeat}{2} \Bbits^2 = |\mathcal{O}_2|$ is computed in $\mathcal{O}(P B^2 + \Dfeat^2 B^2)$ time via the formulas above. Because $\Jstar$ is evaluated at the data, it is built once at the start of training and reused for every gradient step. At $\Dfeat = 8$, $\Bbits = 8$ this is $28 \times 2080 = 5.8 \times 10^4$ Jacobian entries and the build time is a few milliseconds, negligible against the per-step Van den Nest cost.
	
	\subsection{Numerical verification}
	The verification code runs a central-difference finite-difference verification of the full Jacobian against the training data routine at $\Dfeat = 8$, $\Bbits = 2$, $K = 136$, with relative error $5.08 \times 10^{-11}$, at the noise floor of symmetric finite differences with step $\varepsilon = 10^{-5}$ in double precision. 

	\section{Trainability scan, full data}\label{app:bp-supp}
	
	In this appendix we report the full numerical output of the trainability scan of Sec.~\ref{sec:bp-scan}. What follows is empirical data analysis. We do not prove a barren-plateau-free theorem for either \psck{} or the heat-kernel baseline. Whether the flat \psck{} per-gate-variance scaling we see at $\sigma = 0.1$ persists at larger initialization amplitudes is an open question.
	
	\subsection{Scan protocol}
	For each $\nfeat \in \{16, 24, 32, 48, 64\}$ (feature counts $\Dfeat = 8$ and bit counts $\Bbits \in \{2, 3, 4, 6, 8\}$) we perform the following routine.
	\begin{enumerate}
		\item draw a fresh Erd\H{o}s--R\'enyi gate graph $\mathcal{G}_\nfeat$ at average degree $6$ with graph seed $43$
		\item sample $K_{\mathrm{init}} = 200$ independent parameter vectors $\boldsymbol\theta^{(k)} \sim \mathcal{N}(0, \sigma^2 \mathbb{I})$ at $\sigma = 0.1$
		\item for each initialization, evaluate the full loss gradient $\nabla \mathcal{L}$ via $M_{\mathrm{grad}} = 2048$ Van den Nest latents
		\item record the per-gate gradient component variance, the gradient-norm-squared $\|\nabla\mathcal{L}\|^2$, and the loss value $\mathcal{L}(\boldsymbol\theta^{(k)})$.
	\end{enumerate}
	
	All reported statistics use $K_{\mathrm{init}} = 200$ inits per $\nfeat$, for both \psck{} ($\eta = 5$) and the Liu and Wang baseline. The total sample size underlying the per-gate-variance columns of Table~\ref{tab:bp-pgv} is $K_{\mathrm{init}} \cdot |\mathcal{G}_\nfeat|$, from $12\,800$ at $\nfeat = 16$ to $55\,200$ at $\nfeat = 64$.
	
	\begin{table}[h]
		\centering\small
		\begin{tabular}{r r r r r r r}
			\toprule
			$\nfeat$ & $|\mathcal{G}|$ &
			$\overline{\mathrm{pgv}}^{\psck}$ & s.e.m. & median &
			$\overline{\mathrm{pgv}}^{\mathrm{LW}}$ & median \\
			\midrule
			16 & 64 & $3.40 \!\times\! 10^{3}$ & $\pm 2.9 \!\times\! 10^{2}$ & $3.12 \!\times\! 10^{3}$ & $1.91 \!\times\! 10^{1}$ & $2.12 \!\times\! 10^{1}$ \\
			24 & 85 & $5.96 \!\times\! 10^{3}$ & $\pm 5.8 \!\times\! 10^{2}$ & $4.26 \!\times\! 10^{3}$ & $4.60 \!\times\! 10^{1}$ & $5.40 \!\times\! 10^{1}$ \\
			32 & 123 & $5.99 \!\times\! 10^{3}$ & $\pm 5.4 \!\times\! 10^{2}$ & $3.55 \!\times\! 10^{3}$ & $7.55 \!\times\! 10^{1}$ & $8.48 \!\times\! 10^{1}$ \\
			48 & 195 & $4.76 \!\times\! 10^{3}$ & $\pm 4.3 \!\times\! 10^{2}$ & $2.34 \!\times\! 10^{3}$ & $1.59 \!\times\! 10^{2}$ & $1.80 \!\times\! 10^{2}$ \\
			64 & 276 & $4.09 \!\times\! 10^{3}$ & $\pm 3.4 \!\times\! 10^{2}$ & $1.60 \!\times\! 10^{3}$ & $2.62 \!\times\! 10^{2}$ & $2.88 \!\times\! 10^{2}$ \\
			\bottomrule
		\end{tabular}
		\caption{Per-gate gradient variance $\mathrm{Var}_k[\partial \mathcal{L} / \partial \theta_j]$ at $\sigma = 0.1$. $\overline{\mathrm{pgv}}^{\psck}$ is the mean of $\mathrm{Var}_k[\partial \Lpsck / \partial \theta_j]$ across the $|\mathcal{G}|$ gates. The s.e.m.\ is taken across gates at fixed $\nfeat$. Medians are given for reference (the distributions are right-skewed at small $\nfeat$).}
		\label{tab:bp-pgv}
	\end{table}
	
	\subsection{Polynomial fits}
	We fit each $\mathrm{pgv}(\nfeat)$ curve in Table~\ref{tab:bp-pgv} by OLS on the log to log axes, $\log \mathrm{pgv} = \log A + b \log \nfeat$. The fit parameters,
	\begin{equation}
		\begin{array}{lll}
			\psck{} & : & b = +0.055 \pm 0.255, \quad R^2 = 0.015 \\
			\text{LW} & : & b = +1.874 \pm 0.049, \quad R^2 = 0.998.
		\end{array}
	\end{equation}
	The \psck{} slope is indistinguishable from zero at this scan resolution ($|b|/\mathrm{s.e.} = 0.22$). The per-gate variance is consistent with constant behavior across $\nfeat \in \{16, \dots, 64\}$ and the residuals are dominated by graph-realization fluctuations. The Liu and Wang slope is precisely determined ($|b|/\mathrm{s.e.} = 38.2$) and corresponds to a clean polynomial growth $\mathrm{Var} \propto \nfeat^{1.87}$.
	
	\subsection{Ruling out exponential decay}
	
	A barren plateau is canonically an exponential decay $\mathrm{Var} \propto a\, e^{-c\nfeat}$ with $c > 0$. Fitting this hypothesis to the same data gives $c_{\psck} = 7 \times 10^{-4}$ (indistinguishable from zero, $R^2 = 0.003$) and $c_{\mathrm{LW}} = -5.2 \times 10^{-2}$ (negative, i.e.\ growth, which is just polynomial scaling re-fit). Comparison by residual sum-of-squares shows that for LW the polynomial fit beats the exponential by a factor of $26$ ($0.009$ vs $0.229$). For \psck{} all three hypotheses (polynomial, exponential, constant) fit equally well because none of them explains the small residual variance. The canonical barren-plateau signature is not present for either loss in this initialization regime.
	
	\subsection{Gradient-norm and loss-value scaling}
	Table~\ref{tab:bp-gn-loss} lists the total gradient norm squared $\|\nabla\mathcal{L}\|^2$ and the mean loss value $\overline{\mathcal{L}}$ across the same $K_{\mathrm{init}} = 200$ inits. Log to log fits give $\|\nabla \Lpsck\|^2 \propto \nfeat^{1.14 \pm 0.20}$ and $\overline{\Lpsck} \propto \nfeat^{0.88 \pm 0.13}$ for \psck{}, and $\|\nabla \mathcal{L}_{\mathrm{LW}}\|^2 \propto \nfeat^{2.95 \pm 0.01}$ and $\overline{\mathcal{L}_{\mathrm{LW}}} \propto \nfeat^{1.80 \pm 0.02}$ for Liu and Wang.
	
	\begin{table}[h]
		\centering\small
		\begin{tabular}{r r r r r}
			\toprule
			$\nfeat$ & $\overline{\|\nabla\Lpsck\|^2}$ & $\overline{\Lpsck}$
			& $\overline{\|\nabla\mathcal{L}_{\mathrm{LW}}\|^2}$
			& $\overline{\mathcal{L}_{\mathrm{LW}}}$ \\
			\midrule
			16 & $2.17 \!\times\! 10^{5}$ & $4.22 \!\times\! 10^{2}$ & $1.22 \!\times\! 10^{3}$ & $3.28 \!\times\! 10^{1}$ \\
			24 & $5.06 \!\times\! 10^{5}$ & $7.69 \!\times\! 10^{2}$ & $3.91 \!\times\! 10^{3}$ & $7.05 \!\times\! 10^{1}$ \\
			32 & $7.37 \!\times\! 10^{5}$ & $9.94 \!\times\! 10^{2}$ & $9.28 \!\times\! 10^{3}$ & $1.15 \!\times\! 10^{2}$ \\
			48 & $9.28 \!\times\! 10^{5}$ & $1.31 \!\times\! 10^{3}$ & $3.10 \!\times\! 10^{4}$ & $2.39 \!\times\! 10^{2}$ \\
			64 & $1.13 \!\times\! 10^{6}$ & $1.46 \!\times\! 10^{3}$ & $7.24 \!\times\! 10^{4}$ & $4.02 \!\times\! 10^{2}$ \\
			\bottomrule
		\end{tabular}
		\caption{Mean gradient-norm-squared and mean loss value over $K_{\mathrm{init}} = 200$ random initializations, as a function of $\nfeat$, for \psck{} and Liu and Wang.}
		\label{tab:bp-gn-loss}
	\end{table}
	
	\subsection{Absolute versus relative gradient strength}
	The trainability-relevant comparison between \psck{} and Liu and Wang is the scale-invariant relative gradient strength $\|\nabla\mathcal{L}\|^2 / \mathcal{L}^2$ (Table~\ref{tab:bp-relative}), since \psck{} and Liu and Wang have different loss scales (Table~\ref{tab:bp-gn-loss}, columns 3 and 5) and Adam-type optimizers rescale step sizes by running second-moment estimates. The absolute per-gate gradient advantage of Table~\ref{tab:bp-pgv}, ranging from $178\times$ at $\nfeat = 16$ down to $15.6\times$ at $\nfeat = 64$, reflects only this loss-scale difference.
	
	\begin{table}[h]
		\centering\small
		\begin{tabular}{r c c c}
			\toprule
			$\nfeat$ & $\|\nabla\Lpsck\|^2/\Lpsck^{\,2}$
			& $\|\nabla\mathcal{L}_{\mathrm{LW}}\|^2/\mathcal{L}_{\mathrm{LW}}^{\,2}$
			& ratio $\psck{}$/LW \\
			\midrule
			16 & 1.22 & 1.13 & $1.08 \times$ \\
			24 & 0.86 & 0.79 & $1.09 \times$ \\
			32 & 0.75 & 0.71 & $1.06 \times$ \\
			48 & 0.54 & 0.54 & $0.99 \times$ \\
			64 & 0.53 & 0.45 & $1.18 \times$ \\
			\bottomrule
		\end{tabular}
		\caption{Relative gradient strength $\|\nabla\mathcal{L}\|^2 / \mathcal{L}^2$ at $\sigma = 0.1$, $K_{\mathrm{init}} = 200$. The two losses agree to within $\sim\! 20\%$ across all $\nfeat$ tested.}
		\label{tab:bp-relative}
	\end{table}
	
	The \psck{}-over-LW gradient advantage in absolute terms ($15 \times$ to $178 \times$) is essentially a rescaling artifact. In dimensionless terms the two losses have nearly identical gradient-to-loss ratios ($1.0 \times$ to $1.2 \times$) across all $\nfeat$ tested. The optimization benefit of \psck{} comes from gradient direction, not gradient magnitude. The rank-$P$ correction $\eta\, \Jstar^{\!\top}\Jstar$ aligns the descent direction with the correlation-error subspace, so that the same gradient flow acts more efficiently on the quantities relevant to $\maerho$. This is consistent with the observation in Fig.~\ref{fig:rho-err-overlay} that \psck{} keeps descending toward the encoding-fidelity floor while Liu and Wang plateaus at $\maerho \approx 0.10$.
	
	\subsection{Scope of the scan}
	Four conditions define the regime the scan covers. First, it is a single-component diagnostic, run at $\Lcomp = 1$ (one IQP component, no mixture), which isolates the per-gate gradient statistics of the objective from mixture averaging. Second, single graph realization per $\nfeat$. Each $\nfeat$ uses one ER graph at graph-seed $43$, so graph-to-graph fluctuations at fixed $(\nfeat, \langle k \rangle)$ are not sampled. The residual variance in the \psck{} pgv fit ($R^2 = 0.015$) is plausibly dominated by that source. A multi-graph extension would tighten the empirical slope bound on \psck{}, though we would not expect it to change the qualitative ``no exponential decay'' finding.
	
	Third, small-$\sigma$ initialization regime only. The scan is at $\sigma = 0.1$. Lerch et al.~\cite{kumar2026data_dependent_init} show that MMD-type losses hit barren plateaus at $\sigma \sim \pi/2$ (full-angle initialization), which we stay away from. Extrapolation of our $\sigma = 0.1$ results to other amplitudes is not supported by the data we show.
	
	Fourth, fixed $\Lcomp$. The scan runs at $\Lcomp = 1$. The $\Lcomp$-dependence of the gradient-variance scaling is not tested here. For large $\Lcomp$ the \moiqp{} loss becomes a sum of $\Lcomp$ single-IQP gradient variances under independent parameter sampling, for which a $\sqrt{\Lcomp}$ tightening of concentration would naively be expected, but we have not run that experiment.

	\section{Per-seed convergence diagnostics}\label{app:per-seed}
	
	This appendix provides the full per-seed numerical output of the headline $\Dfeat = 8$, $\Bbits = 8$, $\Lcomp = 8$, $1500$-epoch sweep summarized in Sec.~\ref{sec:experiments}, together with a secondary $\Lcomp = 4$, $800$-epoch sweep that serves as a reproducibility cross-check at a reduced configuration. Table~\ref{tab:per-seed-L8} reports training-split and held-out test-split metrics side by side for each of the five seeds at the headline configuration.
	
	\begin{table*}[t]
		\centering\small
		\begin{tabular}{r c c c | c c c c}
			\toprule
			& \multicolumn{3}{c|}{training split (best-loss checkpoint)}
			& \multicolumn{4}{c}{held-out test split} \\
			seed & $\maerho$ & $\rfit$ & $\maez$
			& $\maerho$ & $\rfit$ & $\maez$ & gap \\
			\midrule

			42 & $0.0648$ & $0.9901$ & $0.0112$ & $0.0669$ & $0.9890$ & $0.0154$ & $+0.0022$ \\
			43 & $0.0610$ & $0.9911$ & $0.0104$ & $0.0624$ & $0.9900$ & $0.0145$ & $+0.0014$ \\
			44 & $0.0770$ & $0.9873$ & $0.0124$ & $0.0785$ & $0.9861$ & $0.0164$ & $+0.0015$ \\
			45 & $0.0710$ & $0.9886$ & $0.0116$ & $0.0727$ & $0.9874$ & $0.0158$ & $+0.0016$ \\
			46 & $0.0651$ & $0.9894$ & $0.0103$ & $0.0676$ & $0.9884$ & $0.0143$ & $+0.0025$ \\
			\hline
			mean $\pm$ std & $0.068\pm0.006$ & $0.9893\pm0.0014$ & $0.0112\pm0.0008$ & $0.070\pm0.006$ & $0.9882\pm0.0015$ & $0.0153\pm0.0009$ & $0.0019\pm0.0005$ \\

			\bottomrule
		\end{tabular}
		\caption{Per-seed metrics at the headline configuration $\Lcomp = 8$, $\nfeat = 64$, $1500$ epochs. Left block, training split at the best-training-loss checkpoint (epochs $1481$ to $1494$ across seeds), not the final epoch. Right block, held-out test split. All entries are evaluated by Van den Nest Monte Carlo at $M = 10^{5}$ latents, a noise floor of $M^{-1/2} \approx 3 \times 10^{-3}$. Gap is $\maerho^{\mathrm{test}} - \maerho^{\mathrm{train}}$.}
		\label{tab:per-seed-L8}
	\end{table*}
	
	Training time averages $158.1 \pm 4.2$ min per seed (total sweep $13.2$ CPU-hours).
	
\begin{figure}[t]
		\centering
		\includegraphics[width=1\columnwidth]{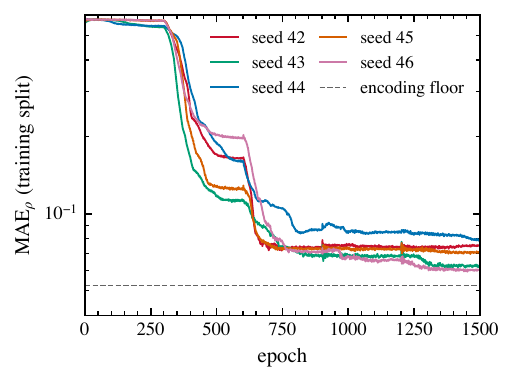}
		\caption{Per-seed training trajectories at $\Lcomp = 8$, $1500$ epochs (headline configuration). The dashed line marks the training-split encoding-fidelity floor ($0.0515$). The sweep is tightly converged by epoch~$1500$.}
		\label{fig:per-seed-trajectories}
	\end{figure}

	\paragraph*{Convergence status.}
	Fig.~\ref{fig:per-seed-trajectories} shows the $\Lcomp = 8$ training trajectories. Defining a convergence indicator $\Delta \equiv \overline{\maerho}_{\,\mathrm{ep}\,1401\text{--}1450} - \overline{\maerho}_{\,\mathrm{ep}\,1451\text{--}1500}$ (positive values indicate the trajectory is still descending), seeds~42, 43, 45, and 46 all satisfy $\Delta \leq 0.0001$ (fully plateaued). Seed~44 has $\Delta = 0.0024$ (still descending mildly but within the same plateau region). All five seeds cross $\maerho < 0.10$ by epoch~$768$. Four out of five cross $\maerho < 0.08$ by epoch~$726$, and seed~44 crosses the same threshold at epoch~$1458$. All five sit on their asymptotic plateau by epoch~$1500$.
	
	\paragraph*{\ciqp{} verification per seed.}
	Running \ciqp{} deployment at $M_{\mathrm{verify}} = 2 \times 10^{5}$ Van den Nest latents on each seed's trained model, the cIQP-vs-\moiqp{} $Z$-correlator MAE divided by the Monte Carlo noise floor $M^{-1/2} \approx 2.24 \times 10^{-3}$ averages $0.600 \pm 0.012$ across the five seeds on the corrected reruns.
	
	\paragraph*{Reduced-configuration reproducibility check.}
	For completeness, Table~\ref{tab:per-seed-L4} reports the per-seed metrics at a reduced configuration ($\Lcomp = 4$, $800$ epochs, same $\nfeat$ and same seeds). This is a reproducibility cross-check, not the paper's headline.
	
	\begin{table*}[t]
		\centering\small
		\begin{tabular}{r c c c c c}
			\toprule
			seed & $\maerho$ & $\rfit$ & $\maez$ & final loss & train time \\
			\midrule
			42 & 0.0676 & 0.9888 & 0.0145 & $1.33$ & $28.4$ min \\
			43 & 0.0629 & 0.9908 & 0.0145 & $1.40$ & $28.0$ min \\
			44 & 0.0942 & 0.9830 & 0.0196 & $3.95$ & $30.6$ min \\
			45 & 0.0816 & 0.9891 & 0.0163 & $1.92$ & $30.1$ min \\
			46 & 0.1232 & 0.9796 & 0.0158 & $2.13$ & $27.5$ min \\
			\midrule
			mean $\pm$ std
			& $0.086 \!\pm\! 0.024$ & $0.986 \!\pm\! 0.005$
			& $0.016 \!\pm\! 0.002$ & $2.15 \!\pm\! 1.06$
			& $28.9 \!\pm\! 1.4$ min \\
			\bottomrule
		\end{tabular}
		\caption{Per-seed final-epoch training-split metrics for the reduced-configuration reproducibility sweep at $\Lcomp = 4$, $800$ epochs, $\nfeat = 64$. Included as a reference point for the compute/quality tradeoff at smaller mixture size.}
		\label{tab:per-seed-L4}
	\end{table*}

	\bibliographystyle{apsrev4-2}
	\section{Hardware deployment details}\label{app:hardware}
	
	This appendix records the details of the Sec.~\ref{sec:hardware} deployment needed to reproduce every quoted number from the shipped shot record.
	
	\subsection{Job inventory}
	
	One dcIQP job on \texttt{ibm\_kingston} uses $67$ physical qubits ($\nfeat = 64$ data, $a = 3$ ancilla), $\Lcomp = 8$ feedforward branches, $5\times10^{4}$ shots, shot-aligned $(\ell, x)$ record. Executed two-qubit budget $138$ CZ per shot, zero SWAP.
	
	\subsection{Intra-job stability}
	
	Splitting the record into its first and second $2.5\times10^{4}$ shots and comparing per-branch, per-observable correlators with binomial shot variances gives $\chi^2/\mathrm{dof} = 1.17$ over $2\,912$ comparisons. The job is internally stable at the shot-noise level, and per-bin histogram spreads between the halves are consistent with pure shot noise. Any drift relevant to reproducing this record is therefore a between-calibration effect, invisible within a single execution window. Its magnitude for this configuration awaits the duplicate job, and the $2.8\%$ per-bin proxy used in the figure error bars is taken from repeated executions of the same protocol class on the same device generation.
	
	\subsection{Ancilla-label diagnostics}
	
	The measured mixture labels deviate from uniform by total variation $0.024$, consistent with per-ancilla assignment asymmetry rather than circuit error, since the ancillas are measured immediately after a single Hadamard, before any two-qubit gate acts. Because the generative marginal averages over $\ell$, a label bias $\delta_\ell$ perturbs pooled correlators only at order $\sum_\ell \delta_\ell (z^{(\ell)}_\beta - \bar z_\beta)$, below the observed correlator scatter.
	
	\subsection{Mitigation status}
	
	No error mitigation of any kind is applied to the numbers reported in Sec.~\ref{sec:hw-results}. No readout unfolding~\cite{nation2021m3} and none of the expectation-value techniques surveyed in Ref.~\cite{cai2023mitigation}. A related caveat applies to the deferred-measurement control of Sec.~\ref{sec:future}. The two arms are not interchangeable at the level of executed gates. The classical-mixing arm admits fractional $ZZ$ rotations, which dynamic circuits on this device generation do not, so unless fractional gates are explicitly disabled the control executes about half the two-qubit budget of dcIQP. Because the per-$(\ell,\beta)$ statistic carries shot-noise variance only, that difference is not absorbed anywhere in the test and would present as an apparent falsification of an identity that Appendix~\ref{app:ciqp-derivation} establishes exactly. We record this as a design requirement rather than a result, since the control has not been run under gate-matched conditions.
	
	The weight-resolved mitigation of this program fits per-weight fidelity factors $F_w$ against the Van den Nest ideal correlators of the deployed model and divides the measured correlators by $F_{|\beta|}$. It is model-referenced by construction and is therefore not applied in this paper, whose deployed-model parameter record is reported separately with the larger-register campaign. The quoted oracle rescaling ($0.647$ slope, $\maerho = 0.075$) bounds what it can achieve on this record.

\end{document}